\documentclass[aps,prd]{revtex4}
\usepackage{graphicx}

\begin{document}
\title{Exotic solutions in General Relativity:\\
Traversable wormholes and ``warp drive'' spacetimes}


\author{Francisco S. N. Lobo}%
\email{flobo@cosmo.fis.fc.ul.pt} \affiliation{Centro de Astronomia
e Astrof\'{\i}sica da Universidade de Lisboa, Campo Grande, Ed. C8
1749-016 Lisboa, Portugal}
%
\affiliation{Institute of Gravitation \& Cosmology, University of
Portsmouth, Portsmouth PO1 2EG, UK}

\date{\today}

\begin{abstract}

The General Theory of Relativity has been an extremely successful
theory, with a well established experimental footing, at least for
weak gravitational fields. Its predictions range from the
existence of black holes, gravitational radiation to the
cosmological models, predicting a primordial beginning, namely the
big-bang. All these solutions have been obtained by first
considering a plausible distribution of matter, i.e., a plausible
stress-energy tensor, and through the Einstein field equation, the
spacetime metric of the geometry is determined.
However, one may solve the Einstein field equation in the reverse
direction, namely, one first considers an interesting and exotic
spacetime metric, then finds the matter source responsible for the
respective geometry. In this manner, it was found that some of
these solutions possess a peculiar property, namely ``exotic
matter,'' involving a stress-energy tensor that violates the null
energy condition. These geometries also allow closed timelike
curves, with the respective causality violations. Another
interesting feature of these spacetimes is that they allow
``effective'' superluminal travel, although, locally, the speed of
light is not surpassed.
These solutions are primarily useful as ``gedanken-experiments''
and as a theoretician's probe of the foundations of general
relativity, and include traversable wormholes and superluminal
``warp drive'' spacetimes. Thus, one may be tempted to denote
these geometries as ``exotic'' solutions of the Einstein field
equation, as they violate the energy conditions and generate
closed timelike curves.
In this article, in addition to extensively exploring interesting
features, in particular, the physical properties and
characteristics of these ``exotic spacetimes,'' we also analyze
other non-trivial general relativistic geometries which generate
closed timelike curves.

\end{abstract}

\maketitle


\tableofcontents


\def\d{{\mathrm{d}}}

\def\ii{{\hat\imath}}
\def\jj{{\hat\jmath}}
\def\kk{{\hat k}}
\def\lll{{\hat l}}
\def\tt{{\hat t}}
\def\xx{{\hat x}}
\def\yy{{\hat y}}
\def\zz{{\hat z}}


\section{Introduction}

\subsection{Review of wormhole physics}

Wormholes act as tunnels from one region of spacetime to another,
possibly through which observers may freely traverse. Interest in
traversable wormholes, as hypothetical shortcuts in spacetime, has
been rekindled by the classical paper by Morris and Thorne
\cite{Morris}. It was first introduced as a tool for teaching
general relativity, as well as an allurement to attract young
students into the field, but it also served to stimulate research
in several branches. These developments culminated with the
publication of the book {\it Lorentzian Wormholes: From Einstein
to Hawking} by Visser \cite{Visser}, where a review on the subject
up to 1995, as well as new ideas are developed and hinted at.
However, it seems that wormhole physics can originally be traced
back to Flamm in 1916~\cite{Flamm}, when he analyzed the then
recently discovered Schwarzschild solution. Paging through history
one finds next that wormhole-type solutions were considered, in
1935, by Einstein and Rosen~\cite{Einstein-Rosen}, where they
constructed an elementary particle model represented by a
``bridge'' connecting two identical sheets. This mathematical
representation of physical space being connected by a
wormhole-type solution was denoted an ``Einstein-Rosen bridge''.
The field laid dormant, until Wheeler revived the subject in the
1950s. Wheeler considered wormholes, such as Reissner-Nordstr\"om
or Kerr wormholes, as objects of the quantum foam connecting
different regions of spacetime and operating at the Planck
scale~\cite{geons,wheeler1}, which were transformed later into
Euclidean wormholes by Hawking \cite{hawking0} and others.
However, these Wheeler wormholes were not traversable, and
furthermore would, in principle, develop some type of singularity
\cite{geroch}. These objects were obtained from the coupled
equations of electromagnetism and general relativity and were
denoted ``geons'', i.e., gravitational-electromagnetic entities.

Geons possess curious properties such as: firstly, the
gravitational mass originates solely from the energy stored in the
electromagnetic field, and in particular, there are no material
masses present (this gave rise to the term ``mass without mass'');
and secondly, no charges are present (``charge without charge'').
These entities were further explored by several authors in
different contexts \cite{geons1}, but due to the extremely
ambitious program and the lack of experimental evidence soon died
out. Nevertheless, it is interesting to note that Misner inspired
in Wheeler's geon representation, found wormhole solutions to the
source-free Einstein equations in 1960~\cite{Misner-worm}. With
the introduction of multi-connected topologies in physics, the
question of causality inevitably arose, as a light signal
travelling through the short-cut, i.e., the wormhole, could
outpace another light signal. Thus, Wheeler and Fuller examined
this situation in the Schwarzschild solution and found that
causality is preserved~\cite{Fuller-Wheeler}, as the Schwarzschild
throat pinches off in a finite time, preventing the traversal of a
signal from one region to another through the wormhole. However,
Graves and Brill~\cite{Graves-Brill}, considering the
Reissner-Nordstr\"{o}m metric also found wormhole-type solutions
between two asymptotically flat spaces, but with an electric flux
flowing through the wormhole. They found that the region of
minimum radius, the ``throat'', contracted, reaching a minimum and
re-expanded after a finite proper time, rather than pinching off
as in the Schwarzschild case. The throat, ``cushioned'' by the
pressure of the electric field through the throat, pulsated
periodically in time. The modern renaissance of wormhole physics
was mainly brought about by the classic Morris-Thorne
paper~\cite{Morris}. Thorne together with his student Morris
\cite{Morris}, understanding that the energy conditions lay on
shaky ground \cite{wheelerredbook,zeldovichnovikovbook},
considered that wormholes, with two mouths and a throat, might be
objects of nature, as stars and black holes are.

Wormhole physics is a specific example of adopting the reverse
philosophy of solving the Einstein field equation, by first
constructing the spacetime metric, then deducing the stress-energy
tensor components. Thus, it was found that these traversable
wormholes possess a stress-energy tensor that violates the null
energy condition \cite{Morris,Visser}. In fact, they violate all
the known pointwise energy conditions and averaged energy
conditions, which are fundamental to the singularity theorems and
theorems of classical black hole thermodynamics. The weak energy
condition (WEC) assumes that the local energy density is
non-negative and states that $T_{\mu\nu}U^\mu U^\nu \geq 0$, for
all timelike vectors $U^\mu$, where $T_{\mu\nu}$ is the stress
energy tensor (in the frame of the matter this amounts to
$\rho\geq 0$ and $\rho+p\geq 0$). By continuity, the WEC implies
the null energy condition (NEC), $T_{\mu\nu}k^\mu k^\nu \geq 0$,
where $k^\mu$ is a null vector. The null energy condition is the
weakest of the energy conditions, and its violation signals that
the other energy conditions are also violated. Although classical
forms of matter are believed to obey these energy
conditions~\cite{hawkingellis}, it is a well-known fact that they
are violated by certain quantum fields, amongst which we may refer
to the Casimir effect and Hawking evaporation (see
\cite{klinkhammer} for a short review). It was further found that
for quantum systems in classical gravitational backgrounds the
weak or null energy conditions could only be violated in small
amounts, and a violation at a given time through the appearance of
a negative energy state, would be overcompensated by the
appearance of a positive energy state soon after. Thus, violations
of the pointwise energy conditions led to the averaging of the
energy conditions over timelike or null geodesics
\cite{Tipler-AEC,roman1}. For instance, the averaged weak energy
condition (AWEC) states that the integral of the energy density
measured by a geodesic observer is non-negative, i.e., $\int
T_{\mu\nu}U^\mu U^\nu \,d\tau \geq 0$, where $\tau$ is the
observer's proper time. Thus, the averaged energy conditions
permit localized violations of the energy conditions, as long as
they hold when averaged along a null or timelike
geodesic~\cite{Tipler-AEC}.

Pioneering work by Ford in the late 1970's on a new set of energy
constraints \cite{ford1}, led to constraints on negative energy
fluxes in 1991 \cite{ford2}. These eventually culminated in the
form of the Quantum Inequality (QI) applied to energy densities,
which was introduced by Ford and Roman in 1995 \cite{fordroman1}.
The QI was proven directly from Quantum Field Theory, in
four-dimensional Minkowski spacetime, for free quantized, massless
scalar fields. The inequality limits the magnitude of the negative
energy violations and the time for which they are allowed to
exist, yielding information on the distribution of the negative
energy density in a finite neighborhood
\cite{fordroman1,fordroman2,fordroman3,PFord}. The basic
applications to curved spacetimes is that these appear flat if
restricted to a sufficiently small region. The application of the
QI to wormhole geometries is of particular interest
\cite{fordroman2,Kuhfittig2}. A small spacetime volume around the
throat of the wormhole was considered, so that all the dimensions
of this volume are much smaller than the minimum proper radius of
curvature in the region. Thus, the spacetime can be considered
approximately flat in this region, so that the QI constraint may
be applied. The results of the analysis is that either the
wormhole possesses a throat size which is only slightly larger
than the Planck length, or there are large discrepancies in the
length scales which characterize the geometry of the wormhole. The
analysis imply that generically the exotic matter is confined to
an extremely thin band, and/or that large red-shifts are involved,
which present severe difficulties for traversability, such as
large tidal forces \cite{fordroman2}. Due to these results, Ford
and Roman concluded that the existence of macroscopic traversable
wormholes is very improbable (see \cite{World-Scientific,Roman}
for a review). It was also shown that, by using the QI, enormous
amounts of exotic matter are needed to support the Alcubierre warp
drive and the superluminal Krasnikov tube
\cite{Springer,PfenningF,Everett}.

Relative to the energy conditions, the situation has changed
drastically, as it has been now shown that even classical systems,
such as those built from scalar fields non-minimally coupled to
gravity, violate all the energy conditions \cite{barcelovisser1}.
It is interesting to note that recent observations in cosmology
strongly suggest that the cosmological fluid violates the strong
energy condition (SEC), and provides tantalizing hints that the
NEC \emph{might} possibly be violated in a classical
regime~\cite{Riess,jerk,rip}. Thus, gradually the weak and null
energy conditions, and with it the other energy conditions, are
losing their status of a kind of law~\cite{VisserEC}. Surely, this
has had implications on the construction of wormholes. In the
original paper \cite{Morris}, Morris and Thorne first provided a
spherically symmetric spacetime metric, then deduced that it
needed exotic matter to sustain the wormhole geometry. The
engineering work was left to an absurdly advanced civilization,
which could manufacture such matter and construct these wormholes.
Then, once it was understood that quantum effects should enter in
the stress-energy tensor, a self-consistent wormhole solution of
semiclassical gravity was found \cite{hochbergetal}, presumably
obeying the quantum inequalities. Thus, it seems that these exotic
spacetimes arise naturally in the quantum regime, as a large
number of quantum systems have been shown to violate the energy
conditions, such as the Casimir effect. Indeed, various wormhole
solutions in semi-classical gravity have been considered in the
literature. For instance, semi-classical wormholes were found in
the framework of the Frolov-Zelnikov approximation for $\langle
T_{\mu\nu}\rangle$ \cite{Sushkov2}. Analytical approximations of
the stress-energy tensor of quantized fields in static and
spherically symmetric wormhole spacetimes were also explored in
Refs. \cite{Popov2}. However, the first self-consistent wormhole
solution coupled to a quantum scalar field was obtained in Ref.
\cite{hochbergetal}. The ground state of a massive scalar field
with a non-conformal coupling on a short-throat flat-space
wormhole background was computed in Ref. \cite{Khusn2}, by using a
zeta renormalization approach. The latter wormhole model, which
was further used in the context of the Casimir effect
\cite{Khab2}, was constructed by excising spherical regions from
two identical copies of Minkowski spacetime, and finally
surgically grafting the boundaries (A more realistic geometry was
considered in Ref. \cite{Khusn22}). Recently, semi-classical
wormholes have also been obtained using a one-loop graviton
contribution approach \cite{Garattini,Garattini2}. Finally with
the realization that nonminimal scalar fields violate the weak
energy condition, a set of self-consistent classical wormholes was
found \cite{barcelovisserPLB99}.

It is fair to say that, though outside this mainstream, classical
wormholes were found by Homer Ellis back in 1973 \cite{homerellis}
and further explored in \cite{homerellis2}, and related
self-consistent solutions were found by Kirill Bronnikov in 1973
\cite{bronikovWH}, Takeshi Kodama in 1978 \cite{kodama}, and
G\'erard Cl\'ement in 1981 \cite{clementWH}. These papers written
much before the wormhole boom originated from Morris and Thorne's
work \cite{Morris} (see \cite{clement1} for a short account of
these previous solutions). A self-consistent Ellis wormhole was
found again by Harris \cite{harris} by solving, through an exotic
scalar field, an exercise for students posed in \cite{Morris}.
Visser \cite{visser1} motivated by the aim of minimizing the
violation of the energy conditions and the possibility of a
traveller not encountering regions of exotic matter in a traversal
through a wormhole, constructed polyhedral wormholes and, in
particular, cubic wormholes. These contained exotic matter
concentrated only at the edges and the corners of the geometrical
structure, and a traveller could pass through the flat faces
without encountering matter, exotic or otherwise. He further
generalized a suggestion of Roman for a configuration with two
wormholes \cite{Morris} into a Roman ring \cite{visserRing}, and
in 1999 analyzed generic dynamical traversable wormhole throats
\cite{hochvisserWHthroats}. Furthermore, in 1999, together with
Barcelo, he found classically consistent solutions with scalar
fields \cite{barcelovisser1}, and in collaboration with Dadhich,
Kar and Mukherjee has also found self-dual solutions
\cite{dadichetalSelfdual}. Other authors have also made
interesting studies and significant contributions.

Before Visser's book we can quote an interesting application to
wormhole physics by Frolov and Novikov, where they relate wormhole
and black hole physics \cite{frolov1}. An interesting application
of wormholes to cosmology was treated by Hochberg and Kephart
relatively to the horizon problem~\cite{Kephart}. These authors
speculated that wormholes allow the two-way passage of signals
between spatially separated regions of spacetime, and could permit
the thermalization of these respective regions. After the Matt
Visser book, Gonz\'alez-D\'{\i}az generalized the static
spherically symmetric traversable wormhole solution to that of a
torus-like topology \cite{gonzalez}. This geometrical construction
was denoted as a ringhole. Gonz\'alez-D\'{\i}az went on to analyze
the causal structure of the solution, i.e., the presence of closed
timelike curves, and has recently studied the ringhole evolution
due to the accelerating expansion of the universe, in the presence
of dark energy \cite{gonzalez2}. Wormhole solutions inside cosmic
strings were found by Cl\'ement \cite{clementWHstrings}, and Aros
and Zamorano \cite{aroszamorano}; wormholes supported by strings
by Schein and Aichelburg \cite{shein1,shein2}; the maintenance of
a wormhole with a scalar field was considered by
Vollick~\cite{Vollick}, solutions with minimal and non-minimal
scalar fields were explored by Kim and Kim~\cite{Kim-Kim} and
exact solutions of charged wormholes considering the back reaction
to the traversable Lorentzian wormhole spacetime by a scalar field
or electric charge were found by Kim and Lee~\cite{Kim-Lee};
rotating wormholes solutions were analyzed by Teo \cite{teo}, and
further generalized by Kuhfittig \cite{Kuhf}; a solution was found
in which the exotic matter is controlled by an external magnetic
field by Parisio~\cite{Parisio}; wormholes with stress-energy
tensor of massless neutrinos and other massless fields were
considered by Krasnikov \cite{kras}; wormholes made of a crossflow
of dust null streams were discussed by Hayward
\cite{haywardnulldust} and Gergely \cite{gergely}; self consistent
charged solutions were found by Bronnikov and Grinyok
\cite{bronnikovgrinyok2}; and the possible existence of wormhole
geometries in the context of nonlinear electrodynamics was also
explored \cite{Arellano}. It is interesting to note that building
on \cite{haywardnulldust}, Hayward and Koyama, using a model of
pure phantom radiation, i.e., pure radiation with negative energy
density, and the idealization of impulsive radiation, considered
analytic solutions describing the theoretical construction of a
traversable wormhole from a Schwarzschild black hole
\cite{Hayward-Koyama}, and the respective enlargement of the
wormhole \cite{Hayward-Koyama2}. More recently, exact solutions of
traversable wormholes were found under the assumption of spherical
symmetry and the existence of a {\it non-static} conformal
symmetry, which presents a more systematic approach in searching
for exact wormhole solutions \cite{Boehmer:2007rm}.

One of the main areas in wormhole research is to try to avoid as
much as possible the violation of the null energy condition. For
static wormholes the null energy condition is violated
\cite{Morris,Visser}, and thus, several attempts have been made to
overcome somehow this problem. In the original
article~\cite{Morris}, Morris and Thorne had already tried to
minimize the violating region by constructing specific examples of
wormhole geometries. As mentioned before, Visser \cite{visser1}
found solutions where observers can pass the throat without
interacting with the exotic matter, which was pushed to the
corners, and Kuhfittig \cite{kuhfittig} has found that the region
made of exotic matter can be made arbitrarily small. For dynamic
wormholes, the null energy condition, more precisely the averaged
null energy condition can be avoided in certain regions
\cite{hochvisserWHthroats,hochvisserPRL98,hochvisserPRD98,kar,kar-sahdev,Kim-evolvWH}.
More recently, Visser {\it et al} \cite{visser2003,Kar2}, noting
the fact that the energy conditions do not actually quantify the
``total amount'' of energy condition violating matter, developed a
suitable measure for quantifying this notion by introducing a
``volume integral quantifier''. This notion amounts to calculating
the definite integrals $\int T_{\mu\nu}U^\mu U^\nu \,dV$ and $\int
T_{\mu\nu}k^\mu k^\nu \,dV$, and the amount of violation is
defined as the extent to which these integrals become negative.
Although the null energy and averaged null energy conditions are
always violated for wormhole spacetimes, Visser {\it et al}
considered specific examples of spacetime geometries containing
wormholes that are supported by arbitrarily small quantities of
averaged null energy condition violating matter.

Some papers have added a cosmological constant to the wormhole
construction analysis. Thin-shell wormhole solutions with
$\Lambda$, in the spirit of Visser \cite{Visser,visser-quantum}
were analyzed in \cite{kimwh,LC-CQG}. Roman \cite{romanLambda}
found a wormhole solution inflating in time to test whether one
could evade the violation of the energy conditions, and Delgaty
and Mann \cite{delgaty} looked for new wormhole solutions with
$\Lambda$. Construction of wormhole solutions by matching an
interior wormhole spacetime to an exterior vacuum solution, at a
junction surface, were also recently analyzed
extensively~\cite{LLQ-PRD,Lobo-CQG,Lobo-GRG}. In particular, a
thin shell around a traversable wormhole, with a zero surface
energy density was analyzed in \cite{LLQ-PRD}, and with generic
surface stresses in \cite{Lobo-CQG}. A similar analysis for the
plane symmetric case, with a negative cosmological constant, is
done in \cite{LL-PRD}. A general class of wormhole geometries with
a cosmological constant and junction conditions was analyzed by
DeBenedictis and Das~\cite{benedectis}, and further explored in
higher dimensions~\cite{deBenedictisDas}.
To know the stability of an object against several types of
perturbation is always an important issue. In particular, the
stability of thin-shell wormholes, constructed using the
cut-and-paste technique, by considering specific equations of
state~\cite{visser-quantum,kimwh,delgaty,VisserPLB90,KKYang,PerryMann},
or by applying a linearized radial perturbation around a stable
solution~\cite{visserNPB,poisson,EiroaRomero,LC-CQG,lake}, were
analyzed. For the Ellis' drainhole \cite{homerellis,homerellis2},
Armend\'ariz-Pic\'on \cite{picon} finds that it is stable against
linear perturbations, whereas Shinkai and Hayward \cite{shinkai}
find this same class unstable to nonlinear perturbations.
Bronnikov and Grinyok \cite{bronnikovgrinyok2,bronnikovgrinyok1}
found that the consistent wormholes of Barcel\'o and Visser
\cite{barcelovisserPLB99} are unstable.

In alternative theories to general relativity wormhole solutions
have been worked out. In higher dimensions, solutions have been
found by Chodos and Detweiler \cite{chodos}, Cl\'ement
\cite{clementhigherD}, and DeBenedictis and Das
\cite{deBenedictisDas}; in the nonsymmetric gravitational theory
traversable wormholes were found by Moffat and
Svoboda~\cite{Moffat-Svoboda}; in Brans-Dicke theory by Agnese and
Camera~\cite{Agnese}, Anchordoqui {\it at al} \cite{Anch-Berg},
Nandi and collaborators \cite{nandi,Bloomfield,nandi2,nandi3}, and
He and Kim~\cite{He-Kim}; in Kaluza-Klein theory by Shen and
collaborators \cite{shen}; in Einstein-Gauss-Bonnet by Bhawal and
Kar \cite{Bhawal}; and Koyama, Hayward and Kim
\cite{koyamahaywardkim} examined wormholes in a two-dimensional
dilatonic theory. Anchordoqui and Bergliaffa found a wormhole
solution in a brane world scenario \cite{anchordoqui}, further
examined by Barcel\'o and Visser \cite{barcelovisserBrane}, and
Bronnikov and Kim considered possible traversable wormhole
solutions in a brane world, by imposing the condition $R=0$, where
$R$ is the four-dimensional scalar curvature \cite{BKim}; the
latter solution was generalized in Ref. \cite{LobobraneWH}. La
Camera using the simplest form of the Randall-Sundrum model,
considered the metric generated by a static, spherically symmetric
distribution of matter on the physical brane, and found that the
solution to the five-dimensional Einstein equations, obtained
numerically, describes a wormhole geometry~\cite{LaCamera}.
Recently, wormhole throats were also analyzed in a higher
derivative gravity model governed by the Einstein-Hilbert
Lagrangian, supplemented with $1/R$ and $R^2$ curvature scalar
terms~\cite{furey-debened}. Using the resulting equations of
motion, it was found that the weak energy condition may be
respected in the throat vicinity, with conditions compatible with
those required for stability~\cite{nojiri-odintsov} and an
acceptable Newtonian limit~\cite{Dick}. The $R^2$ theory was
meticulously studied by Ghoroku and Soma~\cite{Ghoroku-Soma},
where it was concluded that, under the assumption that an
asymptotically flat global solution exists, no weak energy
condition respecting wormhole solution may exist in such a theory.

If it is true that wormholes act as shortcuts between two regions
of spacetime, then it is interesting to note that shortcuts also
exist in the context of brane cosmology~\cite{Cald-Lang}. The
latter model stipulates that our Universe is a three-brane
embedded in a five-dimensional anti-se Sitter spacetime, in which
matter is confined to the brane and gravity exists throughout the
bulk. This implies the causal propagation of light and
gravitational signals is in general different~\cite{Kalbermann}. A
gravitational signal travelling between two points on the brane
may propagate through the bulk, taking a shortcut, and appearing
quicker than a photon which propagates on the brane between the
two respective points~\cite{Abdalla1,Abdalla2}. It is then
expected that these shortcuts would play an important role in
solving the horizon problem~\cite{Abdalla2,Chung}.


An important side effect of wormholes is that they can
theoretically generate closed timelike curves with relative ease,
by performing a sufficient delay to the time of one mouth in
relation to the other (see \cite{Kluwer} for a review on closed
timelike curves). This can be done either by the special
relativistic twin paradox method \cite{mty} or by the general
relativistic redshift way \cite{frolovnovikovTM}. The importance
of wormholes in the study of time machines is that they provide a
non-eternal time machine, where closed timelike curves appear to
the future of some hypersurface, the chronology horizon (a special
case of a Cauchy horizon which is the onset of the nonchronal
region containing closed timelike curves) which is generated in a
compact region in this case~\cite{CauchyCTC,CauchyCTC2}. Since
time travel to the past is in general unwelcome, it is possible to
test whether classical or semiclassical effects will destroy the
time machine. It is found that classically it can be easily
stabilized \cite{mty,Visser}. Semiclassiclaly, there are
calculations that favor the destruction
\cite{kimthorne,hawking,klink}, leading to chronology protection
\cite{hawking}, others that maintain the time machine
\cite{visserRing,KimCTC,lyu}. Other simpler systems that simulate
a wormhole, such as Misner spacetime which is a species of
two-dimensional wormhole, have been studied more thoroughly, with
no conclusive answer. For Misner spacetime the debate still goes
on, favoring chronology protection \cite{hiskonk}, disfavoring it
\cite{gottLI}, and back in favoring \cite{hiscock}. The upshot is
that semiclassical calculations will not settle the issue of
chronology protection \cite{visserCP}, one needs a quantum
gravity, as has been foreseen sometime before by Thorne
\cite{thorneGRG13}.

In a cosmological context, it is extraordinary that recent
observations have confirmed that the Universe is undergoing a
phase of accelerated expansion. Evidence of this cosmological
expansion, coming from measurements of supernovae of type Ia (SNe
Ia)~\cite{Riess2,Perlmutter} and independently from the cosmic
microwave background radiation~\cite{Bennet,Hinshaw}, shows that
the Universe additionally consists of some sort of negative
pressure ``dark energy''. The Wilkinson Microwave Anisotropy Probe
(WMAP), designed to measure the CMB anisotropy with great
precision and accuracy, has recently confirmed that the Universe
is composed of approximately $70$ percent of dark
energy~\cite{Bennet}. Several candidates representing dark energy
have been proposed in the literature, namely, a positive
cosmological constant, the quintessence fields, generalizations of
the Chaplygin gas and so-called tachyon models. A simple way to
parameterize the dark energy is by an equation of state of the
form $\omega\equiv p/\rho$, where $p$ is the spatially homogeneous
pressure and $\rho$ the energy density of the dark
energy~\cite{Cai-Wang}. A value of $\omega<-1/3$ is required for
cosmic expansion, and $\omega=-1$ corresponds to a cosmological
constant~\cite{Carmelli}. A possibility that has been widely
explored, is that of quintessence, where the parameter range is
$-1<\omega<-1/3$. However, a note on the choice of the imposition
$\omega >-1$ is in order. This is considered to ensure that the
null energy condition, $T_{\mu\nu}\,k^\mu\,k^\nu \geq 0$, is
satisfied. If $\omega<-1$~\cite{Melchiorri,Alcaniz,Hoffman}, a
case certainly not excluded by observation, then the null energy
condition is violated, $\rho+p<0$, and consequently all of the
other energy conditions. Matter with the property $\omega<-1$ has
been denoted ``phantom energy'' \cite{phantomenergy}. As the
possibility of phantom energy implies the violation of the null
energy condition, this leads us back to wormhole physics. This
possibility has been explored with wormholes being supported by
phantom energy \cite{phantomWH,phantomWHb,phantomWH2}, the
generalized Chaplygin gas \cite{ChapWH}, and the van der Waals
equation of state \cite{VDWwh}.
An interesting feature is that due to the fact of the accelerated
expansion of the Universe, macroscopic wormholes could naturally
be grown from the submicroscopic constructions that originally
pervaded the quantum foam. In Ref. \cite{gonzalez2} the evolution
of wormholes and ringholes embedded in a background accelerating
Universe driven by dark energy, was analyzed. An interesting
feature is that the wormhole's size increases by a factor which is
proportional to the scale factor of the Universe, and still
increases significantly if the cosmic expansion is driven by
phantom energy. The accretion of dark and phantom energy onto
Morris-Thorne wormholes~\cite{diaz-phantom3,diaz-phantom4}, was
further explored, and it was shown that this accretion gradually
increases the wormhole throat which eventually overtakes the
accelerated expansion of the universe, consequently engulfing the
entire Universe, and becomes infinite at a time in the future
before the big rip. This process was dubbed the ``Big Trip''
\cite{diaz-phantom3,diaz-phantom4}. It was shown that using
$k-$essence dark energy also leads to the big rip \cite{PGDiaz-k},
although, in an interesting article \cite{diaz-phantom},
considering a generalized Chaplygin gas the big rip may be avoided
altogether.

Stars are common for everyone to see, black holes also inhabit the
universe in billions, so one may tentatively assume that
wormholes, formed or constructed from one way or another, can also
appear in large amounts. If they inhabit the cosmological space,
they will produce microlensing effects on point sources at
non-cosmological distances \cite{cramer}, as well as at
cosmological distances, in this case gamma-ray burts could be the
objects microlensed \cite{torres,safonova1}. If peculiarly large,
then wormholes will produce macrolensing effects \cite{safonova2}.
There is now a growing consensus that wormholes are in the same
chain of stars and black holes. For instance, Gonz\'alez-D\'{\i}as
\cite{gonzalez} understood that an enormous pressure on the center
ultimately meant a negative energy density to open up the tunnel;
DeBenedectis and Das \cite{benedectis} mention that the
stress-energy supporting the structure consists of an anisotropic
brown dwarf `star'; and the wormhole joining one
Friedmann-Robertson-Walker universe with Minkowski spacetime or
joining two  Friedmann-Robertson-Walker universes
\cite{hochvisserWHthroats} could be interpreted, after further
matchings,  as a wormhole joining a collapsing (or expanding) star
to Minkowski spacetime or a wormhole joining two dynamical stars,
respectively. It has also been recognized, and emphasized by
Hayward \cite{hayward}, that wormholes and black holes can be
treated in a unified way, the black hole being described by a null
outer trapped surface, and the wormhole by a timelike outer
trapped surface, this surface being the throat where incoming null
rays start to diverge \cite{hochvisserPRD98,hayward}. Thus, it
seems there is a continuum of objects from stars to wormholes
passing through black holes, where stars are made of normal
matter, black holes of vacuum, and wormholes of exotic matter.
Although not so appealing perhaps, wormholes could be called
``exotic stars''.

\subsection{``Warp drive'' spacetimes and superluminal travel}

Much interest has been revived in superluminal travel in the last
few years. Despite the use of the term superluminal, it is not
``really'' possible to travel faster than light, in any
\emph{local} sense. The point to note is that one can make a round
trip, between two points separated by a distance $D$, in an
arbitrarily short time as measured by an observer that remained at
rest at the starting point, by varying one's speed or by changing
the distance one is to cover. Providing a general \emph{global}
definition of superluminal travel is no trivial
matter~\cite{VB,VBL}, but it is clear that the spacetimes that
allow ``effective'' superluminal travel generically suffer from
the severe drawback that they also involve significant negative
energy densities. More precisely, superluminal effects are
associated with the presence of \emph{exotic} matter, that is,
matter that violates the null energy condition (see
\cite{Springer} for a review). In fact, superluminal spacetimes
violate all the known energy conditions, and Ken Olum demonstrated
that negative energy densities and superluminal travel are
intimately related~\cite{Olum}.

Apart from wormholes {\cite{Morris,Visser}}, two spacetimes which
allow superluminal travel are the Alcubierre warp drive
\cite{Alcubierre} and the solution known as the Krasnikov tube
\cite{Krasnikov,Everett}. Alcubierre demonstrated that it is
theoretically possible, within the framework of general
relativity, to attain arbitrarily large velocities
\cite{Alcubierre}. A warp bubble is driven by a local expansion
behind the bubble, and an opposite contraction ahead of it.
However, by introducing a slightly more complicated metric,
Jos\'{e} Nat\'{a}rio~\cite{Natario} dispensed with the need for
expansion. The Nat\'{a}rio version of the warp drive can be
thought of as a bubble sliding through space.

It is interesting to note that Krasnikov \cite{Krasnikov}
discovered a fascinating aspect of the warp drive, in which an
observer on a spaceship cannot create nor control on demand an
Alcubierre bubble, with $v>c$, around the ship \cite{Krasnikov},
as points on the outside front edge of the bubble are always
spacelike separated from the centre of the bubble. However,
causality considerations do not prevent the crew of a spaceship
from arranging, by their own actions, to complete a {\it round
trip} from the Earth to a distant star and back in an arbitrarily
short time, as measured by clocks on the Earth, by altering the
metric along the path of their outbound trip. Thus, Krasnikov
introduced a two-dimensional metric with an interesting property
that although the time for a one-way trip to a distant destination
cannot be shortened, the time for a round trip, as measured by
clocks at the starting point (e.g. Earth), can be made arbitrarily
short. Soon after, Everett and Roman generalized the Krasnikov
two-dimensional analysis to four dimensions, denoting the solution
as the {\it Krasnikov tube}~\cite{Everett}, where they analyzed
the superluminal features, the energy condition violations, the
appearance of closed timelike curves and applied the Quantum
Inequality.

Recently, linearized gravity was applied to warp drive spacetimes,
testing the energy conditions at first and second order of the
non-relativistic warp-bubble velocity~\cite{LV-CQG}, $v\ll 1$.
Thus, attention was not focussed on the ``superluminal'' aspects
of the warp bubble~\cite{Coule}, such as the appearance of
horizons~\cite{Hiscock,Clark,Gonz} and of closed timelike
curves~\cite{EverettCTC}, but rather on a secondary unremarked
effect: The warp drive (\emph{if it can be realised in nature})
appears to be an example of a ``reaction-less drive'' wherein the
warp bubble moves by interacting with the geometry of spacetime
instead of expending reaction mass. A particularly interesting
aspect of this construction is that one may place a finite mass
spaceship at the origin and consequently analyze how the warp
field compares with the mass-energy of the spaceship. This is not
possible in the usual finite-strength warp field, since by
definition the point in the center of the warp bubble moves along
a geodesic and is ``massless''. That is, in the usual formalism
the spaceship is always treated as a test particle, while in the
linearized theory one can treat the spaceship as a finite mass
object.

For warp drive spacetimes, by using the ``quantum inequality''
deduced by Ford and Roman~\cite{fordroman1}, it was soon verified
that enormous amounts of energy are needed to sustain superluminal
warp drive spacetimes~\cite{fordroman2,PfenningF}. To reduce the
enormous amounts of exotic matter needed in the superluminal warp
drive, van den Broeck proposed a slight modification of the
Alcubierre metric which considerably ameliorates the conditions of
the solution~\cite{Broeck1}. It is also interesting to note that,
by using the ``quantum  inequality'', enormous quantities of
negative energy densities are needed to support the superluminal
Krasnikov tube~\cite{Everett}. Gravel and
Plante~\cite{GravelPlante,Gravel} in a way similar in spirit to
the van den Broeck analysis, showed that it is theoretically
possible to lower significantly the mass of the Krasnikov tube.
However, in the linearized analysis, no \emph{a priori}
assumptions as to the ultimate source of the energy condition
violations were made, so that the quantum inequalities were not
used nor needed. This means that the restrictions derived on warp
drive spacetimes are more generic than those derived using the
quantum inequalities -- the restrictions derived in \cite{LV-CQG}
hold regardless of whether the warp drive is assumed to be
classical or quantum in its operation. It was not meant to suggest
that such a ``reaction-less drive'' is achievable with current
technology, as indeed extremely stringent conditions on the warp
bubble were obtained, in the weak-field limit. These conditions
are so stringent that it appears unlikely that the ``warp drive''
will ever prove technologically useful.

\subsection{Closed timelike curves}

As time is incorporated into the proper structure of the fabric of
spacetime, it is interesting to note that general relativity is
contaminated with non-trivial geometries which generate {\it
closed timelike curves}
\cite{Visser,World-Scientific,Springer,Kluwer,LLQ-PRD,Tipler-CTCs,Frolov}.
A closed timelike curve (CTC) allows time travel, in the sense
that an observer which travels on a trajectory in spacetime along
this curve, returns to an event which coincides with the
departure. The arrow of time leads forward, as measured locally by
the observer, but globally he/she may return to an event in the
past. This fact apparently violates causality, opening Pandora's
box and producing time travel paradoxes \cite{Nahin}, throwing a
veil over our understanding of the fundamental nature of Time. The
notion of causality is fundamental in the construction of physical
theories, therefore time travel and it's associated paradoxes have
to be treated with great caution. The paradoxes fall into two
broad groups, namely the {\it consistency paradoxes} and the {\it
causal loops}.

The consistency paradoxes include the classical grandfather
paradox. Imagine travelling into the past and meeting one's
grandfather. Nurturing homicidal tendencies, the time traveller
murders his grandfather, impeding the birth of his father,
therefore making his own birth impossible. In fact, there are many
versions of the grandfather paradox, limited only by one's
imagination. The consistency paradoxes occur whenever
possibilities of changing events in the past arise.

The paradoxes associated to causal loops are related to
self-existing information or objects, trapped in spacetime.
Imagine a time traveller going back to his past, handing his
younger self a manual for the construction of a time machine. The
younger version then constructs the time machine over the years,
and eventually goes back to the past to give the manual to his
younger self. The time machine exists in the future because it was
constructed in the past by the younger version of the time
traveller. The construction of the time machine was possible
because the manual was received from the future. Both parts
considered by themselves are consistent, and the paradox appears
when considered as a whole. One is liable to ask, what is the
origin of the manual, for it apparently surges out of nowhere.
There is a manual never created, nevertheless existing in
spacetime, although there are no causality violations. An
interesting variety of these causal loops was explored by Gott and
Li~\cite{Gott-Li}, where they analyzed the idea of whether there
is anything in the laws of physics that would prevent the Universe
from creating itself. Thus, tracing backwards in time through the
original inflationary state a region of CTCs may be encountered,
giving {\it no} first-cause.

A great variety of solutions to the Einstein Field Equations
(EFEs) containing CTCs exist, but, two particularly notorious
features seem to stand out. Solutions with a tipping over of the
light cones due to a rotation about a cylindrically symmetric
axis; and solutions that violate the Energy Conditions of general
relativity, which are fundamental in the singularity theorems and
theorems of classical black hole thermodynamics
\cite{hawkingellis}. A great deal of attention has also been paid
to the quantum aspects of closed timelike curves
\cite{quantum1,quantum2,quantum3}.

Thus, as shall be shown in Section \ref{secIV}, it is possible to
find solutions to the EFEs, with certain ease, which generate
CTCs, which This implies that if we consider general relativity
valid, we need to include the {\it possibility} of time travel in
the form of CTCs. A typical reaction is to exclude time travel due
to the associated paradoxes. But the paradoxes do not prove that
time travel is mathematically or physically impossible. Consistent
mathematical solutions to the EFEs have been found, based on
plausible physical processes. What they do seem to indicate is
that local information in spacetimes containing CTCs are
restricted in unfamiliar ways. The grandfather paradox, without
doubt, does indicate some strange aspects of spacetimes that
contain CTCs. It is logically inconsistent that the time traveller
murders his grandfather. But, one can ask, what exactly impeded
him from accomplishing his murderous act if he had ample
opportunities and the free-will to do so. It seems that certain
conditions in local events are to be fulfilled, for the solution
to be globally self-consistent. These conditions are denominated
{\it consistency constraints} \cite{Earman}. To eliminate the
problem of free-will, mechanical systems were developed as not to
convey the associated philosophical speculations on free-will
\cite{Echeverria,NovikovCTC}. Much has been written on two
possible remedies to the paradoxes, namely the Principle of
Self-Consistency \cite{CauchyCTC,NovikovCTC,Carlini1,Carlini2} and
the Chronology Protection Conjecture
\cite{hawking,visserCP,GrantCTC}.

One current of thought, led by Igor Novikov, is the Principle of
Self-Consistency, which stipulates that events on a CTC are
self-consistent, i.e., events influence one another along the
curve in a cyclic and self-consistent way. In the presence of CTCs
the distinction between past and future events are ambiguous, and
the definitions considered in the causal structure of well-behaved
spacetimes break down. What is important to note is that events in
the future can influence, but cannot change, events in the past.
The Principle of Self-Consistency permits one to construct local
solutions of the laws of physics, only if these can be prolonged
to a unique global solution, defined throughout non-singular
regions of spacetime. Therefore, according to this principle, the
only solutions of the laws of physics that are allowed locally,
reinforced by the consistency constraints, are those which are
globally self-consistent.

Hawking's Chronology Protection Conjecture \cite{hawking} is a
more conservative way of dealing with the paradoxes. Hawking notes
the strong experimental evidence in favour of the conjecture from
the fact that "we have not been invaded by hordes of tourists from
the future". An analysis reveals that the value of the
renormalized expectation quantum stress-energy tensor diverges in
the imminence of the formation of CTCs. This conjecture permits
the existence of traversable wormoles, but prohibits the
appearance of CTCs. The transformation of a wormhole into a time
machine results in enormous effects of the vacuum polarization,
which destroys it's internal structure before attaining the Planck
scale. Nevertheless, Li has shown given an example of a spacetime
containing a time machine that might be stable against vacuum
fluctuations of matter fields~\cite{Li}, implying that Hawking's
suggestion that the vacuum fluctuations of quantum fields acting
as a chronology protection might break down. There is no
convincing demonstration of the Chronology Protection Conjecture,
but the hope exists that a future theory of quantum gravity may
prohibit CTCs.

Visser still considers the possibility of two other conjectures
\cite{Visser}. The first is the radical reformulation of physics
conjecture, in which one abandons the causal structure of the laws
of physics and allows, without restriction, time travel,
reformulating physics from the ground up. The second is the boring
physics conjecture, in which one simply ceases to consider the
solutions to the EFEs generating CTCs. Perhaps an eventual quantum
gravity theory will provide us with the answers. But, as stated by
Thorne \cite{thorneGRG13}, it is by extending the theory to it's
extreme predictions that one can get important insights to it's
limitations, and probably ways to overcome them. Therefore, time
travel in the form of CTCs, is more than a justification for
theoretical speculation, it is a conceptual tool and an
epistemological instrument to probe the deepest levels of general
relativity and extract clarifying views.



\section{Traversable Lorentzian wormholes}\label{secII}

One adopt the reverse philosophy in solving the Einstein field
equation, namely, one first considers an interesting and exotic
spacetime metric, then finds the matter source responsible for the
respective geometry. In this manner, it was found that some of
these solutions possess a peculiar property, namely ``exotic
matter'', involving a stress-energy tensor that violates the null
energy condition, $T_{\mu\nu}k^\mu k^\nu \geq 0$, where $k^\mu$ is
a null vector. These geometries also allow closed timelike curves,
with the respective causality violations. Another interesting
feature of these spacetimes is that they allow ``effective''
superluminal travel, although, locally, the speed of light is not
surpassed. These solutions are primarily useful as
``gedanken-experiments'' and as a theoretician's probe of the
foundations of general relativity, and include traversable
wormholes, which shall be extensively reviewed in this Section.

\subsection{Spacetime metric}

Consider the following spherically symmetric and static wormhole
solution
\begin{equation}
ds^2=-e ^{2\Phi(r)} \,dt^2+\frac{dr^2}{1- b(r)/r}+r^2 \,(d\theta
^2+\sin ^2{\theta} \, d\phi ^2) \,, \label{metricwormhole}
\end{equation}
where $\Phi(r)$ and $b(r)$ are arbitrary functions of the radial
coordinate $r$. $\Phi(r)$ is denoted the redshift function, for it
is related to the gravitational redshift, and $b(r)$ is denoted
the shape function, because as can be shown by embedding diagrams,
it determines the shape of the wormhole \cite{Morris}. The
coordinate $r$ is non-monotonic in that it decreases from
$+\infty$ to a minimum value $r_0$, representing the location of
the throat of the wormhole, where $b(r_0)=r_0$, and then it
increases from $r_0$ to $+\infty$. The proper circumference of a
circle of fixed $r$ is given by $2\pi r$. Although the metric
coefficient $g_{rr}$ becomes divergent at the throat, which is
signalled by the coordinate singularity, the proper radial
distance
\begin{equation}
 l(r)=\pm\,\int_{r_0}^r{{dr}\over{(1-b(r)/r)^{1/2}}} \,,
                                                 \label{eq:PD}
\end{equation}
is required to be finite everywhere. Note that as $0 \leq 1
-b(r)/r \leq 1$, the proper distance is greater than or equal to
the coordinate distance, i.e., $|l(r)| \geq r - r_0$.   The metric
(\ref{metricwormhole}) may be written in terms of the proper
radial distance as
\begin{equation}
ds^2=-e^{2\Phi(l)}dt^2 + dl^2
 + {r^2}(l)({d\theta}^2+ {\rm sin}^2\theta\,{d\phi}^2)\,.
\end{equation}
The proper distance decreases from $l=+\infty$, in the upper
universe, to $l=0$ at the throat, and then from zero to $-\infty$
in the lower universe. For the wormhole to be traversable it must
have no horizons, which implies that $g_{tt}=-e^{2\Phi(r)}\neq 0$,
so that $\Phi(r)$ must be finite everywhere.

The four-velocity of a static observer is
$U^{\mu}=dx^{\mu}/{d\tau} =(U^{\,t},0,0,0)=(e^{-\Phi(r)},0,0,0)$.
The observer's four-acceleration is $a^{\mu}
=U^{\mu}{}_{;\nu}\,U^{\nu}$, so that taking into account
Eq.~(\ref{metricwormhole}) we have
\begin{eqnarray}
a^t &=& 0 \,,         \nonumber \\
a^r &=& \Gamma^r_{tt}\,\left({dt\over{d\tau}}\right)^2
    ={\Phi}'\,(1-b/r)\,,          \label{radial-acc}
\end{eqnarray}
where the prime denotes a derivative with respect to the radial
coordinate $r$. From the geodesic equation, a radially moving test
particle which starts from rest initially has the equation of
motion
\begin{equation}
{{d^{\,2}r}\over d{\tau}^2}=-\Gamma^r_{tt}\,
\left({dt\over{d\tau}}\right)^2 =-a^r\,.
      \label{radial-accel}
\end{equation}
Therefore, $a^r$ is the radial component of proper acceleration
that an observer must maintain in order to remain at rest at
constant $r,\,\theta,\,\phi$. Note that from
Eq.~(\ref{radial-acc}), a static observer at the throat for
generic $\Phi(r)$ is a geodesic observer. In particular, for a
constant redshift function, $\Phi'(r)=0$, static observers are
also geodesic. It is interesting to note that a wormhole is
``attractive'' if $a^r>0$, i.e., observers must maintain an
outward-directed radial acceleration to keep from being pulled
into the wormhole; and ``repulsive'' if $a^r<0$, i.e., observers
must maintain an inward-directed radial acceleration to avoid
being pushed away from the wormhole. This distinction depends on
the sign of $\Phi'$, as is transparent from
Eq.~(\ref{radial-acc}).

\subsection{The mathematics of embedding and generic static
throat}\label{Sec:embedding}

\begin{figure}
\centering
\includegraphics[width=3.0in]{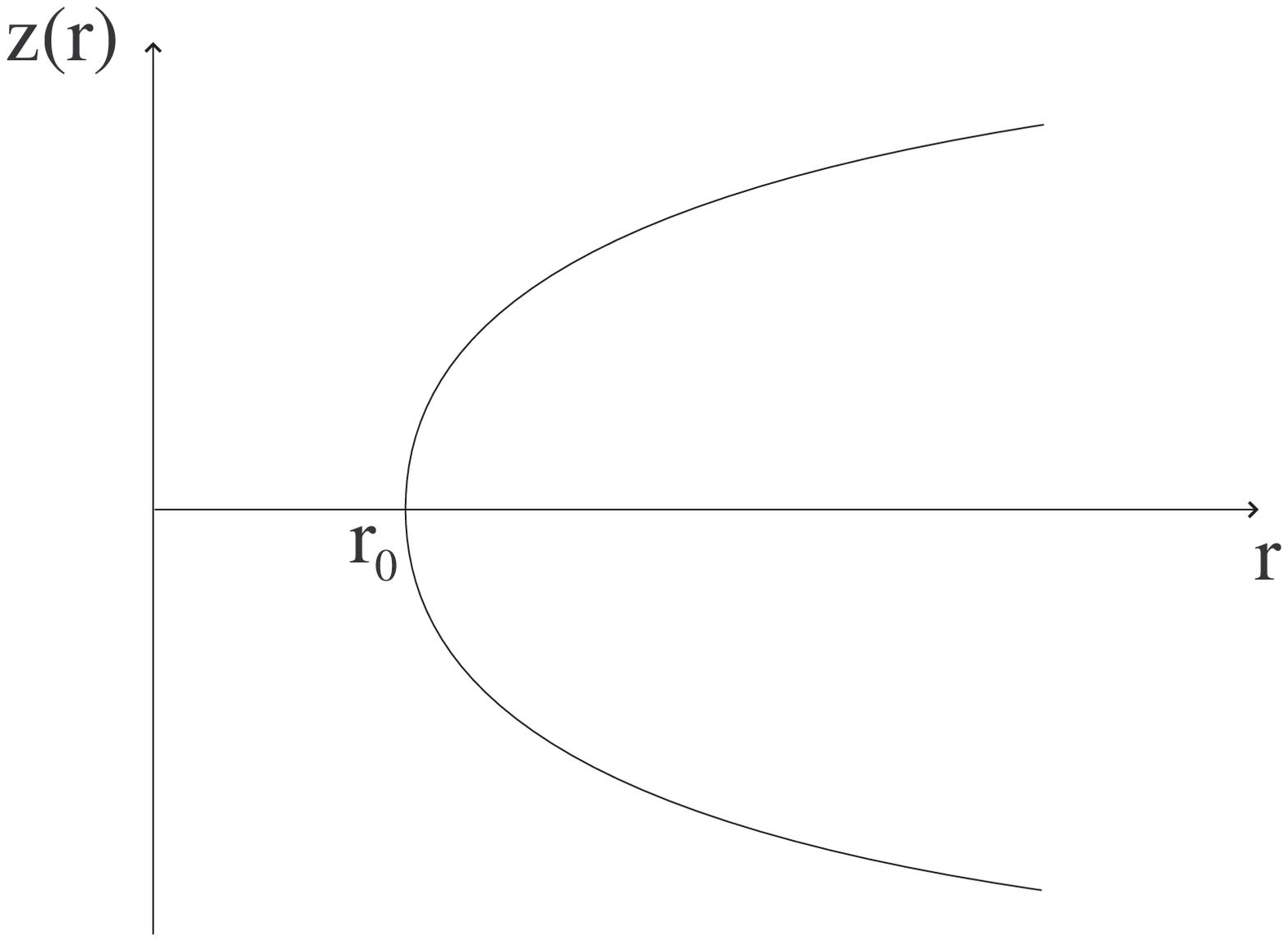}
\hspace{0.25in}
\includegraphics[width=2.4in]{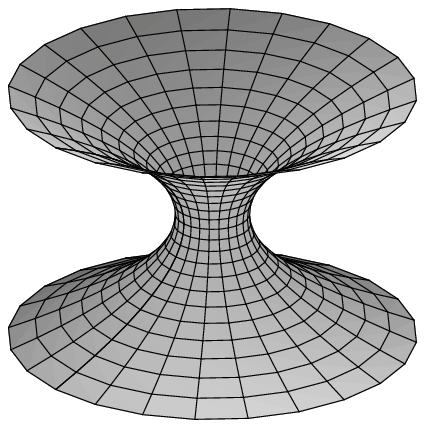}
\caption[Embedding diagrams of traversable wormholes]{The
embedding diagram of a two-dimensional section along the
equatorial plane ($t={\rm const}$, $\theta=\pi/2$) of a
traversable wormhole. For a full visualization of the surface
sweep through a $2\pi$ rotation around the $z-$axis, as can be
seen from the graphic on the right.}\label{fig:embed2}
\end{figure}

We can use embedding diagrams to represent a wormhole and extract
some useful information for the choice of the shape function,
$b(r)$. Due to the spherically symmetric nature of the problem,
one may consider an equatorial slice, $\theta=\pi/2$, without loss
of generality. The respective line element, considering a fixed
moment of time, $t={\rm const}$, is given by
\begin{equation}
ds^2=\frac{dr^2}{1- b(r)/r}+r^2 \, d\phi ^2\,. \label{surface1}
\end{equation}
To visualize this slice, one embeds this metric into
three-dimensional Euclidean space, in which the metric can be
written in cylindrical coordinates, $(r,\phi,z)$, as
\begin{equation}
ds^2=dz^2+dr^2+r^2 \, d\phi ^2 \,.
\end{equation}
Now,  in the three-dimensional Euclidean space the embedded
surface has equation $z=z(r)$, and thus the metric of the surface
can be written as,
\begin{equation}
ds^2=\left [1+\left( \frac{dz}{dr}\right)^2\right] dr^2+r^2 \,
d\phi ^2 \,. \label{surface2}
\end{equation}
Comparing Eq. (\ref{surface2}) with  (\ref{surface1}), we have the
equation for the embedding surface, given by
\begin{equation}
\frac{dz}{dr}=\pm \left(\frac{r}{b(r)}-1 \right)^{-1/2}
\label{lift}\,.
\end{equation}
To be a solution of a wormhole, the geometry has a minimum radius,
$r=b(r)=r_{\rm 0}$, denoted as the throat, at which the embedded
surface is vertical, i.e., $dz/dr \rightarrow \infty$, see Figure
\ref{fig:embed2}. Far from the throat consider that space is
asymptotically flat, $dz/dr \rightarrow 0$ as $r \rightarrow
\infty$.

To be a solution of a wormhole, one needs to impose that the
throat flares out, as in Figure \ref{fig:embed2}. Mathematically,
this flaring-out condition entails that the inverse of the
embedding function $r(z)$, must satisfy $d^2r/dz^2>0$ at or near
the throat $r_{\rm 0}$. Differentiating $dr/dz=\pm
(r/b(r)-1)^{1/2}$ with respect to $z$, we have
\begin{equation}
\frac{d^2r}{dz^2}=\frac{b-b'r}{2b^2}>0   \label{flareout}\,.
\end{equation}
At the throat we verify that the form function satisfies the
condition $b'(r_0)<1$. We will see below that this condition plays
a fundamental role in the analysis of the violation of the energy
conditions.

This treatment has the drawback of being highly coordinate
dependent. However, for a covariant treatment we follow the
analysis by Hochberg and Visser \cite{hochvisserPRL98,Hochberg1}.
Consider a generic static spacetime given by the following metric
\begin{equation}\label{staticmetric}
ds^2=g_{\mu\nu}\,dx^{\mu}\,dx^{\nu}=
-e^{2\Phi(r)}\,dt^2+g_{ij}\,dx^{i}\,dx^{j}\,.
\end{equation}
Consider that Greek indices run from 0 to 3; latin indices
$(i,j,k,...)$ run from 1 to 3; and $(a,b,c,...)$ run from 1 to 2
and refer to the wormhole throat and direction parallel to the
throat.

A wormhole throat $\Sigma$ is defined to be a two-dimensional
hypersurface of minimal area taken in one of the constant-time
spatial slices, and is given by
\begin{equation}\label{def:staticthroat}
A(\Sigma)=\int \, \sqrt{^{(2)}g}\;\;d^2x \,.
\end{equation}
The two-surface is embedded in a three-dimensional space, so that
the definition of the extrinsic curvature is well defined.
Consider Gaussian coordinates $x^i=(x^a,n)$, so that the
hypersurface $\Sigma$ lies at $n=0$, the three-dimensional spatial
metric is given by
\begin{equation}
^{(3)}g_{ij}\,dx^{i}\,dx^{j}=^{(2)}g_{ab}\,dx^{a}\,dx^{b}+dn^2 \,.
\end{equation}
The variation in surface area is given by
\begin{eqnarray}\label{variationthroat}
\delta A(\Sigma)=\int \, \frac{\partial\sqrt{^{(2)}g}}{\partial
n}\;\delta n(x)\;d^2x
      =\int \,\sqrt{^{(2)}g} \;\frac{1}{2} \;g^{ab}\,\frac{\partial
g_{ab}}{\partial n}\;\delta n(x)\;d^2x   \;.
\end{eqnarray}
Using Gaussian normal coordinates the extrinsic curvature is
defined by \cite{Misner}
\begin{equation}
K_{ab}=-\frac{1}{2} \;\frac{\partial g_{ab}}{\partial n}  \,,
\end{equation}
which substituted into Eq. (\ref{variationthroat}), provides
\begin{equation}
\delta A(\Sigma)=-\int \,\sqrt{^{(2)}g} \;\;{\rm tr}(K)\;\delta
n(x)\;d^2x \,,
\end{equation}
where the definition ${\rm tr}(K)=g^{ab}K_{ab}$ is used. The
condition that the area be extremal, for arbitrary $\delta n(x)$,
is then simply ${\rm tr}(K)=0$.

For the area to be minimal, we have the additional requirement
that $\delta^2 A(\Sigma)
>0$. We have then
\begin{eqnarray}\label{throatminimal}
\delta ^2 A(\Sigma)=-\int \,\sqrt{^{(2)}g} \;\left(\frac{\partial
\;{\rm tr}(K)}{\partial n}-{\rm tr}(K)^2 \right) \;\delta
n(x)\;\delta n(x)\;d^2x \;.
\end{eqnarray}
Taking into account the extremal condition, ${\rm tr}(K)=0$,
reduces the minimality constraint to
\begin{eqnarray}
\delta ^2 A(\Sigma)=-\int \,\sqrt{^{(2)}g} \;\left(\frac{\partial
\;{\rm tr}(K)}{\partial n} \right) \;\delta n(x)\;\delta
n(x)\;d^2x \;.
\end{eqnarray}
Considering an arbitrary $\delta n(x)$, at the throat we have the
following important condition
\begin{equation}\label{covariantflareout}
\frac{\partial \;{\rm tr}(K)}{\partial n} <0  \,,
\end{equation}
which is the covariant generalization of the Morris-Thorne
flaring-out condition to arbitrary static wormhole throats.

\subsection{Einstein field equation}

The mathematical analysis and the physical interpretation will be
simplified using a set of orthonormal basis vectors. These may be
interpreted as the proper reference frame of a set of observers
who remain at rest in the coordinate system $(t,r,\theta,\phi)$,
with $(r,\theta,\phi)$ fixed. Denote the basis vectors in the
coordinate system as $({\bf e}_{t},{\bf e}_{r},{\bf
e}_{\theta},{\bf e}_{\phi})$. Thus, the orthonormal basis vectors
are given by
\begin{eqnarray} \label{orthonormalbasis}
\left \{ \begin{array}{l}
{\bf e}_{\hat{t}}=e^{-\Phi} \;{\bf e}_{t}\\
{\bf e}_{\hat{r}}=(1-b/r)^{1/2}\; {\bf e}_{r}\\
{\bf e}_{\hat{\theta}}=r^{-1} \;{\bf e}_{\theta}\\
{\bf e}_{\hat{\phi}}=(r \sin \theta)^{-1}\; {\bf e}_{\phi}\,.
              \end{array} \right.
\end{eqnarray}

The Einstein tensor, given in the orthonormal reference frame by
$G_{\hat{\mu}\hat{\nu}}=R_{\hat{\mu}\hat{\nu}}-\frac{1}{2}R\,
g_{\hat{\mu}\hat{\nu}}$, yields for the metric
(\ref{metricwormhole}), the following non-zero components
\begin{eqnarray}
G_{\hat{t}\hat{t}}&=&\frac{b'}{r^2} \label{Einsteintt}\,, \\
G_{\hat{r}\hat{r}}&=&-\frac{b}{r^3}+ 2 \left(1-\frac{b}{r}
\right) \frac{\Phi'}{r}  \label{Einsteinrr}\,,  \\
G_{\hat{\theta}\hat{\theta}}&=&\left(1-\frac{b}{r}
\right)\left[\Phi ''+ (\Phi')^2- \frac{b'r-b}{2r(r-b)}\Phi'-
\frac{b'r-b}{2r^2(r-b)}+\frac{\Phi'}{r} \right]\,,
\label{Einsteintheta}\\
G_{\hat{\phi}\hat{\phi}}&=&G_{\hat{\theta}\hat{\theta}}\,.
\label{Einsteinphi}
\end{eqnarray}


The Einstein field equation, $G_{\hat{\mu}\hat{\nu}}=8\pi
T_{\hat{\mu}\hat{\nu}}$, stipulates that the stress energy tensor
$T_{\hat{\mu}\hat{\nu}}$ should be proportional to the Einstein
tensor. Thus $T_{\hat{\mu}\hat{\nu}}$ has the same algebraic
structure as $G_{\hat{\mu}\hat{\nu}}$, Eqs.
(\ref{Einsteintt})-(\ref{Einsteinphi}), and the only nonzero
components are precisely the diagonal terms $T_{\hat{t}\hat{t}}$,
$T_{\hat{r}\hat{r}}$, $T_{\hat{\theta}\hat{\theta}}$ and
$T_{\hat{\phi}\hat{\phi}}$. Using the orthonormal basis, these
components carry a simple physical interpretation, i.e.,
\begin{equation}\label{stress-energy-WH}
T_{\hat{t}\hat{t}}=\rho(r)\,,\qquad T_{\hat{r}\hat{r}}=-\tau(r)\,,
\qquad
T_{\hat{\theta}\hat{\theta}}=T_{\hat{\phi}\hat{\phi}}=p(r)\,,
\end{equation}
in which $\rho(r)$ is the energy density, $\tau (r)$ is the radial
tension, with $\tau (r)=-p_r(r)$, i.e., it is the negative of the
radial pressure, $p(r)$ is the pressure measured in the tangential
directions, orthogonal to the radial direction.

Using the Einstein field equation, $G_{\hat{\mu}\hat{\nu}}=8\pi
T_{\hat{\mu}\hat{\nu}}$, we verify the following stress-energy
scenario
\begin{eqnarray}
\rho(r)&=&\frac{1}{8\pi} \, \frac{b'}{r^2}
\label{rhoWH}\,,\\
\tau (r)&=&\frac{1}{8\pi} \, \left[ \frac{b}{r^3}-2
\left(1-\frac{b}{r}
\right) \frac{\Phi'}{r}  \right] \label{tauWH}\,,\\
p(r)&=&\frac{1}{8\pi} \left(1-\frac{b}{r}\right)\left[\Phi ''+
(\Phi')^2- \frac{b'r-b}{2r^2(1-b/r)}\Phi'-
\frac{b'r-b}{2r^3(1-b/r)}+\frac{\Phi'}{r} \right]
\label{pressureWH}\,.
\end{eqnarray}
Evaluated at the throat they assume the following simplified form
\begin{eqnarray}
\rho(r_0)&=&\frac{1}{8\pi} \, \frac{b'(r_0)}{r_0^2}
\,,\\
\tau (r_0)&=&\frac{1}{8\pi r_0^2}  \,,\\
p(r_0)&=&\frac{1}{8\pi}\,\frac{1-b'(r_0)}{2r_0^2}
\;(1+r_0\Phi'(r_0)) \,.
\end{eqnarray}

Integrating Eq. (\ref{rhoWH}), we have
\begin{equation}
b(r)=b(r_0)+\int_{r_0}^r\, 8\pi \,\rho(r')\,r'^2\,dr'=2m(r)  \,.
\end{equation}
This can be expressed in the following manner
\begin{equation}
m(r)=\frac{r_0}{2}+\int_{r_0}^r\, 4\pi \,\rho(r')\,r'^2\,dr'  \,,
\end{equation}
which is the effective mass contained in the interior of a sphere
of radius $r$. Therefore, the form function has an interpretation
which depends on the mass distribution of the wormhole. Moving out
to spatial infinity, we have
\begin{equation}
\lim_{r \rightarrow \infty} m(r)=\frac{r_0}{2}+\int_{r_0}^\infty
\, 4\pi \,\rho(r')\,r'^2\,dr'=M \,.
\end{equation}

By taking the derivative with respect to the radial coordinate
$r$, of Eq. (\ref{tauWH}), and eliminating $b'$ and $\Phi''$,
given in Eq. (\ref{rhoWH}) and Eq. (\ref{pressureWH}),
respectively, we obtain the following equation
\begin{equation}
\tau '=(\rho-\tau )\Phi '-\frac{2}{r}(p+\tau )
\label{tauderivative} \,.
\end{equation}
Equation (\ref{tauderivative}) is the relativistic Euler equation,
or the hydrostatic equation for equilibrium for the material
threading the wormhole, and can also be obtained using the
conservation of the stress-energy tensor,
$T^{\hat{\mu}\hat{\nu}}_{\;\;\;\;;\,\hat{\nu}}=0$, inserting
$\hat{\mu}=r$. The conservation of the stress-energy tensor, in
turn can be deduced from the Bianchi identities,
\begin{equation}\label{Bianchi}
R^{\hat\alpha}{}_{\hat{\beta}[\hat{\lambda}\hat{\mu};\hat\nu]}=0
        \,,
\end{equation}
which are equivalent to $G^{\hat\mu\hat\nu}{}_{;\hat\nu}=0$, also
called the contracted Bianchi identities.

\subsection{Exotic matter}

To gain some insight into the matter threading the wormhole,
Morris and Thorne defined the dimensionless function
$\xi=(\tau-\rho)/|\rho|$ \cite{Morris}. Using equations
(\ref{rhoWH})-(\ref{tauWH}) one finds
\begin{equation}
\xi =\frac{\tau-\rho}{|\rho|}=\frac{b/r-b'- 2r(1-b/r)\Phi'}{|b'|}
\label{exoticity}\,.
\end{equation}
Combining Eq. (\ref{exoticity}) with the flaring-out condition,
Eq. (\ref{flareout}), the exoticity function takes the form
\begin{equation}
\xi =\frac{2b^2}{r|b'|}\;\frac{d^2r}{dz^2}-2r
\left(1-\frac{b}{r}\right)\frac{\Phi'}{|b'|}  \,.
\end{equation}
Considering the finite character of $\rho$, and therefore of $b'$,
and the fact that $(1-b/r)\Phi' \rightarrow 0$ at the throat, we
have the following relationship
\begin{equation}
\xi(r_0) =\frac{\tau_0-\rho_0}{|\rho_0|}>0 \,.
\end{equation}
The restriction $\tau_0>\rho_0$ is an extremely troublesome
condition, as it states that the radial tension at the throat
should exceed the energy density. Thus, Morris and Thorne coined
matter restricted by this condition ``exotic
matter''\cite{Morris}. We shall verify below that this is matter
that violates the null energy condition (in fact, it violates all
the energy conditions)\cite{Morris,Visser,hochvisserWHthroats}.

For instance, consider a specific class of particularly simple
solutions corresponding to the choice of $b=b(r)$ and $\Phi(r)=0$
\cite{Morris}. Equations (\ref{rhoWH})-(\ref{pressureWH}) reduce
to
\begin{eqnarray}
\rho(r)={{b'(r)}\over{8\pi r^2}}\;,
  \qquad
\tau(r)={{b(r)}\over{8\pi r^3}}  \;,
  \qquad
p(r)={{b(r)-b'r}\over{16\pi r^3}}\,.
\end{eqnarray}
Note that the sign of the energy density depends on the sign of
$b'(r)$. In particular, consider the form function given by
$b(r)=r_0^2/ r$. This corresponds to an embedding function $z(r)$
given by $z(r)=r_0 \;{\rm cosh}^{-1}(r/r_0)$, which has the shape
of a catenary, i.e.,
\begin{equation}
{{dz}\over{dr}}={{r_0}\over{\sqrt{r^2-{r_0}^2}}}\,.
\end{equation}
The wormhole material is everywhere exotic, i.e., $\xi>0$
everywhere, extending outward from the throat, with $\rho,\,\tau$,
and $p$ tending to zero as $r\rightarrow +\infty$.

Exotic matter is particularly troublesome for measurements made by
observers traversing through the throat with a radial velocity
close to the speed of light. Consider a Lorentz transformation,
$x^{\hat{\mu}'} = \Lambda^{\hat{\mu}'}{}_{\hat{\nu}} \;
x^{\hat{\nu}}$, with $\Lambda^{\hat{\mu}}{}_{\hat{\alpha}'} \;
\Lambda^{\hat{\alpha}'}{}_{\hat{\nu}} =
\delta^{\hat{\mu}}{}_{\hat{\nu}}$ and
$\Lambda^{\hat{\mu}}{}_{\hat{\nu}'}$ defined as
\begin{equation}
(\Lambda^{\hat{\mu}}{}_{\hat{\nu}'})=\left[
\begin{array}{cccc}
\gamma&0&0&\gamma v \\
0&1&0&0 \\
0&0&1&0 \\
\gamma v&0&0&\gamma
\end{array}
\right]   \label{Lorentzmatrix}\,.
\end{equation}
The energy density measured by these observers is given by
$T_{\hat{0}'\hat{0}'} = \Lambda^{\hat{\mu}}{}_{\hat{0}'}\;
\Lambda^{\hat{\nu}}{}_{\hat{0}'}\; T_{\hat{\mu}\hat{\nu}}$, i.e.,
\begin{eqnarray}
T_{\hat{0}'\hat{0}'}= \gamma^2\,(\rho_0-v^2\tau_0)  \,,
\end{eqnarray}
with $\gamma=(1-v^2)^{-1/2}$. For sufficiently high velocities, $v
\rightarrow 1$, the observer will measure a negative energy
density, $T_{\hat{0}'\hat{0}'}<0$.

This feature also holds for any traversable, nonspherical and
nonstatic wormhole. To see this, one verifies that a bundle of
null geodesics that enters the wormhole at one mouth and emerges
from the other must have a cross-sectional area that initially
increases, and then decreases. This conversion of decreasing to
increasing is due to the gravitational repulsion of matter,
requiring a negative energy density, through which the bundle of
null geodesics traverses.

\subsection{Traversability conditions}

We will be interested in specific solutions for traversable
wormholes and assume that a traveller of an absurdly advanced
civilization, with human traits, begins the trip in a space
station in the lower universe, at proper distance $l=-l_1$, and
ends up in the upper universe, at $l=l_2$. Assume that the
traveller has a radial velocity $v(r)$, as measured by a static
observer positioned at $r$. One may relate the proper distance
travelled $dl$, radius travelled $dr$, coordinate time lapse $dt$,
and proper time lapse as measured by the observer $d\tau$, by the
following relationships
\begin{eqnarray}
v&=&e^{-\Phi}\,\frac{dl}{dt}=\mp\;e^{-\Phi}\,
\left(1-\frac{b}{r}\right)^{-1/2}\frac{dr}{dt}  \,,
             \\
v\,\gamma&=&\frac{dl}{d\tau}=\mp\;
\left(1-\frac{b}{r}\right)^{-1/2}\frac{dr}{d\tau}  \,.
\end{eqnarray}

It is also important to impose certain conditions at the space
stations~\cite{Morris}. Firstly, consider that space is
asymptotically flat at the stations, i.e., $b/r\ll 1$. Secondly,
the gravitational redshift of signals sent from the stations to
infinity should be small, i.e., $\Delta
\lambda/\lambda=e^{-\Phi}-1\approx -\Phi$, so that $|\Phi| \ll 1$.
The condition $|\Phi| \ll 1$ imposes that the proper time at the
station equals the coordinate time. Thirdly, the gravitational
acceleration measured at the stations, given by $g=-(1-b/r)^{-1/2}
\, \Phi' \simeq -\Phi'$, should be less than or equal to the
Earth's gravitational acceleration, $g\leq g_{\oplus}$, so that
the condition $|\Phi'|\leq g_{\oplus}$ is met.

For a convenient trip through the wormhole, certain conditions
should also be imposed~\cite{Morris}. Firstly, the entire journey
should be done in a relatively short time as measured both by the
traveller and by observers who remain at rest at the stations.
Secondly, the acceleration felt by the traveller should not exceed
the Earth's gravitational acceleration, $g_{\oplus}$. Finally, the
tidal accelerations between different parts of the traveller's
body, should not exceed, once again, Earth's gravity.

\begin{description}

\item[{\bf Total time in a traversal}.]


The trip should take a relatively short time, for instance Morris
and Thorne considered one year, as measured by the traveler and
for observers that stay at rest at the space stations, $l=-l_1$
and $l=l_2$, i.e.,
\begin{eqnarray}
\Delta \tau_{\rm traveler} &=&\int_{-l_1}^{+l_2}
\frac{dl}{v\gamma} \leq 1\,\,{\rm year}
\label{travelertime}, \\
\Delta t_{\rm space\,station} &=&\int_{-l_1}^{+l_2} \frac{dl}{v
e^{\Phi}} \leq 1\,\,{\rm year} \label{observertime},
\end{eqnarray}
respectively.

\item[{\bf Acceleration felt by a traveler}.]


An important traversability condition required is that the
acceleration felt by the traveller should not exceed Earth's
gravity \cite{Morris}. Consider an orthonormal basis of the
traveller's proper reference frame, $({\bf e}_{\hat{0}'},{\bf
e}_{\hat{1}'},{\bf e}_{\hat{2}'},{\bf e}_{\hat{3}'})$, given in
terms of the orthonormal basis vectors of Eqs.
(\ref{orthonormalbasis}) of the static observers, by a Lorentz
transformation, i.e.,
\begin{eqnarray}
{\bf e}_{\hat{0}'}=\gamma \,{\bf e}_{\hat{t}}\mp \gamma \,v\,{\bf
e}_{\hat{r}}  \;,
\qquad
{\bf e}_{\hat{1}'}=\mp \,\gamma \,{\bf e}_{\hat{r}} + \gamma
\,v\,{\bf e}_{\hat{t}}  \;,
\qquad
{\bf e}_{\hat{2}'}={\bf e}_{\hat{\theta}}  \;,\qquad {\bf
e}_{\hat{3}'}={\bf e}_{\hat{\phi}}  \;,
\end{eqnarray}
where $\gamma=(1-v^2)^{-1/2}$, and $v(r)$ being the velocity of
the traveller as he passes $r$, as measured by a static observer
positioned there. Thus, the traveller's four-acceleration
expressed in his proper reference frame,
$a^{\hat{\mu}'}=U^{\hat{\nu}'} U^{\hat{\mu}'}_{\;\;\;;
\hat{\nu}'}$, yields the following restriction
\begin{equation}
|\vec{a}|=\left |\left (1-\frac{b}{r}\right)^{1/2}
e^{-\Phi}\,\left(\gamma \,e^{\Phi}\right)' \right|\leq g_{\oplus}
\label{acceleration} \,.
\end{equation}

For the particular case of $\Phi'=0$, this restriction reduces to
\begin{equation}
|\vec{a}|=\left |\left (1-\frac{b}{r}\right)^{1/2} \gamma'c^2
\right|\leq g_{\oplus} \label{acceleration-Phi-const} \,.
\end{equation}
For observers traversing the wormhole with a constant velocity,
$v={\rm const}$, one has $|\vec{a}|=0$, of course!

\item[{\bf Tidal acceleration felt by a traveler}.]


Its important that an observer traversing through the wormhole
should not be ripped apart by enormous tidal forces. Thus, another
of the traversability conditions required is that the tidal
accelerations felt by the traveller should not exceed, for
instance, the Earth's gravitational acceleration \cite{Morris}.
The tidal acceleration felt by the traveller is given by $\Delta
a^{\hat{\mu}'}=-R^{\hat{\mu}'}_{\;\;\hat{\nu}'\hat{\alpha}'\hat{\beta}'}
\,U^{\hat{\nu}'}\eta^{\hat{\alpha}'}U^{\hat{\beta}'}$, where
$U^{\hat{\mu}'}=\delta^{\hat{\mu}'}_{\;\;\hat{0}'}$ is the
traveller's four velocity and $\eta^{\hat{\alpha}'}$ is the
separation between two arbitrary parts of his body. Note that
$\eta^{\hat{\alpha}'}$ is purely spatial in the traveller's
reference frame, as $U^{\hat{\mu}'}\eta_{\hat{\mu}'}=0$, so that
$\eta^{\hat{0}'}=0$. For simplicity, assume that
$|\eta^{\hat{i}'}|\approx 2\,{\rm m}$ along any spatial direction
in the traveller's reference frame. Taking into account the
antisymmetric nature of
$R^{\hat{\mu}'}_{\;\;\hat{\nu}'\hat{\alpha}'\hat{\beta}'}$ in its
first two indices, we verify that $\Delta a^{\hat{\mu}'}$ is
purely spatial with the components
\begin{equation}
\Delta a^{\hat{i}'}=-R^{\hat{i}'}{}_{\hat{0}'\hat{j}'\hat{0}'}
\,\eta^{\hat{j}'}=-R_{\hat{i}'\hat{0}'\hat{j}'\hat{0}'}
\,\eta^{\hat{j}'}. \label{spatial}
\end{equation}

By using a Lorentz transformation of the Riemann tensor components
in the static observer's frame,  $({\bf e}_{\hat{t}},{\bf
e}_{\hat{r}},{\bf e}_{\hat{\theta}},{\bf e}_{\hat{\phi}})$, to the
traveller's frame, $({\bf e}_{\hat{0}'},{\bf e}_{\hat{1}'},{\bf
e}_{\hat{2}'},{\bf e}_{\hat{3}'})$, the nonzero components of
$R_{\hat{i}'\hat{0}'\hat{j}'\hat{0}'}$ are given by
\begin{eqnarray}
R_{\hat{1}'\hat{0}'\hat{1}'\hat{0}'}&=&
R_{\hat{r}\hat{t}\hat{r}\hat{t}}
           \nonumber     \\
&=&-\left(1-\frac{b}{r} \right)\left[-\Phi ''- (\Phi')^2+
\frac{b'r-b}{2r(r-b)}\Phi' \right]\,,
       \\
R_{\hat{2}'\hat{0}'\hat{2}'\hat{0}'}&=&
R_{\hat{3}'\hat{0}'\hat{3}'\hat{0}'} \;=\;\gamma^2 \,
R_{\hat{\theta}\hat{t}\hat{\theta}\hat{t}}+\gamma^2 \,v^2 \,
R_{\hat{\theta}\hat{r}\hat{\theta}\hat{r}}
    \nonumber      \\
&=&\frac{\gamma ^2}{2r^2} \left [v^2\left (b'-\frac{b}{r} \right
)+2(r-b)\Phi ' \right] \label{lateralRiemann} \,.
\end{eqnarray}

Thus, Eq. (\ref{spatial}) takes the form
\begin{eqnarray}
\Delta a^{\hat{1}'}=-R_{\hat{1}'\hat{0}'\hat{1}'\hat{0}'}
\;\eta^{\hat{1}'} , \qquad \Delta
a^{\hat{2}'}=-R_{\hat{2}'\hat{0}'\hat{2}'\hat{0}'}
\;\eta^{\hat{2}'} ,  \qquad \Delta
a^{\hat{3}'}=-R_{\hat{3}'\hat{0}'\hat{3}'\hat{0}'}
\;\eta^{\hat{3}'}  .
\end{eqnarray}
The constraint $|\Delta a^{\hat{\mu}'}|\leq g_{\oplus}$ provides
the tidal acceleration restrictions as measured by a traveller
moving radially through the wormhole, given by the following
inequalities
\begin{eqnarray}
\left |\left (1-\frac{b}{r} \right ) \left [\Phi ''+(\Phi ')^2-
\frac{b'r-b}{2r(r-b)}\Phi' \right] \right
|\,\big|\eta^{\hat{1}'}\big| &\leq & g_\oplus   \,,
    \label{radialtidalconstraint}    \\
\left | \frac{\gamma ^2}{2r^2} \left [v^2\left (b'-\frac{b}{r}
\right )+2(r-b)\Phi ' \right] \right | \,\big|\eta^{\hat{2}'}\big|
&\leq &  g_\oplus    \,.    \label{lateraltidalconstraint}
\end{eqnarray}
The radial tidal constraint, Eq. (\ref{radialtidalconstraint}),
constrains the redshift function, and the lateral tidal
constraint, Eq. (\ref{lateraltidalconstraint}), constrains the
velocity with which observers traverse the wormhole. These
inequalities are particularly simple at the throat, $r_0$,
\begin{eqnarray}
|\Phi '(r_0)| &\leq &
\frac{2g_{\oplus}\,r_0}{(1-b')\,|\eta^{\hat{1}'}|} \,,
          \\
\gamma^2 v^2 &\leq &
\frac{2g_{\oplus}\,r_0^2}{(1-b')\,|\eta^{\hat{2}'}|}    \,,
\end{eqnarray}

For the particular case of a constant redshift function,
$\Phi'=0$, the radial tidal acceleration is zero, and Eq.
(\ref{lateraltidalconstraint}) reduces to
\begin{eqnarray}
\frac{\gamma ^2 v^2}{2r^2}\left |\left (b'-\frac{b}{r} \right )
\right | \,\big|\eta^{\hat{2}'}\big| &\leq & g_\oplus \,,
\label{lateraltidal-Phi-const}
\end{eqnarray}
For this specific case one verifies that stationary observers with
$v=0$ measure null tidal forces.

It is interesting to note that if the tidal forces are velocity
independent, then the wormhole is not traversable. For instance,
consider the lateral tidal constraint, Eq.
(\ref{lateraltidalconstraint}), which is the only component of the
tidal acceleration that is velocity dependent. This velocity
dependence cancels out if and only if
$R_{\hat{\theta}\hat{t}\hat{\theta}\hat{t}}
=-R_{\hat{\theta}\hat{r}\hat{\theta}\hat{r}}$ (see Eq.
(\ref{lateralRiemann})), or
\begin{eqnarray}
b'-\frac{b}{r}=-2r\left(1-\frac{b}{r} \right)\Phi' \,.
\end{eqnarray}
Integrating this restriction yields
\begin{eqnarray}
e^{2\Phi(r)}=e^{2\Phi(\infty)}\left(1-\frac{b}{r} \right) \,,
\end{eqnarray}
which indicates that a horizon is present at $r=r_0$, and that the
wormhole is not traversable.

\end{description}

\subsection{Energy conditions}

\subsubsection{Pointwise energy conditions}

Given the fact that wormhole spacetimes are supported by exotic
matter, we shall specify the energy conditions for the specific
case in which the stress-energy tensor is
diagonal~\cite{hawkingellis}, i.e.,
\begin{equation}
 T^{\mu\nu}={\rm diag}(\rho,p_1,p_2,p_3)\,,
  \label{diagonalT}
\end{equation}
where $\rho$ is the mass density and the $p_j$ are the three
principal pressures. In the case that $p_{1}=p_{2}=p_{3}$ this
reduces to the perfect fluid stress-energy tensor. Although
classical forms of matter are believed to obey these energy
conditions, it is a well-known fact that they are violated by
certain quantum fields, amongst which we may refer to the Casimir
effect.

\begin{description}

\item[{\bf Null energy condition (NEC)}.]

The NEC asserts that for any null vector $k^{\mu}$
\begin{equation}
 \label{eq:nec}
  T_{\mu\nu}k^{\mu}k^{\nu}\geq 0  \,.
\end{equation}
In the case of a stress-energy tensor of the form
(\ref{diagonalT}), we have
\begin{equation}\label{eq:necpf}
\forall \,i\,, \quad \rho+p_{i}\geq 0 \,.
\end{equation}

\item[{\bf Weak energy condition (WEC)}.]

The WEC states that for any timelike vector $U^{\mu}$
\begin{equation}
 \label{eq:wec}
  T_{\mu\nu}U^{\mu}U^{\nu}\geq 0   \,.
\end{equation}
One can physically interpret $T_{\mu\nu}U^{\mu}U^{\nu}$ as the
energy density measured by any timelike observer with
four-velocity $U^{\mu}$. Thus, the WEC requires that this quantity
to be positive.  In terms of the principal pressures this gives
\begin{equation}
 \label{eq:wecpf}
  \rho\geq 0 \quad \mbox{and} \quad
\forall \,i\,, \quad  \rho+p_{i}\geq 0   \,.
\end{equation}
By continuity, the WEC implies the NEC.

\item[{\bf Strong energy condition (SEC)}.]

The SEC asserts that for any timelike vector $U^{\mu}$ the
following inequality holds
\begin{equation}
 \label{eq:sec}
  \left( T_{\mu\nu}-\frac{T}{2}\;g_{\mu\nu}\right)U^{\mu}U^{\nu}\geq
  0  \,,
\end{equation}
where $T$ is the trace of the stress energy tensor.

In terms of the diagonal stress energy tensor (\ref{diagonalT})
the SEC reads
\begin{equation}
 \label{eq:secpf}
 \forall \,i\,, \quad \rho+p_{i}\geq 0 \quad \mbox{and} \quad \rho+\sum_{i}p_{i}\geq 0
   \,.
\end{equation}
The SEC implies the NEC but not necessarily the WEC.

\item[{\bf Dominant energy condition (DEC)}.]

The DEC states that for any timelike vector $U^{\mu}$
\begin{equation}
 \label{eq:dec}
  T_{\mu\nu}U^{\mu}U^{\nu}\geq 0 \quad \mbox{and}
     \quad T_{\mu\nu}U^{\nu}\:\mbox{is not spacelike}
\end{equation}
These conditions imply that the locally observed energy density be
positive and that the energy flux should be timelike or null. The
DEC implies the WEC, and therefore the NEC, but not necessarily
the SEC. In the case of a stress-energy tensor of the form
(\ref{diagonalT}), we have
\begin{equation}
 \label{eq:decpf}
  \rho\geq 0 \quad \mbox{and} \quad \forall \,i\,, \quad p_{i}\in
  [-\rho,+\rho]\,.
\end{equation}

\end{description}

One may readily verify that wormhole spacetimes violate all the
pointwise energy conditions. Taking into account Eqs.
(\ref{rhoWH})-(\ref{tauWH}), we have
\begin{equation}\label{NECthroat}
\rho(r)-\tau(r)= \frac{1}{8\pi}\,\left[\frac{b'r-b}{r^3}+
2\left(1-\frac{b}{r}\right) \frac{\Phi '}{r} \right]  .
\end{equation}
Due to the flaring out condition of the throat deduced from the
mathematics of embedding, Eq. (\ref{flareout}), i.e.,
$(b-b'r)/b^2>0$ \cite{Morris,LLQ-PRD,Visser}, we verify that at
the throat $b(r_0)=r=r_0$, and due to the finiteness of $\Phi(r)$,
from Eq. (\ref{NECthroat}) we have $\rho(r)-\tau(r)<0$. From this
we verify that all the energy conditions are violated. However,
Eq. (\ref{NECthroat}) is precisely the definition of the NEC,
i.e.,
$T_{\hat{\mu}\hat{\nu}}k^{\hat{\mu}}k^{\hat{\nu}}=\rho(r)-\tau(r)$,
with $k^{\hat{\mu}}=(1,1,0,0)$. Matter that violates the NEC is
denoted as ``exotic matter''.

\subsubsection{Averaged energy conditions}

Violations of the pointwise energy conditions led to the averaging
of the energy conditions over timelike or null geodesics
\cite{Tipler-AEC}. The averaged energy conditions are somewhat
weaker than the pointwise energy conditions, as they permit
localized violations of the energy conditions, as long on average
the energy conditions hold when integrated along timelike or null
geodesics.

\begin{description}

\item[{\bf Averaged null energy condition (ANEC)}.]

The ANEC is satisfied along a null curve, $\Gamma$, if the
following holds
\begin{equation}
 \int_{\Gamma} T_{\mu\nu}k^{\mu}k^{\nu}\, d\lambda \geq 0  \,,
\end{equation}
where $\lambda$ is the generalized affine parameter, and $k^{\mu}$
is a null vector. If the curve $\Gamma$ is a null geodesic, then
$\lambda$ is reduced to the ordinary affine parameter.

\item[{\bf Averaged weak energy condition (AWEC)}.]

The AWEC is satisfied along a timelike curve, $\Gamma$, if
\begin{equation}
 \int_{\Gamma} T_{\mu\nu}U^{\mu}U^{\nu}\, ds \geq 0  \,,
\end{equation}
where $s$ is a parameterization, the proper time of the timelike
curve, and $U^{\mu}$ is the respective tangent vector.

\end{description}
It can be shown, under general conditions, that traversable
wormholes violate the ANEC in the region of the throat using the
Raychaudhuri equation for null
geodesics~\cite{Visser,hawkingellis,Wald}.

\subsubsection{Volume Integral Quantifier}

Unfortunately the ANEC involves a line integral, with dimensions
(mass)/(area), not a volume integral, and therefore gives no
useful information regarding the ``total amount'' of
energy-condition violating matter. Therefore, this prompted Visser
{\it et al} \cite{visser2003,Kar2} to propose a ``volume integral
quantifier'' which amounts to calculating the following definite
integrals
\begin{equation}
\int T_{\mu\nu}\,U^{\mu}\,U^{\nu}\;d V  \quad  {\rm and} \quad
\int T_{\mu\nu}\,k^{\mu}\,k^{\nu}\;d V .
\end{equation}
The amount of energy condition violations is then the extent that
these integrals become negative. A more precise measure was
analyzed in Ref. \cite{Nandi:2004ku} by considering the proper
volume in the integral.

To develop the key volume-integral result note that
$T_{\hat{\mu}\hat{\nu}}\,k^{\hat{\mu}}\,k^{\hat{\nu}}$, with the
null vector given by $k^{\hat{\mu}}=(1,1,00)$, can be written in
the following manner
\begin{equation}
\rho-\tau= {1\over8\pi r}  \left( 1 - {b\over r} \right) \left[
\ln\left(\frac{e^{2\Phi}}{1-b/r}\right) \right]'.
\end{equation}
Thus, integrating by parts, we have
\begin{eqnarray}
I_V=\int (\rho-\tau) \;d V  = \left[ (r-b)
\ln\left(\frac{e^{2\Phi}}{1-b/r}\right) \right]_{r_0}^\infty \;\;
- \int_{r_0}^\infty (1-b') \left[
\ln\left(\frac{e^{2\Phi}}{1-b/r}\right) \right] dr.
\label{integral}
\end{eqnarray}
The boundary term at $r_0$ vanishes by the construction of the
wormhole, and the boundary term at infinity also vanishes because
of the assumed asymptotic behaviour.  Thus, Eq. (\ref{integral})
reduces to
\begin{equation}\label{VolInt}
I_V=\int (\rho-\tau) \;d V = - \int_{r_0}^\infty (1-b') \left[
\ln\left(\frac{e^{2\Phi}}{1-b/r}\right) \right] d r.
\label{VolIntQuant}
\end{equation}
This volume-integral theorem provides information about the
``total amount'' of ANEC violating matter in the spacetime, and
one may now consider specific cases, by choosing the form function
and the redshift function.

Consider the solution obtained by setting the following choices
for the redshift function and form function: $\Phi=0$ and
$b={r_0}^2/r$, respectively. In fact, this is a solution obtained
by Homer Ellis \cite{homerellis} in 1973. The properties are
commented in \cite{Morris} and \cite{clement1}. Harris showed that
it is a solution of the EFE with a stress-energy tensor of a
peculiar massless scalar field \cite{harris}. In terms of the
proper radial distance $l(r)$, the metric takes the form
\begin{equation}
ds^2= -dt^2 + dl^2
                + ({r_0}^2 +l^2) \,({d\theta}^2+
{\rm sin}^2\theta\,{d\phi}^2)\,,  \label{Ellisdrainhole}
\end{equation}
where $l=\pm(r^2-{r_0}^2)^{1/2}$. The stress-tensor components are
given by
\begin{equation}
\rho= -\tau= -p=\,- { {r_0}^2 \over {8\pi r^4} }= - { {r_0}^2
\over {8\pi {({r_0}^2 + l^2)}^2 } } \,.   \label{stressdrainhole}
\end{equation}

Suppose now that the wormhole extends from the throat, $r_0$, to a
radius situated at $a$. Evaluating the volume integral, Eq.
(\ref{VolInt}), one deduces
\begin{eqnarray}
I_V=\int (\rho -\tau)\; dV= \frac{1}{a} \left[\left(a^2-r_0^2
\right) \; \ln \left(1-\frac{r_0^2}{a^2} \right) +2r_0 (r_0-a)
\right] .
\end{eqnarray}
Taking the limit as $a\rightarrow r_0^+$, one verifies that $\int
(\rho -\tau)\; dV \rightarrow 0$. Thus, as in the examples
presented in \cite{visser2003,Kar2}, one may construct a
traversable wormhole with arbitrarily small quantities of ANEC
violating matter. The exotic matter threading the wormhole extends
from the throat at $r_0$ to the junction boundary situated at $a$,
where the interior solution is matched to an exterior vacuum
spacetime.

\subsubsection{Quantum Inequality}\label{Sec:QI}

Pioneering work by Ford in the late 1970's on a new set of energy
constraints \cite{ford1}, led to constraints on negative energy
fluxes in 1991 \cite{ford2}. These eventually culminated in the
form of the Quantum Inequality (QI) applied to energy densities,
which was introduced by Ford and Roman in 1995 \cite{fordroman1}.
The QI was proven directly from Quantum Field Theory, in
four-dimensional Minkowski spacetime, for free quantized, massless
scalar fields and takes the following form
\begin{equation}\label{QI}
\frac{\tau_0}{\pi}\int_{-\infty}^{+\infty}\frac{
\langle{T_{\mu\nu}U^{\mu}U^{\nu}}\rangle} {\tau^2+{
\tau^2_0}}d\tau\geq-\frac{3}{32\pi^2\tau^4_0} \,,
\end{equation}
in which, $U^\mu$ is the tangent to a geodesic observer's
wordline; $\tau$ is the observer's proper time and $\tau_0$ is a
sampling time. The expectation value $\langle\rangle$ is taken
with respect to an arbitrary state $|\Psi\rangle$. Contrary to the
averaged energy conditions, one does not average over the entire
wordline of the observer, but weights the integral with a sampling
function of characteristic width, $\tau_0$. The inequality limits
the magnitude of the negative energy violations and the time for
which they are allowed to exist. The physical interpretation of
Eq. (\ref{QI}) is that the more negative the energy density is in
an interval, the shorter must be the duration of the interval.

The basic applications to curved spacetimes is that these appear
flat if restricted to a sufficiently small region. The application
of the QI to wormhole geometries is of particular interest
\cite{fordroman2}. A small spacetime volume around the throat of
the wormhole is considered, so that all the dimensions of this
volume are much smaller than the minimum proper radius of
curvature in the region. Thus, the spacetime can be approximately
flat in this region, so that the QI constraints may be applied.
The sampling time $\tau_0$ is restricted to be small compared to
the local proper radii of curvature and the proper distance to any
boundaries in the spacetime.

The Riemann tensor components will play a fundamental role in the
analysis that follows. At the throat, the components of the
Riemann tensor reduce to
\begin{eqnarray}
R_{\hat{t}\hat{r}\hat{t}\hat{r}}|_{r_0}
&=& { {\Phi'_0} \over {2r_0} } \, (1 - b'_0)  \,,     \\
R_{\hat{t}\hat{\theta}\hat{t}\hat{\theta}}|_{r_0} &=&
R_{\hat{t}\hat{\phi}\hat{t}\hat{\phi}}|_{r_0}
= 0 \,,                       \\
R_{\hat{r}\hat{\theta}\hat{r}\hat{\theta}}|_{r_0} &=&
R_{\hat{r}\hat{\phi}\hat{r}\hat{\phi}}|_{r_0}
= -{1 \over {2 {r_0}^2} } \, (1 - b'_0) \,,     \\
R_{\hat{\theta}\hat{\phi}\hat{\theta}\hat{\phi}}|_{r_0} &=& {1
\over {{r_0}^2} } \,.
\end{eqnarray}
All the other components vanish, except for those related to the
above by symmetry.

Let the magnitude of the maximum curvature component be $R_{\rm
max}$, and the smallest proper radius of curvature be given by
$r_c\approx 1/\sqrt{R_{\rm max}}$. The QI-bound is applied to a
small spacetime volume around the throat of the wormhole such that
all dimensions of this volume are much smaller than $r_c$, the
smallest proper radius of curvature anywhere in the region, so
that in the absence of boundaries, spacetime can be considered to
be approximately Minkowskian in the respective
region~\cite{fordroman2}.

As specific example, consider QI-bound applied to the Ellis
drainhole geometry. Consider the Ellis drainhole, given by
$\Phi=0$ and $b={r_0}^2/r$. The metric is given by Eq.
(\ref{Ellisdrainhole}), and the Riemann curvature components are
\begin{equation}
R_{\hat{\theta}\hat{\phi}\hat{\theta}\hat{\phi}}
=-R_{\hat{l}\hat{\theta}\hat{l}\hat{\theta}}
=-R_{\hat{l}\hat{\phi}\hat{l}\hat{\phi}} ={ {r_0}^2 \over {
{({r_0}^2 + l^2)}^2 } } \,.   \label{Riemdrainhole}
\end{equation}
Note that all the curvature components are equal in magnitude, and
have their maximum magnitude at the throat, i.e., $1/r_0^2$. The
same holds true for the stress-tensor components given by Eq.
(\ref{stressdrainhole}).

Applying the QI-bound to a static observer at $r=r_0$, and as the
energy density seen by this static observer is constant, we have
\begin{equation}
{{\tau_0} \over \pi}\, \int_{-\infty}^{\infty}\, {{\langle
T_{\mu\nu} u^{\mu} u^{\nu}\rangle\, d\tau} \over
{{\tau}^2+{\tau_0}^2}} = \rho_0 \geq - \frac{3}{32\pi^2 \,
{\tau_0}^4}
 \,, \label{QIdrainhole}
\end{equation}
where $\tau$ is the observer's proper time, and $\tau_0$ is the
sampling time. If the sampling time is chosen to be $\tau_0 = f
r_m=f r_0 \ll r_c$, with $f \ll 1$ (recall that the QI is
applicable if $\tau_0$ is smaller than the local proper radius of
curvature), using $\rho_0=-1/(8\pi r_0^2)$ and from
Eq.~(\ref{QIdrainhole}), one finally deduces
\begin{equation}
r_0 \leq { {l_p} \over {2 f^2} } \,, \label{drainrest}
\end{equation}
where $l_p$ is the Planck length. For any reasonable choice of $f$
gives a value of $r_0$ which is not much larger than $l_p$. For
example, for $f \approx 0.01$ one has $r_0 \leq 10^4 \, l_p =
10^{-31} \,{\rm m}$. Note from Eqs.~(\ref{stressdrainhole}) and
{}~(\ref{Riemdrainhole}) that if the spacetime region is such that
$l \ll r_0$, then the curvature and stress-tensor components do
not change very much~\cite{fordroman2}.

Ford and Roman considered more specific examples by choosing
appropriate definitions of length scales. They also found general
bounds on the relative size scales of arbitrary static and
spherically symmetric Morris-Thorne wormholes, i.e., for generic
$\Phi(r)$ and $b(r)$. The results of the analysis is that either
the wormhole possesses a throat size which is only slightly larger
than the Planck length, or there are large discrepancies in the
length scales which characterize the geometry of the wormhole. The
analysis imply that generically the exotic matter is confined to
an extremely thin band, and/or that large red-shifts are involved,
which present severe difficulties for traversability, such as
large tidal forces \cite{fordroman2}.

Due to these results, Ford and Roman argued that the existence of
macroscopic traversable wormholes is very improbable. But, there
are a series of considerations that can be applied to the
QI~\cite{World-Scientific}. Firstly, the QI is only of interest if
one is relying on quantum field theory to provide the exotic
matter to support the wormhole throat. But there are classical
systems (non-minimally coupled scalar fields) that violate the
null and the weak energy conditions \cite{barcelovisser1}, whilst
presenting plausible results when applying the QI. Secondly, even
if one relies on quantum field theory to provide exotic matter,
the QI does not rule out the existence of wormholes, although they
do place serious constraints on the geometry. Thirdly, it may be
possible to reformulate the QI in a more transparent covariant
notation, and to prove it for arbitrary background geometries.

\subsection{Rotating wormholes}


Now, consider the stationary and axially symmetric
$(3+1)-$dimensional spacetime, it possesses a time-like Killing
vector field, which generates invariant time translations, and a
spacelike Killing vector field, which generates invariant
rotations with respect to the angular coordinate $\phi$. We have
the following metric
\begin{equation}\label{3rwh}
  ds^2=-N^2dt^2+e^{\mu}\,dr^2+r^2K^2[d\theta^2+\sin^2\theta(d\phi-\omega\,dt)^2]
\end{equation}
where $N$, $K$, $\omega$ and $\mu$ are functions of $r$ and
$\theta$~\cite{teo}. $\omega(r,\theta)$ may be interpreted as the
angular velocity $ d\phi/ dt$ of a particle that falls freely from
infinity to the point $(r,\theta)$.
For simplicity, we shall consider the definition \cite{teo}
\begin{equation}
e^{-\mu(r,\theta)}=1-\frac{b(r,\theta)}{r}\,,
\end{equation}
which is well suited to describe a traversable wormhole.
Assume that $K(r,\theta)$ is a positive, nondecreasing function of
$r$ that determines the proper radial distance $R$, i.e., $R\equiv
rK$ and $R_r>0$ \cite{teo}, as for the $(2+1)-$dimensional case.
We shall adopt the notation that the subscripts $_r$ and
$_{\theta}$ denote the derivatives in order of $r$ and ${\theta}$,
respectively \cite{teo}.

Note that an event horizon appears whenever $N=0$~\cite{teo}. The
regularity of the functions $N$, $b$ and $K$ are imposed, which
implies that their $\theta$ derivatives vanish on the rotation
axis, $\theta=0,\,\pi$, to ensure a non-singular behavior of the
metric on the rotation axis. The metric (\ref{3rwh}) reduces to
the Morris-Thorne spacetime metric (\ref{metricwormhole}) in the
limit of zero rotation and spherical symmetry
\begin{eqnarray}
N(r,\theta)\rightarrow{\rm e}^{\Phi(r)},\quad
b(r,\theta)\rightarrow b(r)\,,
       \quad
K(r,\theta)\rightarrow1\,, \quad \omega(r,\theta)\rightarrow0\,.
\end{eqnarray}
In analogy with the Morris-Thorne case, $b(r_0)=r_0$ is identified
as the wormhole throat, and the factors $N$, $K$ and $\omega$ are
assumed to be well-behaved at the throat.

The scalar curvature of the space-time (\ref{3rwh}) is extremely
messy, but at the throat $r=r_0$ simplifies to
\begin{eqnarray}\label{rotWHRicciscalar}
R&=&-\frac{1}{r^2K^2}\left(\mu_{\theta\theta}
+\frac{1}{2}\mu_\theta^2\right)
-\frac{\mu_\theta}{Nr^2K^2}\,\frac{(N
\sin\theta)_\theta}{\sin\theta}
           -\frac{2}{Nr^2K^2}\,\frac{(N_{\theta}
\sin\theta)_\theta}{\sin\theta}
-\frac{2}{r^2K^3}\,\frac{(K_\theta \sin\theta)_\theta}{\sin\theta}
           \nonumber    \\
&&+e^{-\mu}\,\mu_r\,\left[\ln(Nr^2K^2)\right]_r
+\frac{\sin^2\theta\,\omega_\theta^2}{2N^2}
+\frac{2}{r^2K^4}\,(K^2+K_\theta^2)  \,.
\end{eqnarray}
The only troublesome terms are the ones involving the terms with
$\mu_\theta$ and $\mu_{\theta\theta}$, i.e.,
\begin{equation}
\mu_\theta=\frac{b_\theta}{(r-b)}\,,   \qquad \mu_{\theta\theta}
+\frac{1}{2}\mu_\theta^2=\frac{b_{\theta\theta}}{r-b}
+\frac{3}{2}{b_\theta{}^2\over(r-b)^2}\,.
\end{equation}
Note that one needs to impose that $b_\theta=0$ and
$b_{\theta\theta}=0$ at the throat to avoid curvature
singularities. This condition shows that the throat is located at
a constant value of $r$.

Thus, one may conclude that the metric (\ref{3rwh}) describes a
rotating wormhole geometry, with an angular velocity $\omega$. The
factor $K$ determines the proper radial distance. $N$ is the
analog of the redshift function in the Morris-Thorne wormhole and
is finite and nonzero to ensure that there are no event horizons
or curvature singularities. $b$ is the shape function which
satisfies $b\leq r$; it is independent of $\theta$ at the throat,
i.e., $b_\theta=0$; and obeys the flaring out condition $b_r<1$.

The analysis is simplified using an orthonormal reference frame,
with the following orthonormal basis vectors
\begin{eqnarray}
{\bf e}_{\hat{t}}=\frac{1}{N}\,{\bf e}_{t}+\frac{\omega}{N}\,{\bf
e}_{\phi} \,,  \quad {\bf
e}_{\hat{r}}=\left(1-\frac{b}{r}\right)^{1/2}\,{\bf e}_{r}  \,,
\quad
{\bf e}_{\hat{\theta}}=\frac{1}{rK}\,{\bf e}_{\theta}  \,, \quad
{\bf e}_{\hat{\phi}}=\frac{1}{rK\sin\theta}\,{\bf e}_{\phi} \,.
\end{eqnarray}
Now the stress-energy tensor components are extremely messy, but
assume a more simplified form using the orthonormal reference
frame and evaluated at the throat. They have the following
non-zero components
\begin{eqnarray}
8\pi T_{\hat{t}\hat{t}}&=&-\frac{(K_\theta
\sin\theta)_\theta}{r^2K^3\sin\theta}
-\frac{\omega_\theta^2\,\sin^2\theta}{4N^2}
+e^{-\mu}\,\mu_r\,\frac{(rK)_r}{rK}
       +\frac{K^2+K_\theta^2}{r^2K^4}
    \,,   \label{rotGtt}
\\
8\pi T_{\hat{r}\hat{r}}&=&\frac{(K_\theta
\sin\theta)_\theta}{r^2K^3\sin\theta}
-\frac{\omega_\theta^2\,\sin^2\theta}{4N^2}
+\frac{(N_\theta \sin\theta)_\theta}{Nr^2K^2\sin\theta}
-\frac{K^2+K_\theta^2}{r^2K^4}
   \,,
\\
8\pi T_{\hat{\theta}\hat{\theta}}&=& \frac{N_\theta(K
\sin\theta)_\theta}{Nr^2K^3\sin\theta}
+\frac{\omega_\theta^2\,\sin^2\theta}{4N^2}
-\frac{\mu_r\,e^{-\mu}(NrK)_r}{2NrK}
   \,,
\\
8\pi T_{\hat{\phi}\hat{\phi}}&=&
-\frac{\mu_r\,e^{-\mu}\,(NKr)_r}{2NKr}-\frac{3\sin^2\theta\,\omega_\theta^2}{4N^2}
+\frac{N_{\theta\theta}}{Nr^2K^2}-\frac{N_{\theta}K_{\theta}}{Nr^2K^3}
\,,    \label{rotGphiphi}
\\
8\pi
T_{\hat{t}\hat{\phi}}&=&\frac{1}{4N^2K^2r}\;\Big(6NK\,\omega_{\theta}\,\cos\theta
+2NK\,\sin \theta\,\omega_{\theta \theta}
       \nonumber     \\
&& -\mu_{r}e^{-\mu}r^2NK^3\,\sin\theta\; \omega_{r}
+4N\,\omega_\theta\,\sin\theta\,K_\theta
-2K\,\sin\theta\,N_{\theta}\,\omega_{\theta}  \Big)   \,.
           \label{rotGtphi}
\end{eqnarray}
The components $T_{\hat{t}\hat{t}}$ and $T_{\hat{i}\hat{j}}$ have
the usual physical interpretations, and in particular,
$T_{\hat{t}\hat{\phi}}$ characterizes the rotation of the matter
distribution. Taking into account the Einstein tensor components
above, the NEC at the throat is given by
\begin{eqnarray}\label{NEC}
8\pi\,T_{\hat{\mu} \hat{\nu}}k^{\hat{\mu}} k^{\hat{\nu}}={\rm
e}^{-\mu}\mu_r{(rK)_r\over rK}
-{\omega_\theta{}^2\sin^2\theta\over2N^2}
+{(N_\theta\sin\theta)_\theta\over(rK)^2N\sin\theta}\,.
\end{eqnarray}

Rather than reproduce the analysis here, we refer the reader to
Ref. \cite{teo}, where it was shown that the NEC is violated in
certain regions, and is satisfied in others. Thus, it is possible
for an infalling observer to move around the throat, and avoid the
exotic matter supporting the wormhole. However, it is important to
emphasize that one cannot avoid the use of exotic matter
altogether.


\subsection{Evolving wormholes in a cosmological background}

Consider the metric element of a wormhole in a cosmological
background given by
\begin{equation}\label{evolvingWHmetric}
ds^{2} = \Omega ^{2}(t) \left[- e ^{2\Phi(r)}\, dt^{2} +
{{dr^{2}}\over {1-kr^2- \frac{b(r)}{r}}} + r^2 \,\left(d\theta
^2+\sin ^2{\theta} \, d\phi ^2 \right) \right]
\end{equation}
where $\Omega ^{2}(t)$ is the conformal factor, which is finite
and positive definite throughout the domain of $t$. It is also
possible to write the metric (\ref{evolvingWHmetric}) using
``physical time'' instead of ``conformal time'', by replacing $t$
by $\tau = \int \Omega (t)dt$ and therefore $\Omega (t)$ by
$R(\tau)$, where the latter is the functional form of the metric
in the $\tau$ coordinate~\cite{kar,kar-sahdev}. When the form
function and the redshift function vanish, $b(r)\rightarrow 0$ and
$\Phi(r)\rightarrow 0$, respectively, the metric
(\ref{evolvingWHmetric}) becomes the FRW metric. As
$\Omega(t)\rightarrow {\rm const}$ and $k\rightarrow 0$, it
approaches the static wormhole metric, Eq. (\ref{metricwormhole}).

The Einstein field equation will be written
$G_{\hat{\mu}\hat{\nu}}=R_{\hat{\mu}\hat{\nu}}
-{1\over2}g_{\hat{\mu}\hat{\nu}}R
={8\pi}T_{\hat{\mu}\hat{\nu}}\,,$, in an orthonormal reference
frame, so that any cosmological constant terms will be
incorporated as part of the stress-energy tensor
$T_{\hat{\mu}\hat{\nu}}$. The components of the stress-energy
tensor $T_{\hat{\mu}\hat{\nu}}$ are given by
\begin{equation}
T_{\hat{t}\hat{t}}=\rho(r,t)\,, \qquad
T_{\hat{r}\hat{r}}=-\tau(r,t) \,, \qquad
T_{\hat{t}\hat{r}}=-f(r,t) \,,  \qquad
T_{\hat{\phi}\hat{\phi}}=T_{\hat{\theta}\hat{\theta}}=p(r,t) \,,
\end{equation}
with
\begin{eqnarray}
\rho(r,t)&=&\frac{1}{8\pi}\,\frac{1}{\Omega^2}\,\left[ 3
e^{-2\Phi}\,\left(\frac{\dot {\Omega}}{\Omega}\right)^2
+\left(3k+\frac{b '}{r^2} \right) \right] \,, \label{dynTtt}
       \\
\tau(r,t)&=&-\frac{1}{8\pi}\,\frac{1}{\Omega ^{2}}\;\left\{
e^{-2\Phi(r)}\,\left[{\left ({{\dot {\Omega}} \over
{\Omega}}\right )^{2}}-2\,\frac{\ddot {\Omega}}{\Omega} \right]
- \left[k+ {b \over
{r^{3}}}-2\,\frac{\Phi'}{r}\,\left(1-kr^2-\frac{b}{r}\right)
\right] \right\}             \,,
       \\
f(r,t)&=&-\frac{1}{8\pi}\,
\left[2\,\frac{\dot{\Omega}}{\Omega^3}\;e^{-\Phi}\,\Phi'
\left(1-kr^2-\frac{b}{r} \right)^{1/2} \right]       \,,
       \\
p(r,t)&=&\frac{1}{8\pi}\, \frac{1}{\Omega ^{2}}\;\Bigg\{
e^{-2\Phi(r)}\,\left[{\left ({{\dot {\Omega}} \over
{\Omega}}\right )^{2}}-2\,\frac{\ddot {\Omega}}{\Omega} \right]
+ \left(1-kr^2-\frac{b}{r}\right) \times
         \nonumber    \\
&&\hspace{1cm}\times\left[\Phi ''+ (\Phi')^2 -
\frac{2kr^3+b'r-b}{2r(r-kr^3-b)}\Phi'-
\frac{2kr^3+b'r-b}{2r^2(r-kr^3-b)}+\frac{\Phi'}{r} \right] \Bigg\}
\,.    \label{dyn-p}
\end{eqnarray}
The overdot denotes a derivative with respect to $t$, and the
prime a derivative with respect to $r$. The physical
interpretation of $\rho(r,t),\,\tau(r,t)\,,f(r,t)$, and $p(r,t)$
are the following: the energy density, the radial tension per unit
area, energy flux in the (outward) radial direction, and lateral
pressures as measured by observers stationed at constant
$r,\,\theta,\,\phi$, respectively. The stress-energy tensor has a
non-diagonal component due to the time dependence of $\Omega(t)$
and/or the dependence of the redshift function on the radial
coordinate. The stress-energy tensor of an imperfect fluid was
analyzed in \cite{Trobo}.

A particularly interesting case of the metric
(\ref{evolvingWHmetric}) is that of a wormhole in a time-dependent
inflationary background, considered by Thomas
Roman~\cite{romanLambda}. The primary goal in the Roman analysis
was to use inflation to enlarge an initially small
\cite{romanLambda}, possibly submicroscopic, wormhole. $\Phi(r)$
and $b(r)$ are chosen to give a reasonable wormhole at $t=0$,
which is assumed to be the onset of inflation. Roman
\cite{romanLambda} went on to explore interesting properties of
the inflating wormholes, in particular, by analyzing constraints
placed on the initial size of the wormhole, if the mouths were to
remain in causal contact throughout the inflationary period; and
the maintenance of the wormhole during and after the decay of the
false vacuum. It is also possible that the wormhole will continue
to be enlarged by the subsequent FRW phase of expansion. One could
perform a similar analysis to ours by replacing the deSitter scale
factor by an FRW scale factor $a(t)$
\cite{kar,kar-sahdev,Kim-evolvWH}. In particular, in
Refs.\cite{kar,kar-sahdev} specific examples for evolving
wormholes that exist only for a finite time were considered, and a
special class of scale factors which exhibit `flashes' of the WEC
violation were also analyzed.

\subsection{Thin shells}

Consider two distinct spacetime manifolds, ${\cal M_+}$ and ${\cal
M_-}$, with metrics given by $g_{\mu \nu}^+(x^{\mu}_+)$ and
$g_{\mu \nu}^-(x^{\mu}_-)$, in terms of independently defined
coordinate systems $x^{\mu}_+$ and $x^{\mu}_-$. The manifolds are
bounded by hypersurfaces $\Sigma_+$ and $\Sigma_-$, respectively,
with induced metrics $g_{ij}^+$ and $g_{ij}^-$. The hypersurfaces
are isometric, i.e., $g_{ij}^+(\xi)=g_{ij}^-(\xi)=g_{ij}(\xi)$, in
terms of the intrinsic coordinates, invariant under the isometry.
A single manifold ${\cal M}$ is obtained by gluing together ${\cal
M_+}$ and ${\cal M_-}$ at their boundaries, i.e., ${\cal M}={\cal
M_+}\cup {\cal M_-}$, with the natural identification of the
boundaries $\Sigma=\Sigma_+=\Sigma_-$.
In particular, assuming the continuity of the four-dimensional
coordinates $x^{\mu}_\pm$ across $\Sigma$, then $g_{\mu
\nu}^-=g_{\mu \nu}^+$ is required, which together with the
continuous derivatives of the metric components $\partial g_{\mu
\nu}/\partial x^\alpha|_-=\partial g_{\mu \nu}/\partial
x^\alpha|_+$, provide the Lichnerowicz conditions~\cite{Lich}.

The three holonomic basis vectors ${\bf e}_{(i)}=\partial
/\partial \xi^i$ tangent to $\Sigma$ have the following components
$e^{\mu}_{(i)}|_{\pm}=\partial x_{\pm}^{\mu}/\partial \xi^i$,
which provide the induced metric on the junction surface by the
following scalar product
\begin{equation}
g_{ij}={\bf e}_{(i)}\cdot {\bf e}_{(j)}=g_{\mu
\nu}e^{\mu}_{(i)}e^{\nu}_{(j)}|_{\pm}.
\end{equation}

We shall consider a timelike junction surface $\Sigma$, defined by
the parametric equation of the form $f(x^{\mu}(\xi^i))=0$. The
unit normal $4-$vector, $n^{\mu}$, to $\Sigma$ is defined as
\begin{equation}\label{defnormal}
n_{\mu}=\pm \,\left |g^{\alpha \beta}\,\frac{\partial f}{\partial
x ^{\alpha}} \, \frac{\partial f}{\partial x ^{\beta}}\right
|^{-1/2}\;\frac{\partial f}{\partial x^{\mu}}\,,
\end{equation}
with $n_{\mu}\,n^{\mu}=+1$ and $n_{\mu}e^{\mu}_{(i)}=0$. The
Israel formalism requires that the normals point from ${\cal M_-}$
to ${\cal M_+}$ \cite{Israel}.

The extrinsic curvature, or the second fundamental form, is
defined as $K_{ij}=n_{\mu;\nu}e^{\mu}_{(i)}e^{\nu}_{(j)}$, or
\begin{eqnarray}\label{extrinsiccurv}
K_{ij}^{\pm}=-n_{\mu} \left(\frac{\partial ^2 x^{\mu}}{\partial
\xi ^{i}\,\partial \xi ^{j}}+\Gamma ^{\mu \pm}_{\;\;\alpha
\beta}\;\frac{\partial x^{\alpha}}{\partial \xi ^{i}} \,
\frac{\partial x^{\beta}}{\partial \xi ^{j}} \right) \,.
\end{eqnarray}
Note that for the case of a thin shell $K_{ij}$ is not continuous
across $\Sigma$, so that for notational convenience, the
discontinuity in the second fundamental form is defined as
$\kappa_{ij}=K_{ij}^{+}-K_{ij}^{-}$. In particular, the condition
that $g_{ij}^-=g_{ij}^+$, together with the continuity of the
extrinsic curvatures across $\Sigma$, $K_{ij}^-=K_{ij}^+$, provide
the Darmois conditions \cite{Darmois}.

Now, the Lanczos equations follow from the Einstein equations for
the hypersurface, and are given by
\begin{equation}
S^{i}_{\;j}=-\frac{1}{8\pi}\,(\kappa ^{i}_{\;j}-\delta
^{i}_{\;j}\kappa ^{k}_{\;k})  \,,
     \label{Lanczos}
\end{equation}
where $S^{i}_{\;j}$ is the surface stress-energy tensor on
$\Sigma$.

The first contracted Gauss-Kodazzi equation or the ``Hamiltonian"
constraint
\begin{eqnarray}
G_{\mu \nu}n^{\mu}n^{\nu}=\frac{1}{2}\,(K^2-K_{ij}K^{ij}-\,^3R)\,,
    \label{1Gauss}
\end{eqnarray}
with the Einstein equations provide the evolution identity
\begin{eqnarray}
S^{ij}\overline{K}_{ij}=-\left[T_{\mu
\nu}n^{\mu}n^{\nu}-\Lambda/8\pi \right]^{+}_{-}\,.
\end{eqnarray}
The convention $\left[X \right]^+_-\equiv
X^+|_{\Sigma}-X^-|_{\Sigma}$ and $\overline{X} \equiv
(X^+|_{\Sigma}+X^-|_{\Sigma})/2$ is used.

The second contracted Gauss-Kodazzi equation or the ``ADM"
constraint
\begin{eqnarray}
G_{\mu \nu}e^{\mu}_{(i)}n^{\nu}=K^j_{i|j}-K,_{i}\,,
    \label{2Gauss}
\end{eqnarray}
with the Lanczos equations gives the conservation identity
\begin{eqnarray}\label{conservation}
S^{i}_{j|i}=\left[T_{\mu \nu}e^{\mu}_{(j)}n^{\nu}\right]^+_-\,.
\end{eqnarray}
The momentum flux term in the right hand side corresponds to the
net discontinuity in the momentum which impinges on the shell.

In particular, considering spherical symmetry considerable
simplifications occur, namely $\kappa ^{i}_{\;j}={\rm diag}
\left(\kappa ^{\tau}_{\;\tau},\kappa ^{\theta}_{\;\theta},\kappa
^{\theta}_{\;\theta}\right)$. The surface stress-energy tensor may
be written in terms of the surface energy density, $\sigma$, and
the surface pressure, ${\cal P}$, as $S^{i}_{\;j}={\rm
diag}(-\sigma,{\cal P},{\cal P})$. The Lanczos equations then
reduce to
\begin{eqnarray}
\sigma &=&-\frac{1}{4\pi}\,\kappa ^{\theta}_{\;\theta} \,,\label{sigma} \\
{\cal P} &=&\frac{1}{8\pi}(\kappa ^{\tau}_{\;\tau}+\kappa
^{\theta}_{\;\theta}) \,. \label{surfacepressure}
\end{eqnarray}

Taking into account the wormhole spacetime metric
(\ref{metricwormhole}) and the Schwarzschild solution, the
non-trivial components of the extrinsic curvature are given by
\begin{eqnarray}
K ^{\tau
\;+}_{\;\;\tau}&=&\frac{\frac{M}{a^2}+\ddot{a}}{\sqrt{1-\frac{2M}{a}+\dot{a}^2}}
\;,  \label{Kplustautau2}\\
K ^{\tau \;-}_{\;\;\tau}&=&\frac{\Phi'
\left(1-\frac{b}{a}+\dot{a}^2
\right)+\ddot{a}-\frac{\dot{a}^2(b-b'a)}{2a(a-b)}}{\sqrt{1-\frac{b(a)}{a}+\dot{a}^2}}
\;, \label{Kminustautau2}
\end{eqnarray}
and
\begin{eqnarray}
K ^{\theta
\;+}_{\;\;\theta}&=&\frac{1}{a}\sqrt{1-\frac{2M}{a}+\dot{a}^2}\;,
 \label{Kplustheta2}\\
K ^{\theta
\;-}_{\;\;\theta}&=&\frac{1}{a}\sqrt{1-\frac{b(a)}{a}+\dot{a}^2}
\;.  \label{Kminustheta2}
\end{eqnarray}

The Lanczos equation, Eq. (\ref{Lanczos}), then provide us with
the following expressions for the surface stresses
\begin{eqnarray}
\sigma&=&-\frac{1}{4\pi a} \left(\sqrt{1-\frac{2M}{a}+\dot{a}^2}-
\sqrt{1-\frac{b(a)}{a}+\dot{a}^2} \, \right)
    \label{surfenergy}   ,\\
{\cal P}&=&\frac{1}{8\pi a} \Bigg[\frac{1-\frac{M}{a}
+\dot{a}^2+a\ddot{a}}{\sqrt{1-\frac{2M}{a}+\dot{a}^2}}
   -\frac{(1+a\Phi') \left(1-\frac{b}{a}+\dot{a}^2
\right)+a\ddot{a}-\frac{\dot{a}^2(b-b'a)}{2(a-b)}}{\sqrt{1-\frac{b(a)}{a}+\dot{a}^2}}
\, \Bigg]         \,,
    \label{surfpressure}
\end{eqnarray}
where $\sigma$ and ${\cal P}$ are the surface energy density and
the tangential surface pressure, respectively.

Using $S^{i}_{\tau|i}=-\left[\dot{\sigma}+2\dot{a}(\sigma +{\cal
P} )/a \right]$, Eq. (\ref{conservation}) provides us with
\begin{equation}
\sigma'=-\frac{2}{a}\,(\sigma+{\cal P})+\Xi
  \,,\label{consequation2}
\end{equation}
where $\Xi$, defined for notational convenience, is given by
\begin{eqnarray}
\Xi=-\frac{1}{4\pi a^2} \left[\frac{b'a-b}{2a\left(1-\frac{b}{a}
\right)}+a\Phi' \right] \sqrt{1-\frac{b}{a}+\dot{a}^2} \,.
       \label{H(a)}
\end{eqnarray}

For self-completeness, we shall also include the $\sigma +{\cal
P}$ term, which is given by
\begin{eqnarray}
\sigma+{\cal P}= \frac{1}{8\pi a}\Bigg[\frac{(1-a\Phi')
\left(1-\frac{b}{a}+\dot{a}^2
\right)-a\ddot{a}+\frac{\dot{a}^2(b-b'a)}{2(a-b)}}
{\sqrt{1-\frac{b(a)}{a}+\dot{a}^2}}
-\frac{1-\frac{3M}{a}
+\dot{a}^2-a\ddot{a}}{\sqrt{1-\frac{2M}{a}+\dot{a}^2}} \, \Bigg]
    \,.
    \label{s+P}
\end{eqnarray}
Thus, taking into account Eq. (\ref{s+P}), and the definition of
$\Xi$, we verify that Eq. (\ref{consequation2}) finally takes the
form
\begin{eqnarray}
\sigma'=\frac{1}{4\pi a^2} \Bigg(\frac{1-\frac{3M}{a}
+\dot{a}^2-a\ddot{a}}{\sqrt{1-\frac{2M}{a}+\dot{a}^2}}
 - \frac{1-\frac{3b}{2a}+\frac{b'}{2}+\dot{a}^2
-a\ddot{a}}{\sqrt{1-\frac{b}{a}+\dot{a}^2}} \, \Bigg) \,,
     \label{sigma'WH}
\end{eqnarray}
which, evaluated at a static solution $a_0$, shall play a
fundamental role in determining the stability regions. Note that
Eq. (\ref{sigma'WH}) can also be deduced by taking the radial
derivative of the surface energy density, Eq. (\ref{surfenergy}).

The construction of dynamic shells in wormholes have been
extensively analyzed in Ref. \cite{LoboCrawford}, where the
stability of generic spherically symmetric thin shells to
linearized perturbations around static solutions were considered,
and applying the analysis to traversable wormhole geometries, by
considering specific choices for the form function, the stability
regions were deduced. It was found that the latter may be
significantly increased by considering appropriate choices for the
redshift function (The linearized stability analysis was also
applied to dark energy stars \cite{Lobo:2005uf}).

\subsection{Late-time cosmic accelerated expansion and traversable wormholes}

In this section, we shall explore the possibility that traversable
wormholes be supported by specific equations of state responsible
for the late time accelerated expansion of the Universe, namely,
phantom energy, the generalized Chaplygin gas, and the van der
Waals quintessence equation of state. Firstly, phantom energy
possesses an equation of state of the form $\omega\equiv
p/\rho<-1$, consequently violating the null energy condition
(NEC), which is a fundamental ingredient necessary to sustain
traversable wormholes. Thus, this cosmic fluid presents us with a
natural scenario for the existence of wormhole
geometries~\cite{phantomWH,phantomWHb,phantomWH2}. Secondly, the
generalized Chaplygin gas (GCG) is a candidate for the unification
of dark energy and dark matter, and is parametrized by an exotic
equation of state given by $p_{ch}=-A/\rho_{ch}^{\alpha}$, where
$A$ is a positive constant and $0<\alpha \leq 1$. Within the
framework of a flat Friedmann-Robertson-Walker cosmology the
energy conservation equation yields the following evolution of the
energy density
$\rho_{ch}=\left[A+Ba^{-3(1+\alpha)}\right]^{1/(1+\alpha)}$, where
$a$ is the scale factor, and $B$ is normally considered to be a
positive integration constant to ensure the dominant energy
condition (DEC). However, it is also possible to consider $B<0$,
consequently violating the DEC, and the energy density is an
increasing function of the scale function \cite{Lopez-Madrid}. It
is in the latter context that we shall explore exact solutions of
traversable wormholes supported by the GCG \cite{ChapWH}.
Thirdly, the van der Waals quintessence equation of state,
$p=\gamma \rho/(1-\beta\rho)-\alpha \rho^2$, is an interesting
scenario for describing the late universe, and seems to provide a
solution to the puzzle of dark energy, without the presence of
exotic fluids or modifications of the Friedmann equations. Note
that $\alpha,\beta \rightarrow 0$ and $\gamma <-1/3$ reduces to
the dark energy equation of state. The existence of traversable
wormholes supported by the VDW equation of state shall also be
explored \cite{VDWwh}.
Despite of the fact that, in a cosmological context, these cosmic
fluids are considered homogeneous, inhomogeneities may arise
through gravitational instabilities, resulting in a nucleation of
the cosmic fluid due to the respective density perturbations.
Thus, the wormhole solutions considered in this work may possibly
originate from density fluctuations in the cosmological
background.

The strategy we shall adopt is to impose an equation of state,
$p_r=p_r(\rho)$, which provides four equations, together with the
Einstein field equations. However, we have five unknown functions
of $r$, i.e., $\rho(r)$, $p_r(r)$, $p_t(r)$, $b(r)$ and $\Phi(r)$.
Therefore, to fully determine the system we impose restricted
choices for $b(r)$ or $\Phi(r)$ \cite{phantomWH,ChapWH,VDWwh}. It
is also possible to consider plausible stress-energy components,
and through the field equations determine the metric
fields\cite{phantomWH2}.

Now, using the equation of state representing phantom energy,
$p_r=\omega \rho$ with $\omega<-1$, and taking into account Eqs.
(\ref{rhoWH}), we have the following condition
\begin{equation}
\Phi'(r)=\frac{b+\omega rb'}{2r^2\left(1-b/r \right)} \,.
            \label{phantEOScondition}
\end{equation}
For instance, consider a constant $\Phi(r)$, so that Eq.
(\ref{phantEOScondition}) provides $b(r)=r_0(r/r_0)^{-1/\omega}$,
which corresponds to an asymptotically flat wormhole geometry. It
was shown that this solution can be constructed, in principle,
with arbitrarily small quantities of averaged null energy
condition violating phantom energy, and the traversability
conditions were explored\cite{phantomWH}. The dynamic stability of
these phantom wormholes were also analyzed\cite{phantomWHb}, and
we refer the reader to \cite{phantomWH,phantomWH2} for further
examples.

Relative to the GCG gas equation of state, $p_r=-A/\rho^{\alpha}$,
using the field equations, we have the following condition
\begin{equation}
2r\left(1-\frac{b}{r} \right)\Phi'(r)=-Ab'\left(\frac{8\pi r^2
}{b'}\right)^{1+\alpha}+\frac{b}{r} \,.
            \label{GCGEOScondition}
\end{equation}
Solutions of the metric (\ref{metricwormhole}), satisfying Eq.
(\ref{GCGEOScondition}) are denoted ``Chaplygin wormholes''. To be
a generic solution of a wormhole, the GCG equation of state
imposes the following restriction $A<(8\pi r_0^2)^{-(1+\alpha)}$,
consequently violating the NEC. However, for the GCG cosmological
models it is generally assumed that the NEC is satisfied, which
implies $\rho \geq A^{1/(1+\alpha)}$. The NEC violation is a
fundamental ingredient in wormhole physics, and it is in this
context that the construction of traversable wormholes, i.e., for
$\rho < A^{1/(1+\alpha)}$, are explored. Note that as emphasized
in \cite{Lopez-Madrid}$\,$, considering $B<0$ in the evolution of
the energy density, one also deduces that $\rho_{ch} <
A^{1/(1+\alpha)}$, which violates the DEC. We refer the reader
to$\,$\cite{ChapWH} for specific examples of Chaplygin wormholes,
where the physical properties and characteristics of these
geometries were analyzed in detail. The solutions found are not
asymptotically flat, and the spatial distribution of the exotic
GCG is restricted to the throat vicinity, so that the dimensions
of these Chaplygin wormholes are not arbitrarily large.

Finally, consider the VDW equation of state for an inhomogeneous
spherically symmetric spacetime, given by $p_r=\gamma
\rho/(1-\beta \rho)-\alpha \rho^2$. The Einstein field equations
provide the following relationship
\begin{equation}
2r\left(1-\frac{b}{r}\right) \Phi'=\frac{b}{r}+\frac{\gamma
b'}{1-\frac{\beta b'}{8\pi r^2}} -\frac{\alpha b'^2}{8\pi r^2} \,.
            \label{vdWEOS}
\end{equation}
It was shown that traversable wormhole solutions may be
constructed using the VDW equation of state, which are either
asymptotically flat or possess finite dimensions, where the exotic
matter is confined to the throat neighborhood \cite{VDWwh}. The
latter solutions are constructed by matching an interior wormhole
geometry to an exterior vacuum Schwarzschild vacuum, and we refer
the reader to$\,$\cite{VDWwh} for further details.

In concluding, it is noteworthy the relative ease with which one
may theoretically construct traversable wormholes with the exotic
fluid equations of state used in cosmology to explain the present
accelerated expansion of the Universe. These traversable wormhole
variations have far-reaching physical and cosmological
implications, namely, apart from being used for interstellar
shortcuts, an absurdly advanced civilization may convert them into
time-machines, probably implying the violation of causality.


\section{``Warp drive'' spacetimes and superluminal travel}\label{secIII}


Much interest has been revived in superluminal travel in the last
few years. Despite the use of the term superluminal, it is not
``really'' possible to travel faster than light, in any
\emph{local} sense. The point to note is that one can make a round
trip, between two points separated by a distance $D$, in an
arbitrarily short time as measured by an observer that remained at
rest at the starting point, by varying one's speed or by changing
the distance one is to cover. Providing a general \emph{global}
definition of superluminal travel is no trivial
matter~\cite{VB,VBL}, but it is clear that the spacetimes that
allow ``effective'' superluminal travel generically suffer from
the severe drawback that they also involve significant negative
energy densities. More precisely, superluminal effects are
associated with the presence of \emph{exotic} matter, that is,
matter that violates the null energy condition [NEC].

In fact, superluminal spacetimes violate all the known energy
conditions, and Ken Olum demonstrated that negative energy
densities and superluminal travel are intimately
related~\cite{Olum}. Although most classical forms of matter are
thought to obey the energy conditions, they are certainly violated
by certain quantum fields~\cite{VisserEC}. Additionally, certain
classical systems (such as non-minimally coupled scalar fields)
have been found that violate the null and the weak energy
conditions~\cite{barcelovisser1,barcelovisserPLB99}. It is also
interesting to note that recent observations in cosmology strongly
suggest that the cosmological fluid violates the strong energy
condition [SEC], and provides tantalizing hints that the NEC
\emph{might} possibly be violated in a classical
regime~\cite{Riess,jerk,rip}.

Apart from wormholes {\cite{Morris,Visser}}, two spacetimes which
allow superluminal travel are the Alcubierre warp drive
\cite{Alcubierre} and the solution known as the Krasnikov tube
\cite{Krasnikov,Everett}. Alcubierre demonstrated that it is
theoretically possible, within the framework of general
relativity, to attain arbitrarily large velocities
\cite{Alcubierre}. A warp bubble is driven by a local expansion
behind the bubble, and an opposite contraction ahead of it.
However, by introducing a slightly more complicated metric,
Jos\'{e} Nat\'{a}rio~\cite{Natario} dispensed with the need for
expansion. Thus, the Nat\'{a}rio version of the warp drive can be
thought of as a bubble sliding through space.

It is interesting to note that Krasnikov \cite{Krasnikov}
discovered a fascinating aspect of the warp drive, in which an
observer on a spaceship cannot create nor control on demand an
Alcubierre bubble, with $v>c$, around the ship \cite{Krasnikov},
as points on the outside front edge of the bubble are always
spacelike separated from the centre of the bubble. However,
causality considerations do not prevent the crew of a spaceship
from arranging, by their own actions, to complete a {\it round
trip} from the Earth to a distant star and back in an arbitrarily
short time, as measured by clocks on the Earth, by altering the
metric along the path of their outbound trip. Thus, Krasnikov
introduced a two-dimensional metric with an interesting property
that although the time for a one-way trip to a distant destination
cannot be shortened, the time for a round trip, as measured by
clocks at the starting point (e.g. Earth), can be made arbitrarily
short. Soon after, Everett and Roman generalized the Krasnikov
two-dimensional analysis to four dimensions, denoting the solution
as the {\it Krasnikov tube}~\cite{Everett}, where they analyzed
the superluminal features, the energy condition violations, the
appearance of closed timelike curves and applied the Quantum
Inequality.

\subsection{``Warp drive'' spacetime metric}

Within the framework of general relativity, Alcubierre
demonstrated that it is in principle possible to warp spacetime in
a small {\it bubble-like} region, in such a way that the bubble
may attain arbitrarily large velocities. Inspired by the
inflationary phase of the early universe, the enormous speed of
separation arises from the expansion of spacetime itself. The
simplest model for hyper-fast travel is to create a local
distortion of spacetime, producing an expansion behind the bubble,
and an opposite contraction ahead of it. Nat\'{a}rio's version of
the warp drive dispensed with the need for expansion at the cost
of introducing a slightly more complicated metric.

The warp drive spacetime metric, in cartesian coordinates, is
given by (with $G=c=1$)
\begin{equation}
d s^2=-d t^2+ [d\vec x - \vec \beta(x,y,z-z_0(t)) \; d t]\cdot
[d\vec x - \vec \beta(x,y,z-z_0(t)) \; d t]\,.
\label{Cartesianwarpmetric-general}
\end{equation}
In terms of the well-known ADM formalism this corresponds to a
spacetime wherein \emph{space} is flat, while the ``lapse
function'' is identically unity, and the only non-trivial
structure lies in the ``shift vector'' $\beta(t,\vec x)$. Thus
warp drive spacetimes can also be viewed as specific examples of
``shift-only'' spacetimes \cite{LV-CQG}. The Alcubierre warp drive
corresponds to taking the shift vector to always lie in the
direction of motion
\begin{equation}
\vec \beta(x,y,z-z_0(t)) =  v(t) \; \hat z \; f(x,y,z-z_0(t)),
\end{equation}
in which $v(t)=dz_0(t)/dt$ is the velocity of the warp bubble,
moving along the positive $z$-axis, whereas in the Nat\'{a}rio
warp drive the shift vector is constrained by being
divergence-free
\begin{equation}
\nabla \cdot \vec \beta(x,y,z) =  0.
\end{equation}

\subsection{Alcubierre warp drive}\label{Alcubierrewarp}

In the Alcubierre warp drive the spacetime metric is
\begin{equation}
d s^2=-d t^2+d x^2+d y^2+\left[d z-v(t)\;f(x,y,z-z_0(t))\; d
t\right]^2 \label{Cartesianwarpmetric}\,.
\end{equation}
The form function $f(x,y,z)$ possesses the general features of
having the value $f=0$ in the exterior and $f=1$ in the interior
of the bubble.  The general class of form functions, $f(x,y,z)$,
chosen by Alcubierre was spherically symmetric: $f(r)$ with
$r=\sqrt{x^2+y^2+z^2}$. Then
\begin{equation}
f(x,y,z-z_0(t)) = f(r(t)) \label{Alcubierreformfunction} \qquad
\hbox{with} \qquad
r(t)=\left\{[(z-z_{0}(t)]^2+x^2+y^2\right\}^{1/2}.
\end{equation}

Whenever a more specific example is required we adopt
\begin{equation}
f(r)=\frac{\tanh\left[\sigma(r+R)\right]
-\tanh\left[\sigma(r-R)\right]}{2\tanh(\sigma R)}\,,
\label{E:form}
\end{equation}
in which $R>0$ and $\sigma>0$ are two arbitrary parameters. $R$ is
the ``radius'' of the warp-bubble, and $\sigma$ can be interpreted
as being inversely proportional to the bubble wall thickness. If
$\sigma$ is large, the form function rapidly approaches a {\it top
hat} function, i.e.,
\begin{equation}
\lim_{\sigma \rightarrow \infty} f(r)=\left\{ \begin{array}{ll}
1, & {\rm if}\; r\in[0,R],\\
0, & {\rm if}\; r\in(R,\infty).
\end{array}
\right.
\end{equation}


It can be shown that observers with the four velocity
\begin{equation}
U^{\mu}=\left(1,0,0,vf\right), \qquad\qquad
U_{\mu}=\left(-1,0,0,0\right).
\end{equation}
move along geodesics, as their $4$-acceleration is zero,
\emph{i.e.}, $a^{\mu} = U^{\nu}\; U^{\mu}{}_{;\nu}=0$. They were
denoted Eulerian observers by Alcubierre. The spaceship, which in
the original formulation is treated as a test particle which moves
along the curve $z=z_0(t)$, can easily be seen to always move
along a timelike curve, regardless of the value of $v(t)$. One can
also verify that the proper time along this curve equals the
coordinate time, by simply substituting $z=z_0(t)$ in Eq.
(\ref{Cartesianwarpmetric}). This reduces to $d\tau=d t$, taking
into account $d x=d y=0$ and $f(0)=1$.

\begin{figure}
\centering
\includegraphics[width=3.4in]{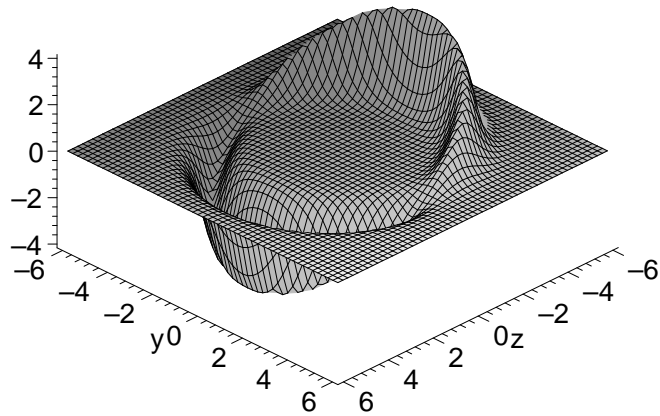}
\caption[The expansion of the volume elements for the Alcubierre
warp drive]{The expansion of the volume elements. These are
expanding behind the spaceship, and contracting in front of
it.}\label{Alcubierre-expansion}
\end{figure}

Consider a spaceship placed within the Alcubierre warp bubble. The
expansion of the volume elements, $\theta=U^{\mu}{}_{;\mu}$, is
given by $\theta=v\;\left({\partial f}/{\partial z} \right)$.
Taking into account Eq. (\ref{E:form}), we have (for Alcubierre's
version of the warp bubble)
\begin{equation}
\theta=v\;\frac{z-z_0}{r}\;\frac{d f(r)}{d r}.
\end{equation}
The center of the perturbation corresponds to the spaceship's
position $z_0(t)$. The volume elements are expanding behind the
spaceship, and contracting in front of it, as shown in Figure
\ref{Alcubierre-expansion}.

\subsection{The violation of the energy conditions}\label{warpENviolations}

If we attempt to treat the spaceship as more than a test particle,
we must confront the fact that by construction we have forced
$f=0$ outside the warp bubble. [Consider, for instance, the
explicit form function of Eq. (\ref{E:form}) in the limit
$r\to\infty$.] This implies that the spacetime geometry is
asymptotically Minkowski space, and in particular the ADM mass
(defined by taking the limit as one moves to spacelike infinity
$i^0$) is zero. That is, the ADM mass of the spaceship and the
warp field generators must be exactly compensated by the ADM mass
due to the stress-energy of the warp-field itself. Viewed in this
light it is now patently obvious that there must be massive
violations of the classical energy conditions (at least in the
original version of the warp-drive spacetime), and the interesting
question becomes ``Where are these energy condition violations
localized?''.

One of our tasks in the current Section will be to see if we can
first avoid this exact cancellation of the ADM mass, and second,
to see if we can make qualitative and quantitative statements
concerning the localization and ``total amount'' of energy
condition violations. (A similar attempt at quantification of the
``total amount'' of energy condition violation in traversable
wormholes was recently presented in~\cite{visser2003,Kar2}). By
using the Einstein field equation, $G_{\mu\nu}=8\pi \;
T_{\mu\nu}$, we can make rather general statements regarding the
nature of the stress energy required to support a warp bubble.

\subsubsection{The violation of the WEC}

The WEC states $T_{\mu\nu} \, U^{\mu} \, U^{\nu}\geq0$, in which
$U^{\mu}$ is a timelike vector and $T_{\mu\nu}$ is the
stress-energy tensor.  Its physical interpretation is that the
local energy density is positive.  By continuity it implies the
NEC. We verify that for the warp drive metric, the WEC is
violated, \emph{i.e.},
\begin{equation}
T_{\mu\nu} \; U^{\mu} \; U^{\nu}= -\frac{v^2}{32\pi}\; \left[
\left (\frac {\partial f}{\partial x} \right )^2 + \left (\frac
{\partial f}{\partial y} \right )^2 \right]  <0 \,,
\end{equation}
or by taking into account the Alcubierre form function
(\ref{E:form}), we have
\begin{equation}
T_{\mu\nu} \; U^{\mu}\; U^{\nu}= -\frac{1}{32\pi}\frac{v^2
(x^2+y^2)}{r^2} \left( \frac{d f}{d r} \right)^2<0
\label{WECviolation} \,.
\end{equation}

\begin{figure}
\centering
\includegraphics[width=3.8in]{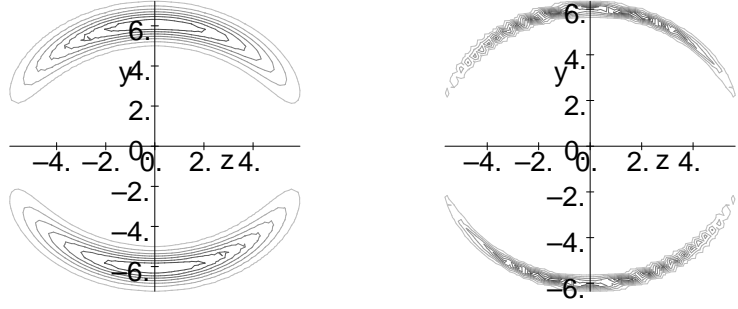}
\caption[Distribution of the energy density for the Alcubierre
warp drive]{The energy density is distributed in a toroidal region
perpendicular to the direction of travel of the spaceship, which
is situated at $z_0(t)$. We have considered the following values,
$v=2$ and $R=6$, with $\sigma=1$ and $\sigma=4$ in $(a)$ and
$(b)$, respectively.}\label{Alcub-energydensity}
\end{figure}

By considering an orthonormal basis, we verify that the energy
density of the warp drive spacetime is given by
$T_{\hat{t}\hat{t}} = T_{\hat{\mu}\hat{\nu}} \; U^{\hat{\mu}} \;
U^{\hat{\nu}}$, that is, Eq. (\ref{WECviolation}). It is easy to
verify that the energy density is distributed in a toroidal region
around the $z$-axis, in the direction of travel of the warp
bubble~\cite{PfenningF}, as may be verified from Figure
\ref{Alcub-energydensity}.  It is perhaps instructive to point out
that the energy density for this class of spacetimes is nowhere
positive. That the total ADM mass can nevertheless be zero is due
to the intrinsic nonlinearity of the Einstein equations.

It is interesting to note that the inclusion of a generic lapse
function $\alpha(x,y,z,t)$, decreases the negative energy density,
which is given by
\begin{equation}
T_{\hat{t}\hat{t}}= -\frac{v^2}{32\pi\,\alpha^2}\; \left[ \left
(\frac {\partial f}{\partial x} \right )^2 + \left (\frac
{\partial f}{\partial y} \right )^2 \right]  \,.
\end{equation}
Now, $\alpha$ may be taken as unity in the exterior and interior
of the warp bubble, so proper time equals coordinate time. In
order to significantly decrease the negative energy density in the
bubble walls, one may impose an extremely large value for the
lapse function. However, the inclusion of the lapse function
suffers from an extremely severe drawback, as proper time as
measured in the bubble walls becomes absurdly large,
$d\tau=\alpha\,dt$, for $\alpha \gg 1$.

We can (in analogy with the definitions in~\cite{visser2003,Kar2})
quantify the ``total amount'' of energy condition violating matter
in the warp bubble by defining
\begin{eqnarray}
M_\mathrm{warp} &=& \int \rho_\mathrm{warp} \; d^3 x = \int
T_{\mu\nu} \; U^{\mu}\; U^{\nu}  \; d^3 x
       \nonumber      \\
&=& - {v^2\over32\pi} \int \frac{x^2+y^2}{r^2} \left( \frac{d f}{d
r} \right)^2 \; r^2 \; d r \; d^2\Omega = -{v^2\over12} \int
\left( \frac{d f}{d r} \right)^2 \; r^2 \; d r.
\end{eqnarray}
This is emphatically not the total mass of the spacetime, but it
characterizes how much (negative) energy one needs to localize in
the walls of the warp bubble. For the specific shape function
(\ref{E:form}) we can estimate
\begin{equation}
M_\mathrm{warp} \approx - v^2 \; R^2 \; \sigma.
\end{equation}
(The integral can be done exactly, but the exact result in terms
of {\sf polylog} functions is unhelpful.) Note that the energy
requirements for the warp bubble scale quadratically with bubble
velocity, quadratically with bubble size, and inversely as the
thickness of the bubble wall \cite{LV-CQG}.

\subsubsection{The violation of the NEC}

The NEC states that $T_{\mu\nu} \, k^{\mu} \, k^{\nu}\geq0$, where
$k^{\mu}$ is \emph{any} arbitrary null vector and $T_{\mu\nu}$ is
the stress-energy tensor. The NEC for a null vector oriented along
the $\pm \hat z$ directions takes the following form
\begin{equation}
T_{\mu\nu} \; k^{\mu} \; k^{\nu}= -\frac{v^2}{8\pi}\, \left[ \left
(\frac{\partial f}{\partial x} \right )^2 + \left (\frac {\partial
f}{\partial y} \right )^2 \right] \pm \frac{v}{8\pi}\left( \frac
{\partial^{2}f}{\partial {x}^{2}} +
 \frac {\partial^{2}f}{\partial {y}^{2}}
\right)  \,.
\end{equation}
In particular if we average over the $\pm \hat z$ directions we
have
\begin{equation}
{1\over2} \left\{ T_{\mu\nu} \; k^{\mu}_{+\hat z} \;
k^{\nu}_{+\hat z} + T_{\mu\nu} \; k^{\mu}_{-\hat z} \;
k^{\nu}_{-\hat z} \right\} = -\frac{v^2}{8\pi}\, \left[ \left
(\frac{\partial f}{\partial x} \right )^2 + \left (\frac {\partial
f}{\partial y} \right )^2 \right],
\end{equation}
which is manifestly negative, and so the NEC is violated for all
$v$. Furthermore, note that even if we do not average, the
coefficient of the term linear in $v$ must be nonzero
\emph{somewhere} in the spacetime. Then at low velocities this
term will dominate and at low velocities the un-averaged NEC will
be violated in either the $+\hat z$ or $-\hat z$ directions.

To be a little more specific about how and where the NEC is
violated consider the Alcubierre form function. We have
\begin{equation}\label{NEC:Alcubierre-form}
T_{\mu\nu} \; k^{\mu}_{\pm\hat z} \; k^{\nu}_{\pm\hat z}=
-\frac{1}{8\pi} \frac{v^2(x^2+y^2)}{r^2} \left( \frac{d f}{d r}
\right)^2 \pm \frac{v}{8\pi}\left[
\frac{x^2+y^2+2(z-z_0(t))^2}{r^3} \; \frac{d f}{d
r}+\frac{x^2+y^2}{r^2}\, \frac{d^2 f}{d r^2}\right] \,.
\end{equation}
The first term is manifestly negative everywhere throughout the
space. As $f$ decreases monotonically from the center of the warp
bubble, where it takes the value of $f=1$, to the exterior of the
bubble, with $f\approx 0$, we verify that ${d f}/{d r}$ is
negative in this domain. The term ${d ^2 f}/{d r^2}$ is also
negative in this region, as $f$ attains its maximum in the
interior of the bubble wall. Thus, the term in square brackets
unavoidably assumes a negative value in this range, resulting in
the violation of the NEC.

\begin{figure}[h]
\centering
  \includegraphics[width=1.9in]{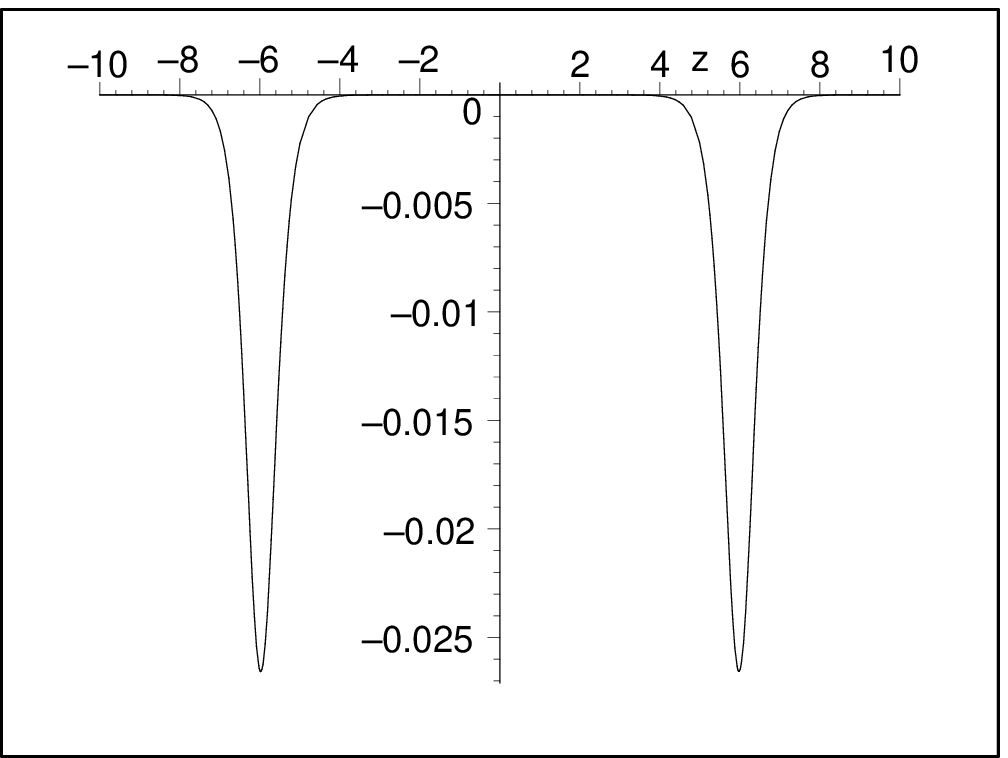}
  \hspace{0.1in}
  \includegraphics[width=1.9in]{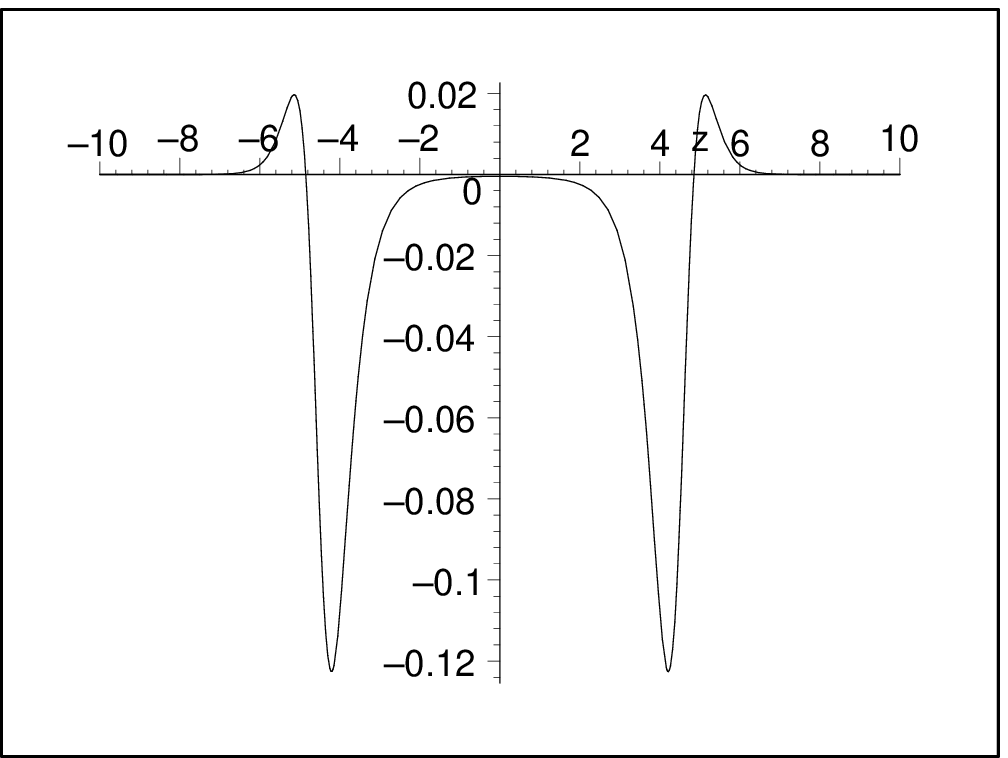}
  \hspace{0.1in}
  \includegraphics[width=1.9in]{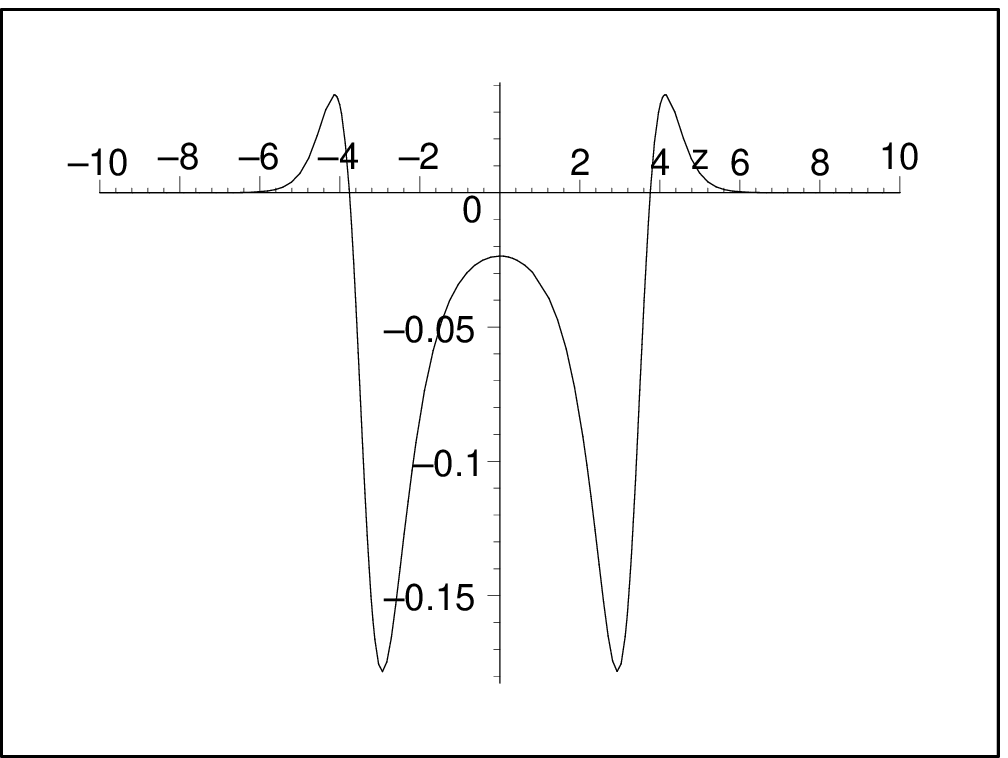}
  \caption[The NEC for a null vector oriented
  along the $+\hat z$ direction]{The NEC for a null vector oriented
  along the $+\hat z$ direction. Taking into account the
  Alcubierre form function, we have considered the parameters
  $v=2$, $\sigma=2$ and $R=6$. Considering the definition $\rho=\sqrt{x^2+y^2}$,
  the plots have the respective values of $\rho=0$, $\rho=4$ and
  $\rho=5$.}\label{Alc:NEC+z0}
\end{figure}

\begin{figure}[h]
\centering
  \includegraphics[width=1.9in]{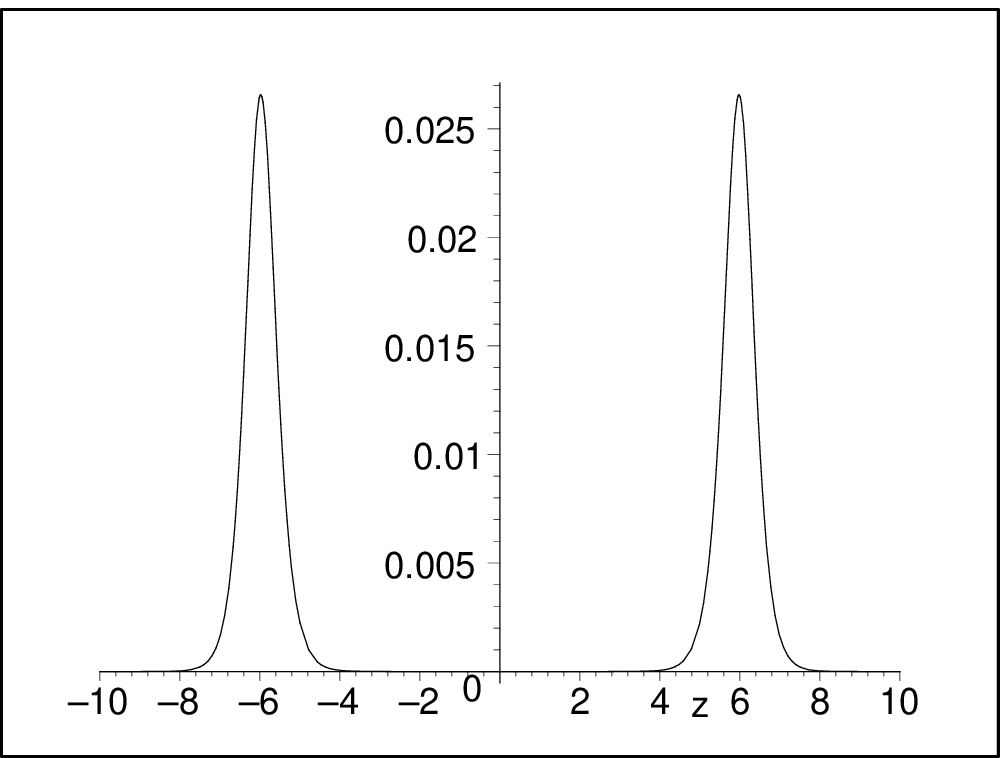}
  \hspace{0.1in}
  \includegraphics[width=1.9in]{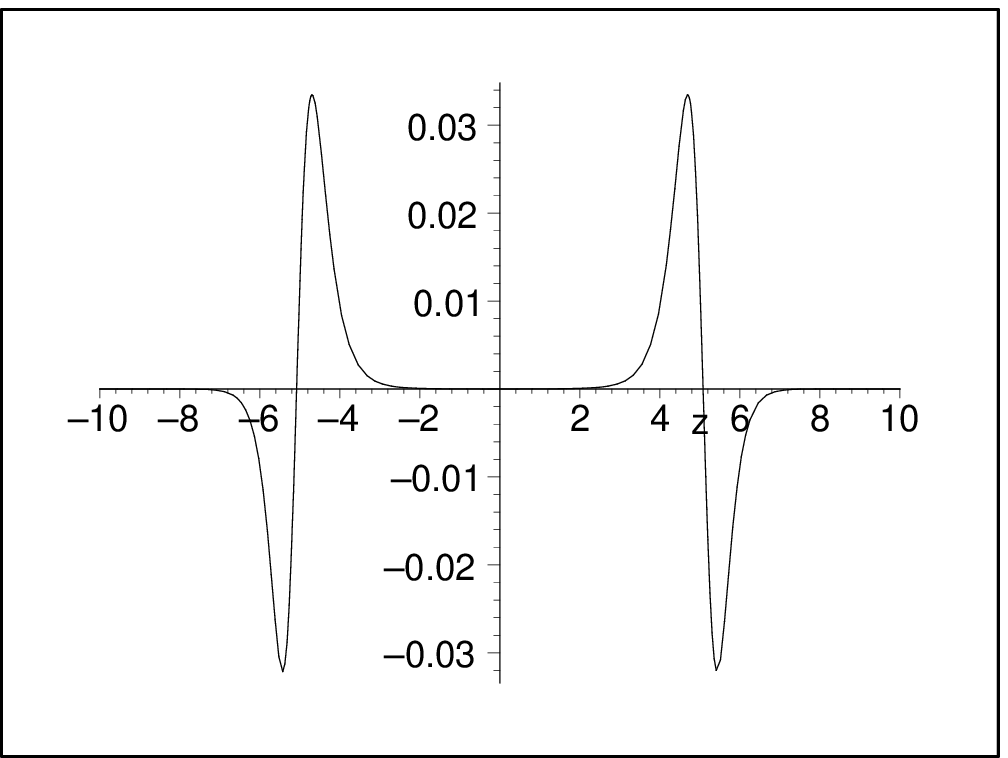}
  \hspace{0.1in}
  \includegraphics[width=1.9in]{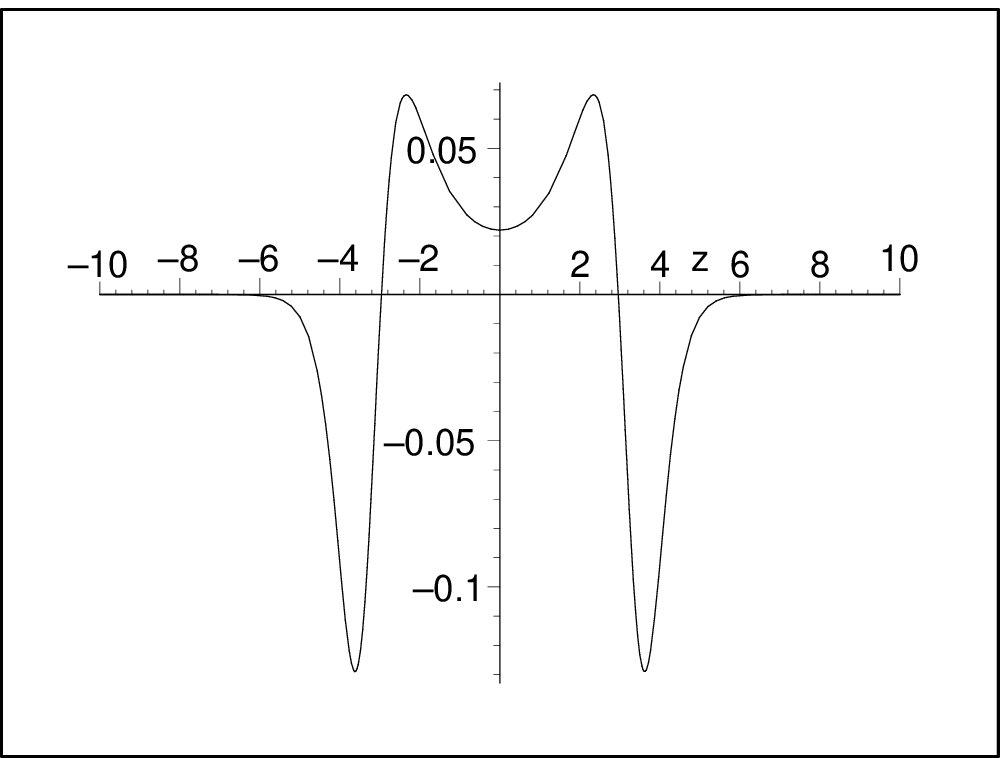}
  \caption[The NEC for a null vector oriented
  along the $-\hat z$ direction]{The NEC for a null vector oriented
  along the $-\hat z$ direction. Taking into account the
  Alcubierre form function, we have considered the parameters
  $v=2$, $\sigma=2$ and $R=6$. Considering the definition $\rho=\sqrt{x^2+y^2}$,
  the plots have the respective values of $\rho=0$, $\rho=3$ and
  $\rho=5$.}\label{Alc:NEC-z0}
\end{figure}

Equation (\ref{NEC:Alcubierre-form}) is plotted in Figures
\ref{Alc:NEC+z0} and \ref{Alc:NEC-z0}, for various values of the
parameters. Figure \ref{Alc:NEC+z0} represents the null vector
oriented along the $+\hat z$ direction, and Figure
\ref{Alc:NEC-z0} along the $-\hat z$ direction. We have considered
the following values of $v=2$, $\sigma=2$ and $R=6$, for the
parameters.

For a null vector oriented perpendicular to the direction of
motion (for definiteness take $\hat k = \pm \hat x$) the NEC takes
the following form
\begin{equation}
T_{\mu\nu} \; k^{\mu}_{\pm\hat x} \; k^{\nu}_{\pm\hat x}=
-\frac{v^2}{8\pi}\, \left[ {1\over2} \left (\frac{\partial
f}{\partial y} \right )^2 + \left(\frac {\partial f}{\partial z}
\right )^2 - (1-f) {\partial^2 f\over \partial z^2} \right] \mp
\frac{v}{8\pi}\left( \frac{\partial^{2}f}{\partial {x}\partial z}
\right)  \,.
\end{equation}
Again, note that the coefficient of the term linear in $v$ must be
nonzero \emph{somewhere} in the spacetime. Then at low velocities
this term will dominate, and at low velocities the NEC will be
violated in one or other of the transverse directions. Upon
considering the specific form of the spherically symmetric
Alcubierre form function, we have
\begin{eqnarray}
T_{\mu\nu} \; k^{\mu}_{\pm\hat x} \; k^{\nu}_{\pm\hat x}&=&
-\frac{v^2}{8\pi} \left[ \frac{y^2+2(z-z_0(t))^2}{2r^2} \left(
\frac{d f}{d r} \right)^2 -(1-f) \left(\frac{x^2+y^2}{r^3}
\;\frac{d f}{d r} +\frac{(z-z_0(t))^2}{r^2} \; \frac{d^2f}{d r^2}
\right) \right]
      \nonumber  \\
&&\mp \frac{v}{8\pi}\frac{x\,(z-z_0(t))}{r^2} \left( \frac{d^2
f}{d r^2}-\frac{1}{r}\frac{d f}{d r} \right) \,.
\end{eqnarray}
Again, the message to take from this is that localized NEC
violations are ubiquitous and persist to arbitrarily low warp
bubble velocities.

Using the ``volume integral quantifier'' (as defined
in~\cite{visser2003,Kar2}), we may estimate the ``total amount''
of averaged null energy condition violating matter in this
spacetime, given by
\begin{eqnarray}
\int T_{\mu\nu} \;k^{\mu}_{\pm\hat z} \; k^{\nu}_{\pm\hat z}  \;
d^3 x \approx \int T_{\mu\nu} \;k^{\mu}_{\pm\hat x}  \;
k^{\nu}_{\pm\hat x}  \; d^3 x  \approx - v^2\; R^2 \; \sigma
\approx M_{\mathrm{warp}}\,.
\end{eqnarray}
The key things to note here are that the net volume integral of
the $O(v)$ term is zero, and that the net volume average of the
NEC violations is approximately the same as the net volume average
of the WEC violations, which are $O(v^2)$.

\subsection{The Quantum Inequality applied to the ``warp drive''}

It is of a particular interest to apply the Quantum Inequality
(QI), outlined in Section \ref{Sec:QI}, to ``warp drive''
spacetimes \cite{PfenningF}. Inserting the energy density, Eq.
(\ref{WECviolation}), into the QI, Eq. (\ref{QI}), one gets
\begin{equation}
t_0 \int_{-\infty}^{+\infty} {v(t)^2 \over r^2} \left({df\over
dr}\right)^2 {dt\over {t^2+t_0^2}} \leq {3\over \rho^2 \,t_0^4}
\,, \label{expr_QIwarp}
\end{equation}
where $\rho=\left(x^2+y^2 \right)^{1/2}$ is defined for notational
convenience.

The warp bubble's velocity can be considered roughly constant,
$v_s(t) \approx v_b$, if the time scale of the sampling is
sufficiently small compared to the time scale over which the
bubble's velocity is changing. Taking into account the small
sampling time, the $(t^2 +t_0^2)^{-1}$ term becomes strongly
peaked, so that only a small portion of the geodesic is sampled by
the QI integral. Consider that the observer is at the equator of
the warp bubble at $t=0$ \cite{PfenningF}, so that the geodesic is
approximated by
\begin{equation}
x(t) \approx f(\rho) v_b t \; ,
\end{equation}
so that we have $r(t) = \left[ (v_b t)^2 (f(\rho) - 1)^2 + \rho^2
\right]^{1/2}$.

Without a significant loss of generality, one may consider a
piece-wise continuous form of the shape function given by
\begin{equation}
f_{p.c.}(r) = \left\{\matrix{1 & r < R-{\Delta\over2}\cr
-{1\over\Delta}(r -R-{\Delta\over2})\qquad & R-{\Delta\over2}<r<
R+{\Delta\over2}\cr 0&r > R+{\Delta\over2}}\right.
\label{eq:pointwise}
\end{equation}
where $R$ is the radius of the bubble, and $\Delta$ the bubble
wall thickness \cite{PfenningF}. $\Delta$ is related to the
Alcubierre parameter $\sigma$ by setting the slopes of the
functions $f(r)$ and $f_{p.c.}(r)$ to be equal at $r=R$, which
provides the following relationship
\begin{equation}
\Delta = {\left[1+\tanh^2(\sigma R)\right]^2\over
 2\;\sigma\;\tanh(\sigma R)}\;,
\end{equation}
Note that in the limit of large $\sigma R$ one obtains the
approximation $\Delta \simeq 2/\sigma$. The QI-bound then becomes
\begin{equation}
 t_0 \int_{-\infty}^{+\infty} {dt \over (t^2 + \beta^2) (t^2+t_0^2)}
\leq {3 \Delta^2 \over v_b^2 \; t_0^4 \; \beta^2}
\end{equation}
where
\begin{equation}
\beta = {\rho \over {v_b \left[1 - f(\rho) \right] }}\,,
\end{equation}
and yields the following inequality
\begin{equation}
{\pi \over 3} \leq {\Delta^2 \over {v_b^2 \; t_0^4}} \left[{v_b
t_0 \over \rho} (1-f(\rho)) +1\right] \; . \label{wd-QI-bound}
\end{equation}

It is important to emphasize that the above inequality is only
valid for sampling times on which the spacetime may be considered
approximately flat. Considering the Riemann tensor components in
an orthonormal frame \cite{PfenningF}, the largest component is
given by
\begin{equation}
|R_{{\hat t}{\hat y}{\hat t}{\hat y}}| = {3 v_b^2\; y^2 \over
4\;\rho^2}\left[ {d f(\rho) \over d\rho} \right]^2  \,,
\end{equation}
which yields $r_{\rm min} \equiv 1/ \sqrt{|R_{{\hat t}{\hat
y}{\hat t}{\hat y}}|} \sim {2\Delta \over {\sqrt 3}\; v_b}$, when
$y= \rho$ and the piece-wise continuous form of the shape function
is used.  The sampling time must be smaller than this length
scale, so that one my define
\begin{equation}
t_0=\alpha{2\Delta \over {\sqrt 3}\; v_b}\,, \qquad 0<\alpha\ll
1\,.
\end{equation}
Considering $\Delta / \rho \sim v_b\,t_0/\rho \ll 1$, the term
involving $1-f(\rho)$ in Eq. (\ref{wd-QI-bound}) may be neglected,
which provides
\begin{equation}
\Delta \leq {3\over 4}\sqrt{3\over\pi}\;{v_b \over \alpha^2}\, .
\end{equation}
Now, for instance considering $\alpha = 1/10$, one obtains
\begin{equation}
\Delta \leq 10^2\, v_b\; L_{\rm Planck}\, ,
\label{eq:wall_thickness}
\end{equation}
where $L_{\rm Planck}$ is the Planck length.  Thus, unless $v_b$
is extremely large, the wall thickness cannot be much above the
Planck scale.

It is also interesting to find an estimate of the total amount of
negative energy that is necessary to maintain a warp metric. It
was found that the energy required for a warp bubble is on the
order of
\begin{equation}
E \leq - 3 \times 10^{20} \; M_{\rm galaxy} \; v_b\; ,
\end{equation}
which is an absurdly enormous amount of negative energy, roughly
ten orders of magnitude greater than the total mass of the entire
visible universe \cite{PfenningF}.

\subsection{Linearized warp drive}

To ever bring a ``warp drive'' into a strong-field regime, any
highly-advanced civilization would first have to take it through
the ``weak-field'' regime. The central point of this Section is to
demonstrate that there are significant problems that already arise
even in the weak-field regime, and long before strong field
effects come into play. In the weak-field regime, applying
linearized theory, the physics is much simpler than in the
strong-field regime and this allows us to ask and answer questions
that are difficult to even formulate in the strong field regime
\cite{LV-CQG}. Our goal now  is to try to build a more realistic
model of a warp drive spacetime where the warp bubble is
interacting with a finite mass spaceship. To do so we first
consider the linearized theory applied to warp drive spacetimes,
for non-relativistic velocities, $v\ll 1$.

\subsubsection{The WEC violation to first order of $v$}

It is interesting to consider the specific case of an observer
which moves with an arbitrary velocity, $\tilde\beta$, along the
positive $z$ axis measure a negative energy density [at $O(v)$].
That is, $T_{\hat{0}\hat{0}}<0$. The $\tilde\beta$ occurring here
is completely independent of the shift vector
$\beta(x,y,z-z_0(t))$, and is also completely independent of the
warp bubble velocity $v$.

\begin{figure}[h]
\centering
  \includegraphics[width=1.9in]{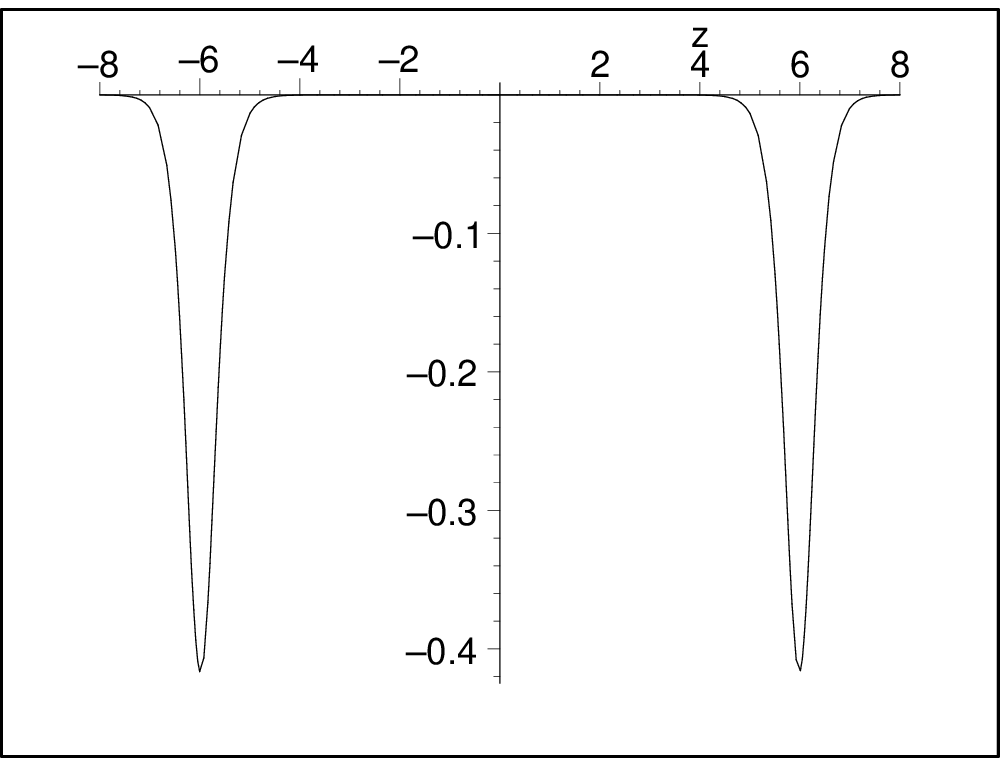}
  \hspace{0.1in}
  \includegraphics[width=1.9in]{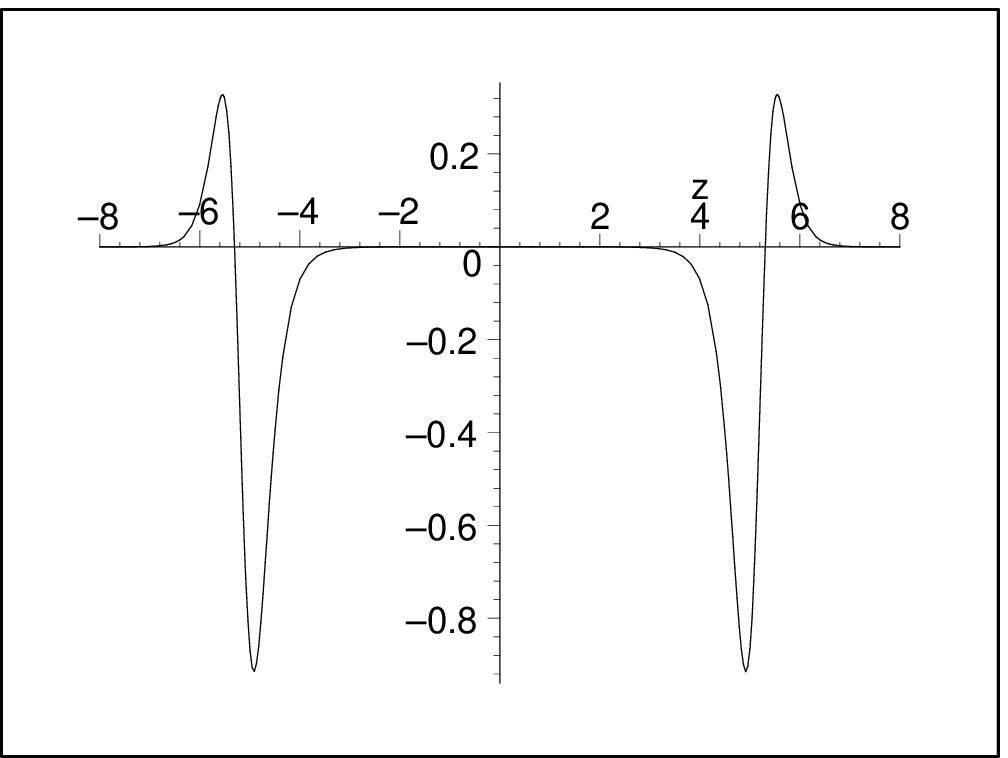}
  \hspace{0.1in}
  \includegraphics[width=1.9in]{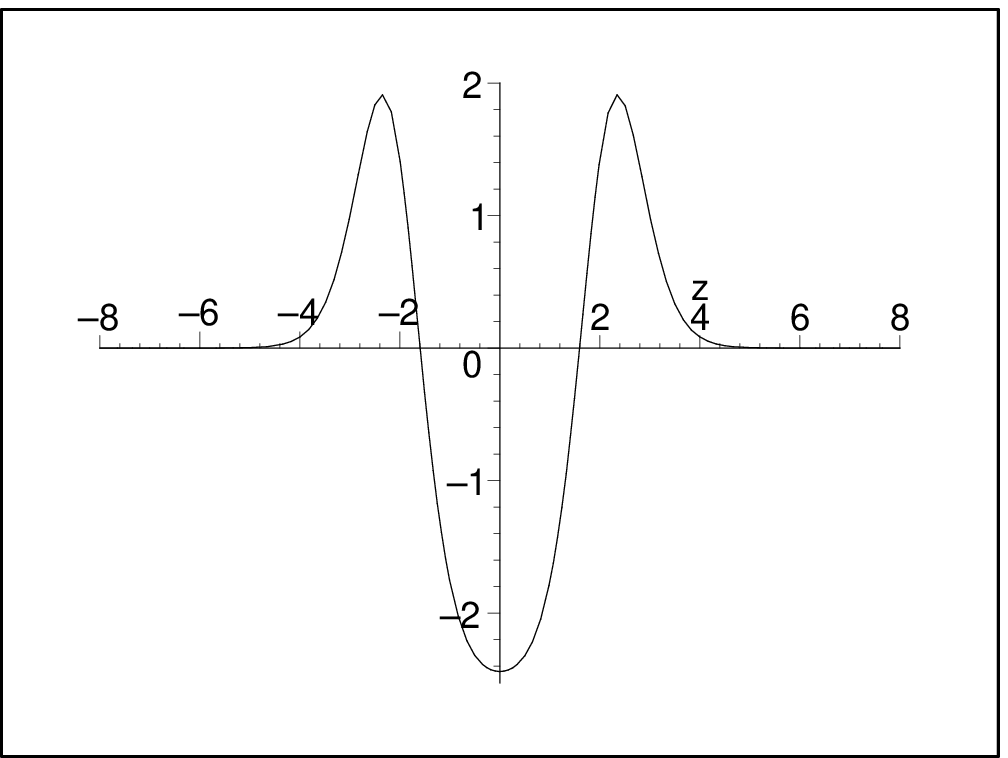}
  \caption[WEC violations for arbitrarily low warp bubble velocities]
  {The term in square brackets of Eq.
  (\ref{negativeenergydensity}) is plotted as a function of
  the $z$ coordinate. Taking into account the
  Alcubierre form function, we have considered the following
  values for the parameters $\sigma=2.5$ and $R=6$.
  Considering the definition $\rho=\sqrt{x^2+y^2}$,
  the plots have the respective values of $\rho=0$, $\rho=3$ and
  $\rho=5.8$.}\label{linearWEC}
\end{figure}

We have
\begin{equation}
T_{\hat{0}\hat{0}}= \frac{\gamma^{2}\tilde\beta v}{8\pi} \left[
\left(\frac{x^2+y^2}{r^2}\right) \; \frac{d^2 f}{d r^2} +
\left(\frac{x^2+y^2+2(z-z_0(t))^2}{r^3}\right) \; \frac{d f}{d r}
\right]  + O(v^2) \label{negativeenergydensity}\,.
\end{equation}
See Ref. \cite{LV-CQG} for details. A number of general features
can be extracted from the terms in square brackets, without
specifying an explicit form of $f$. In particular, $f$ decreases
monotonically from its value at $r=0$, $f=1$, to $f\approx 0$ at
$r\geq R$, so that ${d f}/{d r}$ is negative in this domain. The
form function attains its maximum in the interior of the bubble
wall, so that ${d^2 f}/{d r^2}$ is also negative in this region.
Therefore there is a range of $r$ in the immediate interior
neighbourhood of the bubble wall that necessarily provides
negative energy density, as seen by the observers considered
above. Again we find that WEC violations persist to arbitrarily
low warp bubble velocities. The negative character of the energy
density can be seen from Figure \ref{linearWEC}.


\subsubsection{Spaceship within the warp bubble}

Consider now a spaceship in the interior of an Alcubierre warp
bubble, which is moving along the positive $z$ axis with a
non-relativistic constant velocity \cite{LV-CQG}, i.e., $v\ll 1$.
The metric is given by
\begin{eqnarray}
\lefteqn{ d s^2=-d t^2+d x^2+d y^2+\left[d z-v\;f(x,y,z-vt)\,d t
\right]^2 }
\nonumber   \\
&&-2\Phi(x,y,z-vt)\, \left[d t^2+d x^2+d y^2+(d
z-v\;f(x,y,z-vt)\,d t)^2\right] \label{warpspaceshipmetric}  \,.
\end{eqnarray}
If $\Phi =0$, the metric (\ref{warpspaceshipmetric}) reduces to
the warp drive spacetime of Eq. (\ref{Cartesianwarpmetric}). If
$v=0$, we have the metric representing the gravitational field of
a static source.


Consider the approximation in which we keep the exact $v$
dependence but linearize in the gravitational field of the
spaceship $\Phi$.

The WEC is given by
\begin{eqnarray}
T_{\hat{\mu}\hat{\nu}} \; U^{\hat{\mu}} \; U^{\hat{\nu}}=\rho -
\frac{v^2}{32\pi}\left[\left(\frac{\partial f}{\partial
x}\right)^2 + \left(\frac{\partial f}{\partial y}\right)^2\right]
+O(\Phi^2)
 \,,
\end{eqnarray}
or by taking into account the Alcubierre form function, we have
\begin{equation}
T_{\mu\nu} \; U^{\mu}\; U^{\nu}=\rho -\frac{1}{32\pi}\frac{v^2
(x^2+y^2)}{r^2} \left( \frac{d f}{d r} \right)^2 +O(\Phi^2)
 \,.
\end{equation}

Once again, using the ``volume integral quantifier'', we find the
following estimate
\begin{eqnarray}
\int T_{\hat{\mu}\hat{\nu}} \; U^{\hat{\mu}} \; U^{\hat{\nu}} \;
d^3 x =  M_{\rm ship} -v^2 \;R^2\;\sigma + \int O(\Phi^2)  \;d^3
x\,,
\end{eqnarray}
which we can recast as
\begin{equation}
M_\mathrm{ADM} = M_\mathrm{ship} + M_\mathrm{warp} + \int
O(\Phi^2)  \; d^3 x\,.
\end{equation}
Now suppose we demand that the volume integral of the WEC at least
be positive, then
\begin{equation}
v^2 \;R^2\;\sigma  \leq M_{\rm ship} .
\end{equation}
This equation is effectively the quite reasonable  condition that
the net total energy stored in the warp field be less than the
total mass-energy of the spaceship itself, which places a powerful
constraint on the velocity of the warp bubble. Re-writing this in
terms of the size of the spaceship $R_\mathrm{ship}$ and the
thickness of the warp bubble walls $\Delta = 1/\sigma$, we have
\begin{equation}
v^2 \leq {M_{\rm ship}\over R_\mathrm{ship}}\; {R_\mathrm{ship} \;
\Delta\over R^2}.
\end{equation}
For any reasonable spaceship this gives extremely low bounds on
the warp bubble velocity.

\medskip

In a similar manner, the NEC, with $k^\mu=(1,0,0,\pm 1)$, is given
by
\begin{eqnarray}
T_{\hat{\mu}\hat{\nu}} \; k^{\hat{\mu}} \; k^{\hat{\nu}} &=& \rho
\pm \frac{v}{8\pi}\left(\frac{\partial^2 f}{\partial x^2} +
\frac{\partial^2 f}{\partial y^2}\right) -
\frac{v^2}{8\pi}\left[\left(\frac{\partial f}{\partial x}\right)^2
+ \left(\frac{\partial f}{\partial y} \right)^2\right] + O(\Phi^2)
\,.
\end{eqnarray}
Considering the ``volume integral quantifier'', we verify that, as
before, the exact solution in terms of polylogarithmic functions
is unhelpful, although we may estimate that
\begin{equation}
\int T_{\hat{\mu}\hat{\nu}} \; k^{\hat{\mu}} \; k^{\hat{\nu}} \;
d^3 x = M_{\rm ship}  -v^2 \, R^2 \, \sigma   + \int O(\Phi^2) \;
d^3 x\,,
\end{equation}
which is [to order $O(\Phi^2)$] the same integral we encountered
when dealing with the WEC. This volume integrated NEC is now
positive if
\begin{equation}
v^2 \;R^2\;\sigma  \leq M_{\rm ship} .
\end{equation}

Finally, considering a null vector oriented perpendicularly to the
direction of motion (for definiteness take $\hat k = \pm \hat x$),
the NEC takes the following form
\begin{equation}
T_{\hat{\mu}\hat{\nu}} \; k^{\hat{\mu}} \; k^{\hat{\nu}}=\rho
-\frac{v^2}{8\pi}\, \left[ {1\over2} \left (\frac{\partial
f}{\partial y} \right )^2 + \left(\frac {\partial f}{\partial z}
\right )^2 - (1-f) {\partial^2 f\over \partial z^2} \right] \mp
\frac{v}{8\pi}\left( \frac{\partial^{2}f}{\partial {x}\partial z}
\right) + O(\Phi^2) \,.
\end{equation}

Once again, evaluating the ``volume integral quantifier'', we have
\begin{equation}
\int T_{\hat{\mu}\hat{\nu}} \; k^{\hat{\mu}} \; k^{\hat{\nu}} \;
d^3 x =M_{\rm ship} -{v^2\over4}  \int \left( \frac{d f}{d r}
\right)^2 \; r^2 \; d r  + \frac{v^2}{6} \int
(1-f)\,\left(2r\frac{d f}{d r}+r^2\frac{d^2f}{d r^2} \right) \, d
r    + \int O(\Phi^2) d^3 x \,,
\end{equation}
which, as before, may be estimated as
\begin{equation}
\int T_{\hat{\mu}\hat{\nu}} \; k^{\hat{\mu}} \; k^{\hat{\nu}} \;
d^3 x \approx  M_{\rm ship} -v^2 R^2\, \sigma     + \int O(\Phi^2)
\; d^3 x\,.
\end{equation}
If we do not want the total NEC violations in the warp field to
exceed the mass of the spaceship itself we must again demand
\begin{equation}
v^2 \;R^2\;\sigma  \leq M_{\rm ship} .
\end{equation}
This places an extremely stringent condition on the warp drive
spacetime, namely, that for all conceivably interesting situations
the bubble velocity should be absurdly low, and it therefore
appears unlikely that, by using this analysis, the warp drive will
ever prove to be technologically useful. Finally, we point out
that any attempt at building up a ``strong-field'' warp drive
starting from an approximately Minkowski spacetime will inevitably
have to pass through a weak-field regime. Since the weak-field
warp drives are already so tightly constrained, the analysis above
implies additional difficulties for developing a ``strong field''
warp drive. See Ref. \cite{LV-CQG} for more details.

\subsection{Interesting aspects of the Alcubierre spacetime}
\label{sec:krasnikov-analyis}

\subsubsection{Superluminal travel in the warp drive}

To demonstrate that it is possible to travel to a distant point
and back in an arbitrary short time interval, let us consider two
distant stars, $A$ and $B$, separated by a distance $D$ in flat
spacetime. Suppose that, at the instant $t_0$, a spaceship
initiates it's movement using the engines, moving away from $A$
with a velocity $v<1$. It comes to rest at a distance $d$ from
$A$. For simplicity, assume that $R\ll d\ll D$. It is at this
instant that the perturbation of spacetime appears, centered
around the spaceship's position. The perturbation pushes the
spaceship away from $A$, rapidly attaining a constant
acceleration, $a$. Half-way between $A$ and $B$, the perturbation
is modified, so that the acceleration rapidly varies from $a$ to
$-a$. The spaceship finally comes to rest at a distance, $d$, from
$B$, in which the perturbation disappears. It then moves to $B$ at
a constant velocity in flat spacetime. The return trip to $A$ is
analogous.

If the variations of the acceleration are extremely rapid, the
total coordinate time, $T$, in a one-way trip will be
\begin{equation}
T=2\left( \frac{d}{v}+\sqrt{\frac{D-2d}{a}} \right) \,.
\end{equation}
The proper time of the stars are equal to the coordinate time,
because both are immersed in flat spacetime. The proper time
measured by observers within the spaceship is given by:
\begin{equation}
\tau=2\left( \frac{d}{\gamma v}+\sqrt{\frac{D-2d}{a}} \right)  \,,
\end{equation}
with $\gamma =(1-v^2)^{-1/2}$. The time dilation only appears in
the absence of the perturbation, in which the spaceship is moving
with a velocity $v$, using only it's engines in flat spacetime.

Using $R\ll d\ll D$, we can then obtain the following
approximation
\begin{equation}
\tau\approx T\approx 2\sqrt{\frac{D}{a}}  \,.
\end{equation}
Note that $T$ can be made arbitrarily short, by increasing the
value of $a$. The spaceship may travel faster than the speed of
light. However, it moves along a spacetime temporal trajectory,
contained within it's light cone, as light suffers the same
distortion of spacetime \cite{Alcubierre}.

\subsubsection{The Krasnikov analysis}

Krasnikov discovered a fascinating aspect of the warp drive, in
which an observer on a spaceship cannot create nor control on
demand an Alcubierre bubble, with $v>c$, around the ship
\cite{Krasnikov}. It is easy to understand this, as an observer at
the origin (with $t=0$), cannot alter events outside of his future
light cone, $|r|\leq t$, with $r=(x^2+y^2+z^2)^{1/2}$. Applied to
the warp drive, points on the outside front edge of the bubble are
always spacelike separated from the centre of the bubble.

The analysis is simplified in the proper reference frame of an
observer at the centre of the bubble. Using the transformation
$z'=z-z_{0}(t)$, the metric is given by
\begin{equation}
ds^2=-dt^2+dx^2+dy^2+\left[dz'+(1-f)vdt\right]^2  \,.
\end{equation}
Consider a photon emitted along the $+Oz$ axis (with
$ds^2=dx=dy=0$):
\begin{equation}
\frac{dz'}{dt}=1-(1-f)v  \,.
\end{equation}
If the spaceship is at rest at the center of the bubble, then
initially the photon has $dz/dt = v + 1$ or $dz'/dt = 1$ (because
$f=1$ in the interior of the bubble). However, at some point
$z'=z'_c$, with $f=1-1/v$, we have $dz'/dt=0$ \cite{Everett}. Once
photons reach $z'_c$, they remain at rest relative to the bubble
and are simply carried along with it. Photons emitted in the
forward direction by the spaceship never reach the outside edge of
the bubble wall, which therefore lies outside the forward light
cone of the spaceship. The bubble thus cannot be created (or
controlled) by any action of the spaceship crew. This behaviour is
reminiscent of an {\it event horizon}. This does not mean that
Alcubierre bubbles, if it were possible to create them, could not
be used as a means of superluminal travel. It only means that the
actions required to change the metric and create the bubble must
be taken beforehand by some observer whose forward light cone
contains the entire trajectory of the bubble.

\subsubsection{Reminiscence of an Event Horizon}

The appearance of an event horizon becomes evident in the
2-dimensional model of the Alcubierre space-time
\cite{Hiscock,Clark,Gonz}. The axis of symmetry coincides with the
line element of the spaceship. The metric, Eq.
(\ref{Cartesianwarpmetric}), reduces to
\begin{equation}
ds^2=-(1-v^2 f^2)dt^2-2vfdzdt+dz^2  \,.\label{2+1warpmetric}
\end{equation}

For simplicity, we consider the velocity of the bubble constant,
$v(t)=v_{b}$, and we have $r=[(z-v_{b}t)^2]^{1/2}$. If $z>v_{b}t$,
we consider the transformation $r=(z-v_{b}t)$. Note that the
metric components of Eq. (\ref{2+1warpmetric}) only depend on $r$,
which may be adopted as a coordinate.

Using the transformation, $dz=dr+v_{b}\,dt$, the metric, Eq.
(\ref{2+1warpmetric}) is given by
\begin{equation}
ds^2=-A(r)\left[dt-\frac{v_{b}(1-f(r))}{A(r)}\;dr\right]^2+\frac{dr^2}{A(r)}
\,.     \label{Hiscockwarp}
\end{equation}
The function $A(r)$, denoted by the Hiscock function, is given by
\begin{equation}
A(r)=1-v_{b}^2\,\left[1-f(r)\right]^2   \,.
\end{equation}
Its possible to represent the metric, Eq. (\ref{Hiscockwarp}), in
a diagonal form, using a new time coordinate
\begin{equation}
d\tau=dt-\frac{v_{b}\,\left[1-f(r)\right]}{A(r)}\;dr  \,,
\end{equation}
with which Eq. (\ref{Hiscockwarp}) reduces to
\begin{equation}
ds^2=-A(r)\,d\tau^2+\frac{dr^2}{A(r)}   \,.
\end{equation}

This form of the metric is manifestly static. The $\tau$
coordinate has an immediate interpretation in terms of an observer
on board of a spaceship: $\tau$ is the proper time of the
observer, because $A(r)\rightarrow 1$ in the limit $r\rightarrow
0$. We verify that the coordinate system is valid for any value of
$r$, if $v_{b}<1$. If $v_{b}>1$, we have a coordinate singularity
and an event horizon at the point $r_{0}$ in which
$f(r_{0})=1-1/v_b$ and $A(r_{0})=0$.

\subsection{Superluminal subway: The Krasnikov
tube}\label{Sec:Krasnikov}

It was pointed out in Section \ref{sec:krasnikov-analyis}, that
Krasnikov discovered an interesting aspect of the warp drive, in
which an observer on a spaceship cannot create nor control on
demand an Alcubierre bubble, i.e., points on the outside front
edge of the bubble are always spacelike separated from the centre
of the bubble. However, causality considerations do not prevent
the crew of a spaceship from arranging, by their own actions, to
complete a {\it round trip} from the Earth to a distant star and
back in an arbitrarily short time, as measured by clocks on the
Earth, by altering the metric along the path of their outbound
trip. Thus, Krasnikov introduced a metric with an interesting
property that although the time for a one-way trip to a distant
destination cannot be shortened, the time for a round trip, as
measured by clocks at the starting point (e.g. Earth), can be made
arbitrarily short, as will be demonstrated below.

\subsubsection{The 2-dimensional Krasnikov solution}

\begin{figure}[h]
\centering
  \includegraphics[width=2.0in]{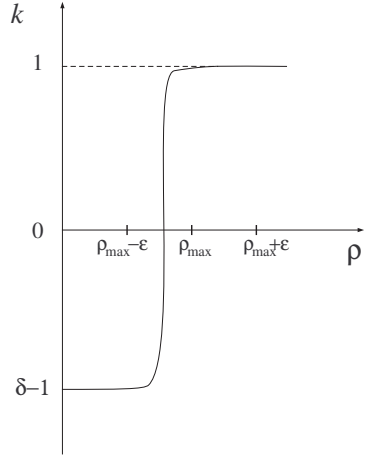}
  \caption[Graph of the Krasnikov form function]
  {Graph of the Krasnikov form function $k$ vs $\rho$
  at constant $x$ and $t$, considering only the region of $\epsilon<x<D-\epsilon$.}
  \label{fig:Kras-form}
\end{figure}

The 2-dimensional metric is given by
\begin{eqnarray}
ds^2&=&-(dt-dx)(dt+k(t,x)dx)
     \nonumber   \\
&=&-dt^2+\left[1-k(x,t)\right]\,dx\,dt+k(x,t)\,dx^2 \,.
       \label{kras:2d-metric}
\end{eqnarray}
The form function $k(x,t)$ is defined by
\begin{equation}
k(t,x)=1-(2-\delta)
\theta_{\varepsilon}(t-x)\left[\theta_{\varepsilon}(x)-
\theta_{\varepsilon}(x+\varepsilon-D)\right]   \,,
     \label{def:k}
\end{equation}
where $\delta$ and $\varepsilon$ are arbitrarily small positive
parameters. $\theta_{\varepsilon}$ denotes a smooth monotone
function
\[
\theta_{\varepsilon}(\xi)=\left\{ \begin{array}{ll}
                     1,   & {\rm if}\; \xi>\varepsilon \,, \\
                     0,    & {\rm if}\; \xi<0  \,,
                     \end{array}
                     \right.
\]
which is depicted in Figure \ref{fig:Kras-form}.

There are three distinct regions in the Krasnikov two-dimensional
spacetime, which we shall summarize in the following manner.

\begin{description}

\item[{\bf The outer region}.]


The outer region is given by the following set
\begin{equation}
\{x<0\}\cup\{x>D\}\cup\{x>t\}   \,.
\end{equation}
The two time-independent $\theta_\epsilon$-functions between the
brackets in Eq.~(\ref{def:k}) vanish for $x < 0$ and cancel for $x
> D$, ensuring $k=1$ for all $t$ except between $x=0$ and $x=D$.
When this behavior is combined with the effect of the factor
$\theta_\epsilon(t-x)$, one sees that the metric
~(\ref{kras:2d-metric}) is flat, $k=1$, and reduces to the
Minkowski spacetime everywhere for $t<0$ and at all times outside
the range $0<x<D$. Future light cones are generated by the
vectors:
\[
\left\{ \begin{array}{ll}
                     r_{O}=\partial_{t}+\partial_{x}   \,,\\
                     l_{O}=\partial_{t}-\partial_{x}  \,.
                     \end{array}
                     \right.
\]

\item[{\bf The inner region}.]


The inner region is given by the following set
\begin{equation}
\{x<t-\varepsilon\}\cap\{\varepsilon<x<D-\varepsilon\}  \,,
\end{equation}
so that the first two $\theta_\epsilon$-functions in
Eq.~(\ref{def:k}) both equal $1$, while $\theta_\epsilon(x +
\epsilon - D) = 0$, giving $k = \delta -1$ everywhere within this
region. This region is also flat, but the light cones are {\it
more open}, being generated by the following vectors
\[
\left\{ \begin{array}{ll}
                     r_{I}=\partial_{t}+\partial_{x} \\
                     l_{I}=-(1-\delta)\partial_{t}-\partial_{x} \,.
                     \end{array}
                     \right.
\]

\item[{\bf The transition region}.]


The transition region is a narrow curved strip in spacetime, with
width $\sim \varepsilon$. Two spatial boundaries exist between the
inner and outer regions. The first lies between $x=0$ and
$x=\varepsilon$, for $t>0$. The second lies between
$x=D-\varepsilon$ and $x=D$, for $t>D$. It is possible to view
this metric as being produced by the crew of a spaceship,
departing from point $A$ ($x=0$), at $t=0$, travelling along the
$x$-axis to point $B$ ($x=D$) at a speed, for simplicity,
infinitesimally close to the speed of light, therefore arriving at
$B$ with $t\approx D$.

\end{description}

\begin{figure}[h]
\centering
  \includegraphics[width=2.5in]{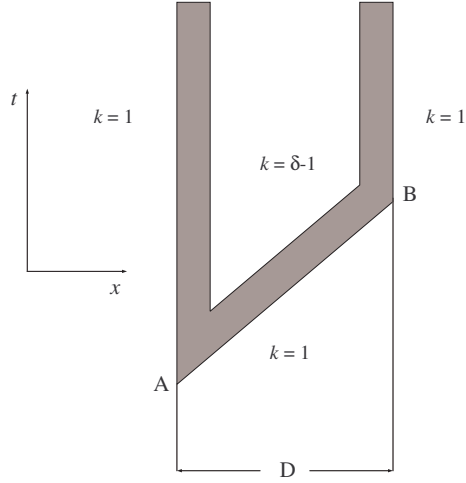}
  \caption[The Krasnikov spacetime in the $(x,t)$ plane]
  {The Krasnikov spacetime in the $(x,t)$ plane. The vertical lines
  $A$ and $B$ are the world lines of the stars $A$ and $B$,
  respectively. The world line of the spaceship is approximately represented
  by the line segment $AB$.}\label{fig:Kras-2d}
\end{figure}

The metric is modified by changing $k$ from $1$ to $\delta-1$
along the $x$-axis, in between $x=0$ and $x=D$, leaving a
transition region of width $\sim \varepsilon$ at each end for
continuity. But, as the boundary of the forward light cone of the
spaceship at $t=0$ is $|x|=t$, it is not possible for the crew to
modify the metric at an arbitrary point $x$ before $t=x$. This
fact accounts for the factor $\theta_{\varepsilon}(t-x)$ in the
metric, ensuring a transition region in time between the inner and
outer region, with a duration of $\sim \varepsilon$, lying along
the wordline of the spaceship, $x\approx t$. The geometry is shown
in the $(x,t)$ plane in Figure \ref{fig:Kras-2d}.


\subsubsection{Superluminal travel within the Krasnikov tube}

\begin{figure}[h]
\centering
  \includegraphics[width=5.0in]{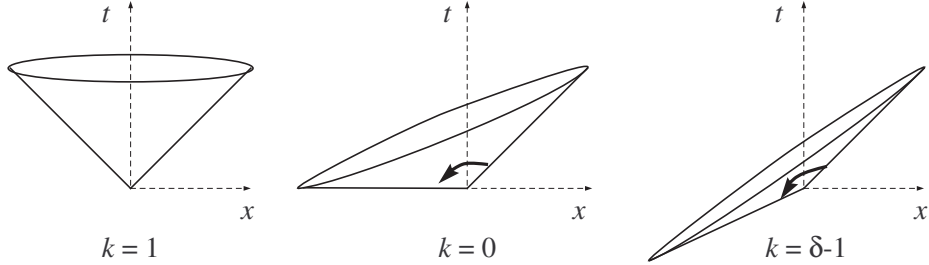}
  \caption[Forward light cones in the 2-dimensional Krasnikov
  spacetime]{Forward light cones in the 2-dimensional Krasnikov
  spacetime for $k=1$, $k=0$ and $k=\delta-1$.}\label{fig:Kras-lightcones}
\end{figure}

The properties of the modified metric with $\delta-1\leq k \leq 1$
can be easily seen from the factored form of $ds^2=0$. The two
branches of the forward light cone in the $(t,x)$ plane are given
by $dx/dt=1$ and $dx/dt=-k$. As $k$ becomes smaller and then
negative, the slope of the left-hand branch of the light cone
becomes less negative and then changes sign, i. e., the light cone
along the negative $x$-axis ``opens out''. See Figure
\ref{fig:Kras-lightcones}.


The inner region, with $k=\delta -1$, is flat because the metric,
Eq. (\ref{kras:2d-metric}), may be cast into the Minkowski form,
applying the following coordinate transformations
\begin{equation}
dt'=dt+\left( \frac{\delta}{2}-1 \right) dx\,,\qquad dx'=\left(
\frac{\delta}{2} \right) dx  \,.
     \label{Mink-tranf}
\end{equation}
The transformation is singular at $\delta=0$, i.e., $k=-1$. Note
that the left branch of the region is given by $dx'/dt'=-1$.

From the above equations, one may easily deduce the following
expression
\begin{equation}
\frac{dt}{dt'}=1+\left( \frac{2-\delta}{\delta}
\right)\frac{dx'}{dt'}  \,.
     \label{kras:time}
\end{equation}
For an observer moving along the positive $x'$ and $x$ directions,
with $dx'/dt'<1$, we have $dt'>0$ and consequently $dt>0$, if
$0<\delta \leq 2$. However, if the observer is moving sufficiently
close to the left branch of the light cone, given by $dx'/dt'=-1$,
Eq. (\ref{kras:time}) provides us with $dt/dt'<0$, for $\delta<
1$. Therefore $dt<0$, the observer traverses backward in time, as
measured by observers in the outer region, with $k=1$.

The superluminal travel analysis is as follows. Imagine a
spaceship leaving star $A$ and arriving at star $B$, at the
instant $t\approx D$. The crew of the spaceship modify the metric,
so that $k\approx -1$, for simplicity, along the trajectory. Now
suppose the spaceship returns to star $A$, travelling with a
velocity arbitrarily close to the speed of light, i.e.,
$\frac{dx'}{dt'}\approx -1$. Therefore, from Eq.
(\ref{Mink-tranf}), one obtains the following relation
\begin{equation}
v_{\rm return}=\frac{dx}{dt}\approx
-\frac{1}{k}=\frac{1}{1-\delta}\approx 1
\end{equation}
and $dt<0$, for $dx<0$. The return trip from star $B$ to $A$ is
done in an interval of $\Delta t_{\rm return}=-D/v_{\rm
return}=D/(\delta -1)$. The total interval of time, measured at
$A$, is given by $T_A =D+\Delta t_{\rm return}=D \delta$. For
simplicity, consider $\varepsilon$ negligible. Superluminal travel
is implicit, because $|\Delta t_{\rm return}|<D$, if $\delta >0$,
i.e., we have a spatial spacetime interval between $A$ and $B$.
Note that $T_A$ is always positive, but may attain a value
arbitrarily close to zero, for an appropriate choice of $\delta$.

Note that for the case $\delta<1$, it is always possible to choose
an allowed value of $dx'/dt'$ for which $dt/dt' = 0$, meaning that
the return trip is instantaneous as seen by observers in the
external region. This follows easily from Eq.~(\ref{kras:time}),
which implies that $dt/dt' = 0$ when $dx'/dt'$ satisfies
\begin{equation}
\frac{dx'}{dt'} = - \frac{\delta}{(2- \delta)} \,,
\label{eq:DX'/DT'I}
\end{equation}
which lies between $0$ and $-1$ for $0< \delta < 1$.

\subsubsection{The 4-dimensional generalization}

Soon after the Krasnikov two-dimensional solution, Everett and
Roman~\cite{Everett} generalized the analysis to four dimensions,
denoting the solution as the {\it Krasnikov tube}. Consider that
the 4-dimensional modification of the metric begins along the path
of the spaceship, which is moving along the $x$-axis, occurring at
position $x$ at time $t \approx x$, the time of passage of the
spaceship. Also assume that the disturbance in the metric
propagates radially outward from the $x$-axis, so that causality
guarantees that at time $t$ the region in which the metric has
been modified cannot extend beyond $\rho = t - x$, where $\rho
={(y^2 + z^2)}^{1/2}$. The modification in the metric should also
not extend beyond some maximum radial distance $\rho_{max} \ll D$
from the $x$-axis. Thus, the metric in the 4-dimensional
spacetime, written in cylindrical coordinates, is given by
\cite{Everett}
\begin{equation}
ds^2=-dt^2+(1-k(t,x,\rho))dx dt+k(t,x,\rho)dx^2+d\rho^2+\rho^2
d\phi^2   \,,
   \label{4d-Krasnikov-metric}
\end{equation}
with
\begin{equation}
k(t,x,\rho)=1-(2-\delta)\theta_{\varepsilon}(\rho_{max}-\rho)
\theta_{\varepsilon}(t-x-\rho)[\theta_{\varepsilon}(x)-
\theta_{\varepsilon}(x+\varepsilon-D)]   \,.
    \label{4d:form}
\end{equation}
For $t\gg D+\rho_{max}$ one has a tube of radius $\rho_{max}$
centered on the $x$-axis, within which the metric has been
modified. This structure is denoted by the {\it Krasnikov tube}.
In contrast with the Alcubierre spacetime metric, the metric of
the Krasnikov tube is static once it has been created.

The stress-energy tensor element $T_{tt}$ given by
\begin{equation}
T_{tt}=\frac{1}{32
\pi(1+k)^2}\left[-\frac{4(1+k)}{\rho}\frac{\partial k}{\partial
\rho}+3\left(\frac{\partial k}{\partial
\rho}\right)^2-4(1+k)\frac{\partial^2 k}{\partial \rho^2}\right]
\,,
\end{equation}
can be shown to be the energy density measured by a static
observer \cite{Everett}, and violates the WEC in a certain range
of $\rho$, i.e., $T_{\mu\nu}U^{\mu}U^{\nu}<0$.

To verify the violation of the WEC, consider the energy density in
the middle of the tube and at a time long after it's formation,
i.e., $x=D/2$ and $t\gg x+\rho +\varepsilon $, respectively. In
this region we have $\theta_{\varepsilon}(x)=1$,
$\theta_{\varepsilon}(x+\varepsilon-D)=0$ and
$\theta_{\varepsilon}(t-x-\rho)=1$. With this simplification the
form function, Eq. (\ref{4d:form}), reduces to
\begin{equation}
k(t,x,\rho)=1-(2-\delta)\theta_{\varepsilon}(\rho_{max}-\rho)  \,.
     \label{4d:midtube-form}
\end{equation}
Consider the following specific form for
$\theta_{\varepsilon}(\xi)$ \cite{Everett} given by
\begin{equation}
\theta_{\varepsilon}(\xi)=\frac{1}{2}\left \{ \tanh \left
[2\left(\frac{2\xi}{\varepsilon}-1 \right )\right]+1 \right \} \,,
\end{equation}
so that the form function of Eq. (\ref{4d:midtube-form}) is
provided by
\begin{equation}
k=1-\left (1-\frac{\delta}{2}\right)\left \{ \tanh \left
[2\left(\frac{2\xi}{\varepsilon}-1 \right )\right]+1 \right \} \,.
\end{equation}

\begin{figure}
\centering
\includegraphics[width=7cm]{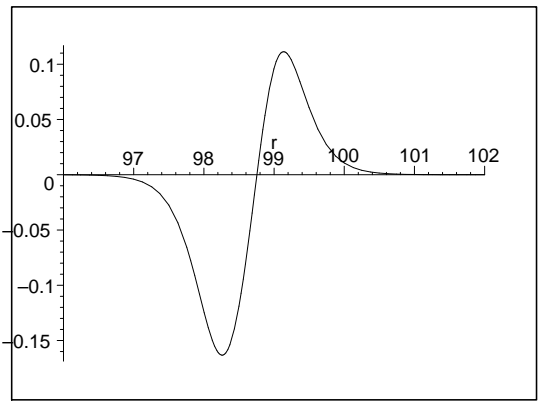}
\caption[Energy density of the Krasnikov tube]{Graph of the energy
density, $T_{tt}$, as a function of $\rho$ at the middle of the
Krasnikov tube, $x=D/2$, and long after it's formation, $t\gg
x+\rho +\varepsilon $. We consider the following values for the
parameters: $\delta =0.1$, $\varepsilon =1$ and
$\rho_{max}=100\varepsilon =100$. }\label{fig3-superluminal}
\end{figure}

Choosing the following values for the parameters: $\delta =0.1$,
$\varepsilon =1$ and $\rho_{max}=100\varepsilon =100$, the
negative character of the energy density is manifest in the
immediate inner vicinity of the tube wall, as shown in Figure
\ref{fig3-superluminal}.


\section{Closed timelike curves and causality
violation}\label{secIV}

As time is incorporated into the proper structure of the fabric of
spacetime, it is interesting to note that general relativity is
contaminated with non-trivial geometries which generate {\it
closed timelike curves}
\cite{Visser,World-Scientific,Springer,Kluwer,LLQ-PRD,Tipler-CTCs}.
A closed timelike curve (CTC) allows time travel, in the sense
that an observer which travels on a trajectory in spacetime along
this curve, returns to an event which coincides with the
departure. The arrow of time leads forward, as measured locally by
the observer, but globally he/she may return to an event in the
past. This fact apparently violates causality, opening Pandora's
box and producing time travel paradoxes \cite{Nahin}. The notion
of causality is fundamental in the construction of physical
theories, therefore time travel and it's associated paradoxes have
to be treated with great caution. A great variety of solutions to
the Einstein Field Equations (EFEs) containing CTCs exist, but,
two particularly notorious features seem to stand out. Solutions
with a tipping over of the light cones due to a rotation about a
cylindrically symmetric axis; and solutions that violate the
Energy Conditions of general relativity, which are fundamental in
the singularity theorems and theorems of classical black hole
thermodynamics \cite{hawkingellis}.

\subsection{Stationary and axisymmetric solutions generating CTCs}

The tipping over of light cones seem to be a generic feature of
some solutions with a rotating cylindrical symmetry. The general
metric for a stationary, axisymmetric solution with rotation is
given by \cite{Visser,Wald}
\begin{equation}
ds^2=-F(r)\,dt^2+H(r)\,dr^2+L(r)\,d\phi^2+2\,M(r)\,d\phi \,dt+H
(r)\,dz^2  \,,   \label{stationarymetric}
\end{equation}
The metric components are functions of $r$ alone. It is clear that
the determinant, $g={\rm det}(g_{\mu\nu})=-(FL+M^2)H^2$ is
Lorentzian, provided that $(FL+H^2)>0$.

Due to the periodic nature of the angular coordinate, $\phi$, an
azimuthal curve with $\gamma =\{t={\rm const},r={\rm const},z={\rm
const}\}$ is a closed curve of invariant length $s_{\gamma}^2
\equiv L(r)(2\pi)^2$. If $L(r)$ is negative then the integral
curve with $(t,r,z)$ fixed is a CTC. If $L(r)=0$, then the
azimuthal curve is a closed null curve, CNC. Alternatively,
consider a null azimuthal curve in the $(\phi,t)$ plane with
$(r,z)$ fixed. It is not necessarily a geodesic, nor will it be a
closed curve. The null condition, $ds^2=0$, implies
\begin{equation}
0=-F+2M\dot{\phi}+L\dot{\phi}^2  \,,
\end{equation}
with $\dot{\phi}=d\phi/dt$. Solving the quadratic, we have
\begin{equation}
\frac{d\phi}{dt}=\dot{\phi}=\frac{-M\pm \sqrt{M^2+FL}}{L}  \,.
\end{equation}

In virtue of the Lorentzian signature constraint, $FL+H^2>0$, the
roots are real. If $L(r)<0$ then the light cones are tipped over
sufficiently far to permit a trip to the past. By going once
around the azimuthal direction, the total backward time-jump for a
null curve is
\begin{equation}
\Delta T=\frac{2\pi |L|}{-M+ \sqrt{M^2-F|L|}}   \,.
\end{equation}
Roughly, light cones which are tilted over are generic features of
spacetimes which contain CTCs.

If $L(r)<0$ for even a single value of $r$, then there is a closed
causal curve passing through every point of the spacetime. To
visualize this consider a null curve beginning at an arbitrary
$x$, reaching $r$ such that $L(r)<0$, Then follow the null curve
that wraps around the azimuth a total of $N$ times. The total
backward time-jump is then $N\Delta T$. Finally, follow an
ordinary null curve back to the starting point $x$. So, if
$L(r)<0$ for even a single value of $r$, the chronology-violation
region covers the entire spacetime. See Ref. \cite{Visser} for
more details.

The present Section is far from making an exhaustive search of all
the EFE solutions generating CTCs with these features, but the
best known spacetimes will be briefly analyzed, namely, the van
Stockum spacetime, the G\"{o}del universe, the spinning cosmic
strings and the Gott two-string time machine, which is a variation
on the theme of the spinning cosmic string.

\subsubsection{Van Stockum spacetime}

The earliest solution to the EFEs containing CTCs, is probably
that of the van Stockum spacetime. It is a stationary,
cylindrically symmetric solution describing a rapidly rotating
infinite cylinder of dust, surrounded by vacuum. The centrifugal
forces of the dust are balanced by the gravitational attraction.
The metric, assuming the respective symmetries, takes the form of
Eq. (\ref{stationarymetric}). Consider a frame in which the matter
is at rest, and it can be shown that the source is simply positive
density dust, implying that all of the energy condition are
satisfied \cite{Visser}.

The metric for the interior solution $r<R$, where the surface of
the cylinder is located at $r=R$, is
\begin{equation}
ds^2=-dt^2+2\omega r^2 d\phi dt+r^2(1-\omega^2
r^2)d\phi^2+\exp(-\omega^2 r^2)(dr^2+dz^2)
\end{equation}
where $\omega$ is the angular velocity of the cylinder. It is
readily verified that CTCs arise if $\omega r>1$, i.e., for
$r>1/\omega$ the azimuthal curves with $(t,r,z)$ fixed are CTCs.
The condition $M^2+FL=\omega^2 r^4+r^2(1-\omega^2 r^2)=r^2>0$ is
imposed.

The causality violation region could be eliminated by requiring
that the boundary of the cylinder to be at $r=R<1/a$. The interior
solution would then be joined to an exterior solution, which would
be causally well-behaved. The resulting upper bound to the
``velocity'' $\omega R$ would be $1$, although the orbits of the
particles creating the field are timelike for all $r$.

Van Stockum also developed a procedure which generates an exterior
solution for all $\omega R>0$~\cite{Tipler}. It can be shown that
the causality violation is avoided for $\omega R\leq 1/2$, but in
the region $\omega R> 1/2$, CTCs appear. Consider the exterior
metric components for this range, $\omega R>1/2$:
\begin{eqnarray*}
&&H(r)=\exp(-\omega^2 r^2) \left (r/R \right )^{-2\omega^2 r^2},
\qquad
L(r)=\frac{Rr \sin(3\beta +\gamma)}{2\sin(2\beta)
\cos(\beta)},\\
&&M(r)=\frac{r \sin(\beta +\gamma)}{\sin(2\beta)},
\qquad
F(r)=\frac{r \sin(\beta -\gamma)}{R\sin(\beta)}  \,,
\end{eqnarray*}
with
\begin{eqnarray*}
\gamma=\gamma(r)=(4\omega^2 R^2-1)^{1/2} \ln \left(r/R \right )
\qquad {\rm and} \qquad
\beta=\beta(r)=\arctan(4\omega^2 R^2-1)^{1/2} \,.
\end{eqnarray*}
As is the interior solution, one may verify that $FL+M^2=r^2$, so
that the metric signature is Lorentzian for $R\leq r<\infty $.

\begin{figure}[h]
  \centering
  \includegraphics[width=4.0in]{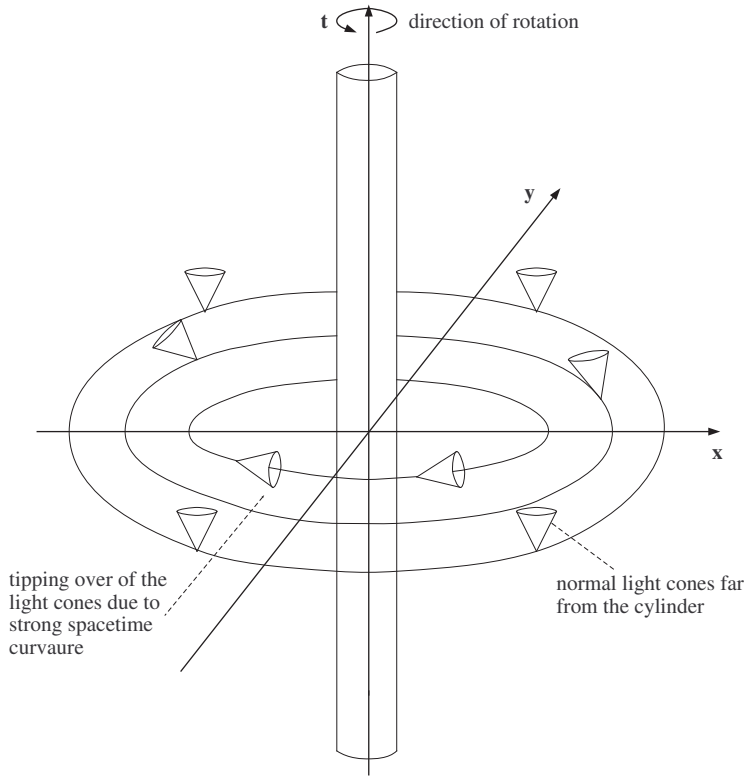}
  \caption[Tipping over of light cones in the van Stockum spacetime]
  {Van Stockum spacetime showing the tipping over of
  light cones close to the cylinder, due to the strong curvature of
  spacetime, which induce closed timelike curves.}\label{fig:stockum}
\end{figure}

The causality violations arise from the sinusoidal factors of the
metric components. Thus, causality violation occur in the
matter-free space surrounding a rapidly rotating infinite
cylinder, as shown in Figure \ref{fig:stockum}. However, it is not
clear that the properties of such a cylinder also hold for
realistic cylinders.

The van Stockum spacetime is not asymptotically flat. But, the
gravitational potential of the cylinder's Newtonian analog also
diverges at radial infinity. Shrinking the cylinder down to a
``ring'' singularity, one ends up with the Kerr solution, which
also has CTCs (The causal structure of the Kerr spacetime has been
extensively analyzed by de Felice and
collaborators~\cite{Felice1,Felice2,Felice3,Felice4,Felice5}).

In summary, the van Stockum solution contains CTC provided $\omega
R>1/2$. The causality-violating region covers the entire
spacetime. Reactions to the van Stockum solution is that it is
unphysical, as it applies to an infinitely long cylinder and it is
not asymptotically flat.

\subsubsection{The G\"{o}del Universe}

Kurt G\"{o}del in $1949$ discovered an exact solution to the EFEs
of a uniformly rotating universe containing dust and a nonzero
cosmological constant~\cite{Godel}. It can be shown that the null,
weak and dominant energy conditions are satisfied. However, the
dominant energy condition is in the imminence of being violated.

Consider a set of alternative coordinates, which explicitly
manifest the rotational symmetry of the solution, around the axis
$r=0$, and suppressing the irrelevant $z$ coordinate
\cite{hawkingellis,Godel}, the metric of the G\"{o}del solution is
provided by
\begin{equation}
ds^2=2w^{-2} (-dt'^2+dr^2-(\sinh ^4r-\sinh ^2r)\,d\phi
^2+2(\sqrt{2})\sinh ^2r \,d\phi \,dt)  \,.
\end{equation}
Moving away from the axis, the light cones open out and tilt in
the $\phi$-direction. The azimuthal curves with $\gamma =\{t={\rm
const},r={\rm const},z={\rm const}\}$ are CTCs if the condition
$r>\ln (1+\sqrt{2})$ is satisfied \cite{Visser}.

It is interesting to note that in the G\"{o}del spacetime, closed
timelike curves are not geodesics. However, Novello and
Rebou\c{c}as \cite{Nov-Reb} discovered a new generalized solution
of the G\"{o}del metric, of a shear-free nonexpanding rotating
fluid, in which successive concentric causal and noncausal regions
exist, with closed timelike curves which are geodesics. A complete
study of geodesic motion in G\"{o}del's universe, using the method
of the effective potential was further explored by Novello {\it et
al}~\cite{Nov-Soar}. Much interest has been aroused in time travel
in the G\"{o}del spacetime, from which we may mention the analysis
of the geodesical and non-geodesical motions considered by
Pfarr~\cite{Pfarr} and Malament~\cite{Malament1,Malament2}.

\subsubsection{Spinning Cosmic String}

Consider an infinitely long straight string that lies and spins
around the $z$-axis. The symmetries are analogous to the van
Stockum spacetime, but the asymptotic behavior is different
\cite{Visser,Jensen}. We restrict the analysis to an infinitely
long straight string, with a delta-function source confined to the
$z$-axis. It is characterized by a mass per unit length, $\mu$; a
tension, $\tau$, and an angular momentum per unit length, $J$. For
cosmic strings, the mass per unit length is equal to the tension,
$\mu=\tau$.

In cylindrical coordinates the metric takes the following form
\begin{equation}
ds^2=-\left [d(t+4GJ\varphi) \right
]^2+dr^2+(1-4G\mu)^2\,r^2\;d\varphi^2+dz^2 \,,
\end{equation}
with the following coordinate range
\begin{equation}
-\infty <t<+\infty,  \qquad   0<r<\infty,  \qquad   0\leq\varphi
\leq 2\pi,  \qquad  -\infty <z<+\infty   \,.
\end{equation}

Consider an azimuthal curve, i.e., an integral curve of $\varphi$.
Closed timelike curves appear whenever
\begin{equation}
r<\frac{4GJ}{1-4G\mu}  \,.
\end{equation}
These CTCs can be deformed to cover the entire spacetime,
consequently, the chronology-violating region covers the entire
manifold \cite{Visser}.

\subsubsection{Gott Cosmic String time machine}

An extremely elegant model of a time-machine was constructed by
Gott \cite{GottCTC}. The Gott time-machine is an exact solution of
the EFE for the general case of two moving straight cosmic strings
that do not intersect \cite{GottCTC}. This solution produces CTCs
even though they do not violate the WEC, have no singularities and
event horizons, and are not topologically multiply-connected as
the wormhole solution. The appearance of CTCs relies solely on the
gravitational lens effect and the relativity of simultaneity.

It is also interesting to verify whether the CTCs in the Gott
solution appear at some particular moment, i.e., when the strings
approach each other's neighborhood, or if they already pre-exist,
i.e., they intersect any spacelike hypersurface. These questions
are particularly important in view of Hawking's Chronology
Protection Conjecture \cite{hawking}. This conjecture states that
the laws of physics prevent the creation of CTCs. If correct, then
the solutions of the EFE which admit CTCs are either unrealistic
or are solutions in which the CTCs are pre-existing, so that the
time -machine is not created by dynamical processes. Amos Ori
proved that in Gott's spacetime, CTCs intersect every $t={\rm
const}$ hypersurface \cite{Ori:CTCstring}, so that it is not a
counter-example to the Chronology Protection Conjecture.

The global structure of the Gott spacetime was further explored by
Cutler~\cite{Cutler-CTC}, and it was shown that the closed
timelike curves are confined to a certain region of the spacetime,
and that the spacetime contains complete spacelike and achronal
hypersurfaces from which the causality violating regions evolve.
Grant also examined the global structure of the two-string
spacetime and found that away from the strings, the space is
identical to a generalized Misner space~\cite{Grant-CTC}. The
vacuum expectation value of the energy-momentum tensor for a
conformally coupled scalar field was then calculated on the
respective generalized Misner space, which was found to diverge
weakly on the chronology horizon, but diverge strongly on the
polarized hypersurfaces. Thus, the back reaction due to the
divergent behaviour around the polarized hypersurfaces are
expected to radically alter the structure of spacetime, before
quantum gravitational effects become important, suggesting that
Hawking's chronology protection conjecture holds for spaces with a
noncompactly generated chronology horizon. Soon after,
Laurence~\cite{Laurence-CTC} showed that the region containing
CTCs in Gott's two-string spacetime is identical to the regions of
the generalized Misner space found by Grant, and constructed a
family of isometries between both Gott's and Grant's regions. This
result was used to argue that the slowly diverging vacuum
polarization at the chronology horizon of the Grant space carries
over without change to the Gott space. Furthermore, it was shown
that the Gott time machine is unphysical in nature, for such an
acausal behaviour cannot be realized by physical and timelike
sources \cite{Deser,Deser2,Deser3,Farhi,Farhi2}.

\subsection{Solutions violating the Energy Conditions}

The traditional manner of solving the EFEs, $G_{\mu \nu}=8\pi G
T_{\mu \nu}$, consists in considering a plausible stress-energy
tensor, $T_{\mu \nu}$, and finding the geometrical structure,
$G_{\mu\nu}$. But one can run the EFE in the reverse direction by
imposing an exotic metric $g_{\mu\nu}$, and eventually finding the
matter source for the respective geometry. In this fashion,
solutions violating the energy conditions have been obtained.
Adopting the reverse philosophy, solutions such as traversable
wormholes, the warp drive, the Krasnikov tube and the Ori-Soen
spacetime have been obtained. These solutions violate the energy
conditions and with simple manipulations generate CTCs.

\subsubsection{Conversion of traversable wormholes into time
machines}

Much interest has been aroused in traversable wormholes since the
classical article by Morris and Thorne \cite{Morris}. A wormhole
is a hypothetical tunnel which connects different regions in
spacetime. These solutions are multiply-connected and probably
involve a topology change, which by itself is a problematic issue.
One of the most fascinating aspects of wormholes is their apparent
ease in generating CTCs \cite{mty}. There are several ways to
generate a time machine using multiple wormholes \cite{Visser},
but a manipulation of a single wormhole seems to be the simplest
way \cite{mty,visser-babyuniv}. The basic idea is to create a time
shift between both mouths. This is done invoking the time dilation
effects in special relativity or in general relativity, i.e., one
may consider the analogue of the twin paradox, in which the mouths
are moving one with respect to the other, or simply the case in
which one of the mouths is placed in a strong gravitational field.

\begin{figure}[h]
\centering
  \includegraphics[width=2.6in]{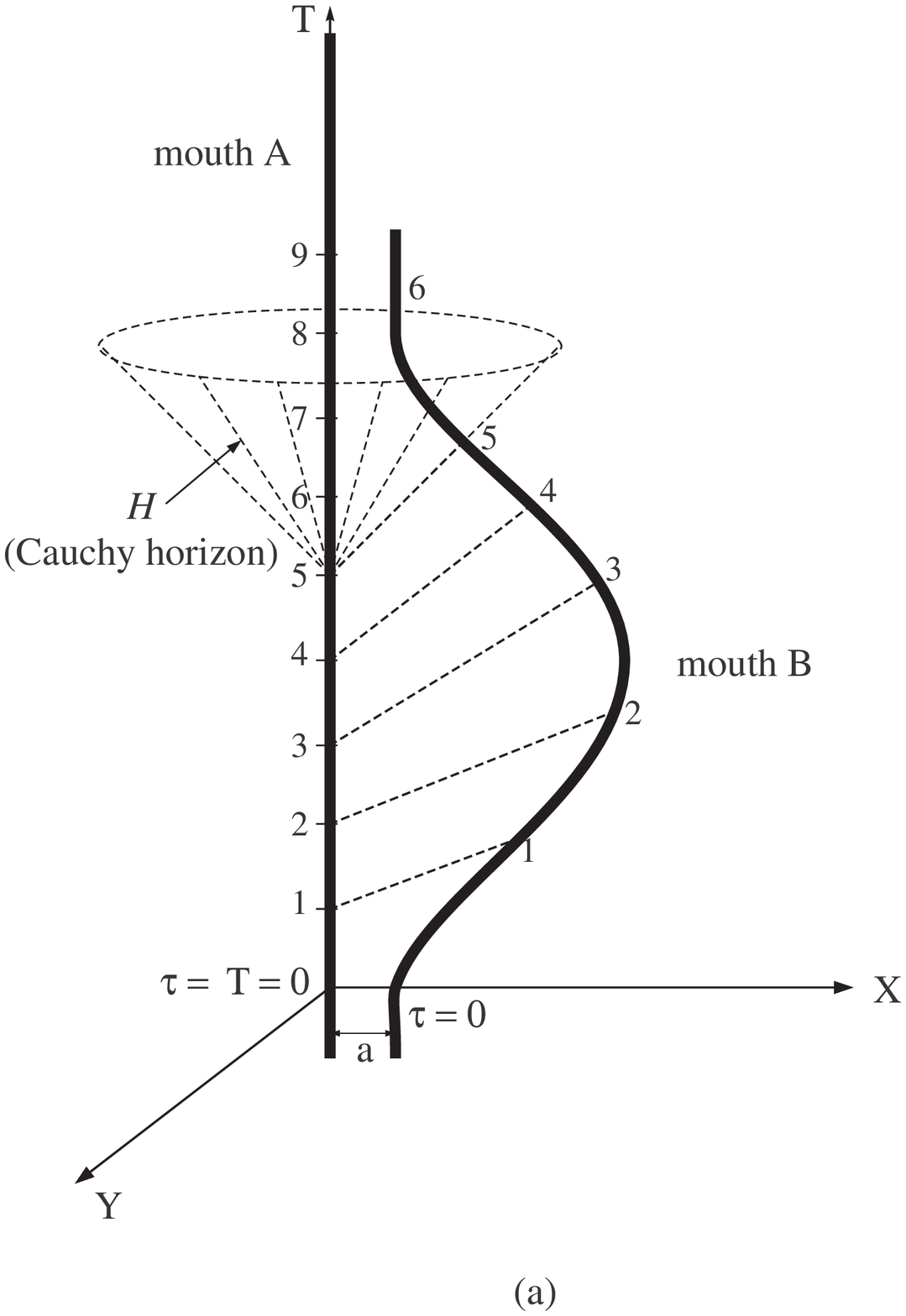}
  \hspace{0.4in}
  \includegraphics[width=2.4in]{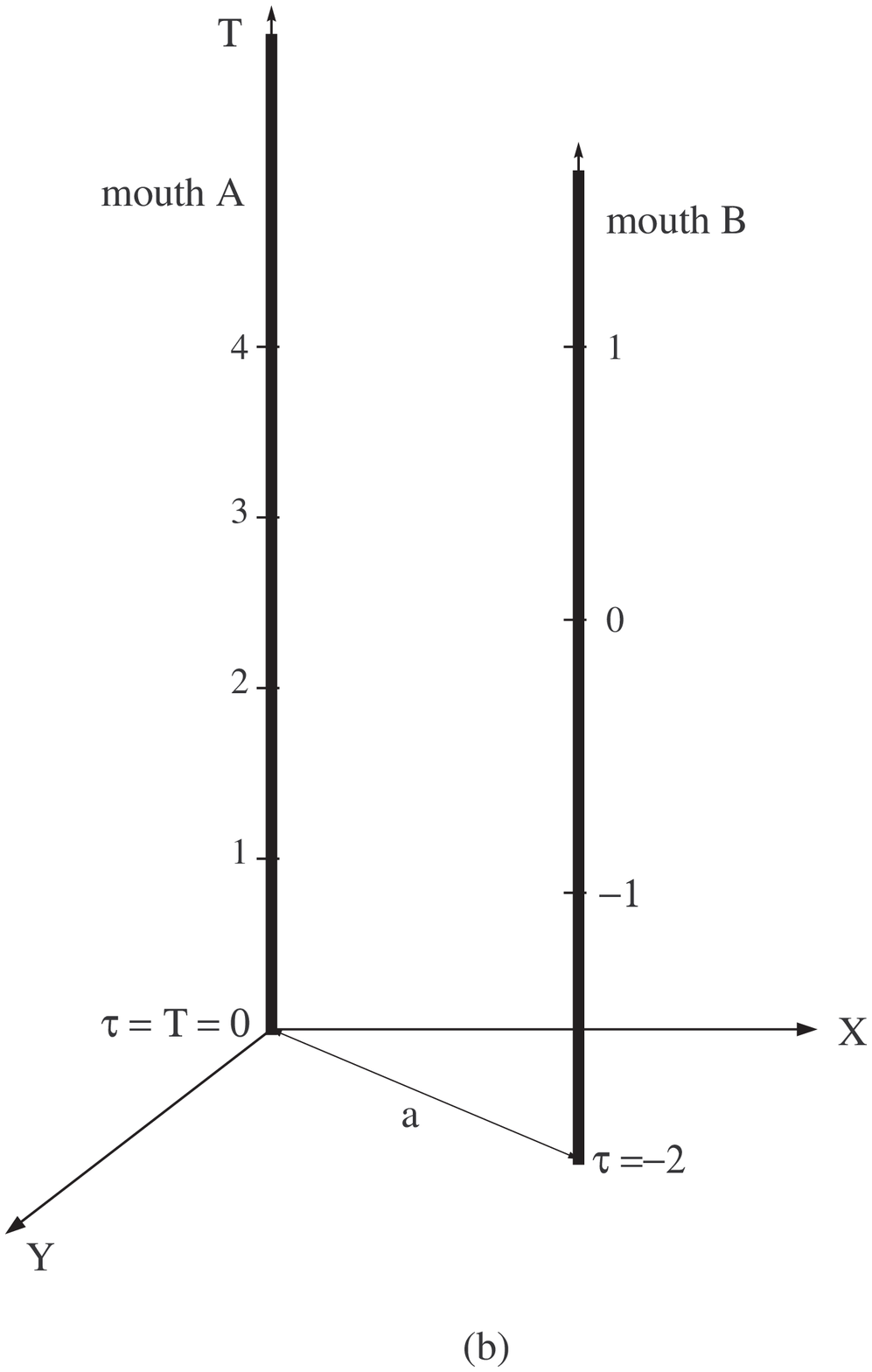}
  \caption[Wormhole spacetimes with closed timelike curves]
  {Depicted are two examples of wormhole spacetimes with closed
  timelike curves. The wormholes tunnels are arbitrarily short, and
  its two mouths move along two world tubes depicted as thick
  lines in the figure. Proper time $\tau$ at the wormhole throat is
  marked off, and note that identical values are the same event as seen
  through the wormhole handle. In Figure $(a)$, mouth $A$ remains at rest,
  while mouth $B$ accelerates from $A$ at a high velocity, then
  returns to its starting point at rest. A time shift is induced
  between both mouths, due to the time dilation effects of special
  relativity. The light cone-like hypersurface ${\it H}$ shown is
  a Cauchy horizon. Through every event to the future of ${\it H}$
  there exist CTCs, and on the other hand there are no CTCs to the past
  of ${\it H}$. In Figure $(b)$, a time shift between both mouths
  is induced by placing mouth $B$ in strong gravitational field.
  See text for details.}
  \label{fig:WH-time-machine}
\end{figure}

To create a time shift using the twin paradox analogue, consider
that the mouths of the wormhole may be moving one with respect to
the other in external space, without significant changes of the
internal geometry of the handle. For simplicity, consider that one
of the mouths $A$ is at rest in an inertial frame, whilst the
other mouth $B$, initially at rest practically close by to $A$,
starts to move out with a high velocity, then returns to its
starting point. Due to the Lorentz time contraction, the time
interval between these two events, $\Delta T_B$, measured by a
clock comoving with $B$ can be made to be significantly shorter
than the time interval between the same two events, $\Delta T_A$,
as measured by a clock resting at $A$. Thus, the clock that has
moved has been slowed by $\Delta T_A-\Delta T_B$ relative to the
standard inertial clock. Note that the tunnel (handle), between
$A$ and $B$ remains practically unchanged, so that an observer
comparing the time of the clocks through the handle will measure
an identical time, as the mouths are at rest with respect to one
another. However, by comparing the time of the clocks in external
space, he will verify that their time shift is precisely $\Delta
T_A-\Delta T_B$, as both mouths are in different reference frames,
frames that moved with high velocities with respect to one
another. Now, consider an observer starting off from $A$ at an
instant $T_0$, measured by the clock stationed at $A$. He makes
his way to $B$ in external space and enters the tunnel from $B$.
Consider, for simplicity, that the trip through the wormhole
tunnel is instantaneous. He then exits from the wormhole mouth $A$
into external space at the instant $T_0-(\Delta T_A-\Delta T_B)$
as measured by a clock positioned at $A$. His arrival at $A$
precedes his departure, and the wormhole has been converted into a
time machine. See Figure \ref{fig:WH-time-machine}.

For concreteness, following the Morris {\it et al}
analysis~\cite{mty}, consider the metric of the accelerating
wormhole given by
\begin{equation}
ds^2=-(1+glF(l)\cos\theta)^2\;e^{2\Phi(l)}\;dt^2+dl^2+r^2(l)\,(d\theta^2
+\sin^2\theta\,d\phi^2)   \,,
    \label{accerelatedWH}
\end{equation}
where the proper radial distance, $dl=(1-b/r)^{-1/2}\,dr$, is
used. $F(l)$ is a form function that vanishes at the wormhole
mouth $A$, at $l\leq 0$, rising smoothly from 0 to 1, as one moves
to mouth $B$; $g=g(t)$ is the acceleration of mouth $B$ as
measured in its own asymptotic rest frame. Consider that the
external metric to the respective wormhole mouths is $ds^2 \cong
-dT^2+dX^2+dY^2+dZ^2$. Thus, the transformation from the wormhole
mouth coordinates to the external Lorentz coordinates is given by
\begin{equation}
T=t\,, \qquad Z=Z_A+l\,\cos\theta\,, \qquad
X=l\,\sin\theta\,\cos\phi \,, \qquad  X=l\,\sin\theta\,\sin\phi
\,,
\end{equation}
for mouth $A$, where $Z_A$ is the time-independent $Z$ location of
the wormhole mouth $A$, and
\begin{equation}
T=T_B+v\gamma \,l\,\cos\theta\,, \qquad
Z=Z_B+\gamma\,l\,\cos\theta\,, \qquad X=l\,\sin\theta\,\cos\phi
\,, \qquad X=l\,\sin\theta\,\sin\phi \,,
\end{equation}
for the accelerating wormhole mouth $B$. The world line of the
center of mouth $B$ is given by $Z=Z_B(t)$ and $T=T_B(t)$ with
$ds^2=dT_B^2-dZ_B^2$; $v(t)\equiv dZ_B/dT_B$ is the velocity of
mouth $B$ and $\gamma=(1-v^2)^{-1/2}$ the respective Lorentz
factor; the acceleration appearing in the wormhole metric is given
$g(t)=\gamma^2\;dv/dt$~\cite{Misner}.

Novikov considered other variants of inducing a time shift through
the time dilation effects in special relativity, by using a
modified form of the metric (\ref{accerelatedWH}), and by
considering a circular motion of one of the mouths with respect to
the other~\cite{Novikov-CTCWH}. Another interesting manner to
induce a time shift between both mouths is simply to place one of
the mouths in a strong external gravitational field, so that times
slows down in the respective mouth. The time shift will be given
by $T=\int_{i}^{f}\,(\sqrt{g_{tt}(x_A)}-\sqrt{g_{tt}(x_A)}\;)\;dt$
~\cite{Visser,frolovnovikovTM}.

\subsubsection{The Ori-Soen time machine}

A time-machine model was also proposed by Amos Ori and Yoav Soen
which significantly ameliorates the conditions of the EFE's
solutions which generate CTCs
\cite{Ori,Soen-Ori1,Soen-Ori2,Olum-SoenOriTM}. The Ori-Soen model
presents some notable features. It was verified that CTCs evolve
from a well-defined initial slice, a partial Cauchy surface, which
does not display causality violation. The partial Cauchy surface
and spacetime are asymptotically flat, contrary to the Gott
spacetime, and topologically trivial, contrary to the wormhole
solutions. The causality violation region is constrained within a
bounded region of space, and not in infinity as in the Gott
solution. The WEC is satisfied up until and beyond a time slice
$t=1/a$, on which the CTCs appear. Ameliorated versions were
recently proposed \cite{Ori2}.

\subsubsection{Warp drive and closed timelike curves}

Within the framework of general relativity, it is possible to warp
spacetime in a small {\it bubblelike} region \cite{Alcubierre}, in
such a way that the bubble may attain arbitrarily large
velocities, $v(t)$. Inspired in the inflationary phase of the
early Universe, the enormous speed of separation arises from the
expansion of spacetime itself. The model for hyperfast travel is
to create a local distortion of spacetime, producing an expansion
behind the bubble, and an opposite contraction ahead of it. See
Section \ref{Alcubierrewarp} for details.

One may consider a hypothetical spaceship immersed within the
bubble, moving along a timelike curve, regardless of the value of
$v(t)$. Due to the arbitrary value of the warp bubble velocity,
the metric of the warp drive permits superluminal travel, which
raises the possibility of the existence of CTCs. Although the
solution deduced by Alcubierre by itself does not possess CTCs,
Everett demonstrated that these are created by a simple
modification of the Alcubierre metric \cite{Everett}, by applying
a similar analysis as in tachyons.

\subsubsection{The Krasnikov tube and closed timelike curves}

Krasnikov discovered an interesting feature of the warp drive, in
which an observer in the center of the bubble is causally
separated from the front edge of the bubble. Therefore he/she
cannot control the Alcubierre bubble on demand. Krasnikov proposed
a two-dimensional metric \cite{Krasnikov}, which was later
extended to a four-dimensional model \cite{EverettCTC}, as
outlined in Section \ref{Sec:Krasnikov}. One Krasnikov tube in two
dimensions does not generate CTCs. But the situation is quite
different in the 4-dimensional generalization. Using two such
tubes it is a simple matter, in principle, to generate CTCs. The
analysis is similar to that of the warp drive, so that it will be
treated in summary.

Imagine a spaceship travelling along the $x$-axis, departing from
a star, $S_1$, at $t=0$, and arriving at a distant star, $S_2$, at
$t=D$. An observer on board of the spaceship constructs a
Krasnikov tube along the trajectory. It is possible for the
observer to return to $S_1$, travelling along a parallel line to
the $x$-axis, situated at a distance $\rho_0$, so that $D\gg
\rho_0\gg 2\rho_{max}$, in the exterior of the first tube. On the
return trip, the observer constructs a second tube, analogous to
the first, but in the opposite direction, i.e., the metric of the
second tube is obtained substituting $x$ and $t$, for $X=D-x$ and
$T=t-D$, respectively in Eq. (\ref{4d-Krasnikov-metric}). The
fundamental point to note is that in three spatial dimensions it
is possible to construct a system of two non-overlapping tube
separated by a distance $\rho_0$.

After the construction of the system, an observer may initiate a
journey, departing from $S_1$, at $x=0$ and $t=2D$. One is only
interested in the appearance of CTCs in principle, therefore the
following simplifications are imposed: $\delta$ and $\varepsilon$
are infinitesimal, and the time to travel between the tubes is
negligible. For simplicity, consider the velocity of propagation
close to that of light speed. Using the second tube, arriving at
$S_2$ at $x=D$ and $t=D$, then travelling through the first tube,
the observer arrives at $S_1$ at $t=0$. The spaceship has
completed a CTC, arriving at $S_1$ before it's departure.

\section{Conclusion}

In this Chapter we have considered two particular spacetimes that
violate the energy conditions of general relativity, namely,
traversable wormholes and ``warp drive'' spacetimes. It is
important to emphasize that these solutions are primarily useful
as ``gedanken-experiments'' and as a theoretician's probe of the
foundations of general relativity. They have also been important
to stimulate research in the issues of the energy condition
violations, closed timelike curves and the associated causality
violations and ``effective'' superluminal travel.

In the Introduction, we have outlined a review of wormhole physics
dating from the ``Einstein-Rosen'' bridge, the revival of the
issue by Wheeler with the introduction of the ``geon'' concept in
the 1960s, the full renaissance of the subject by Thorne and
collaborators in the late 1980s, culminating in the monograph by
Visser, and detailed the issues that branched therefrom to the
present date. In Section \ref{secII}, we have presented a
mathematical overview of the Morris-Thorne wormhole, paying close
attention to the pointwise and averaged energy condition
violations, the concept of the Quantum Inequality and the
respective constraints on wormhole geometries, and the
introduction of the ``Volume Integral Quantifier'' which in a
certain measure quantifies the amount of energy condition
violating matter needed to sustain wormhole spacetimes. We then,
treated rotating wormholes and evolving wormholes in a
cosmological background, focussing mainly on the energy condition
violations, and in particular, on the inflating wormhole and the
respective evolution in a flat FRW universe.

In Section \ref{secIII}, we have considered the superluminal
``warp drive'' spacetimes, which also violate the energy
conditions and generate closed timelike curves with slight
modifications to the spacetime metric. In particular, we have
analyzed the Alcubierre and Nat\'{a}rio spacetimes, paying close
attention to the energy condition violations, reviewing the
application of the Quantum Inequality and the superluminal
features of these spacetimes, in particular, the appearance of
horizons. The discovery that the outer frontal regions of the warp
bubble is not in causal contact with a hypothetical observer
placed in the center on the bubble, and thus cannot be controlled
on demand, inspired the solution known as the Krasnikov spacetime.
This spacetime in two and four dimensions were also briefly
analyzed.

Using linearized theory, we have also considered a more realistic
model of the warp drive spacetime where the warp bubble interacts
with a finite mass spaceship. We have tested and quantified  the
energy conditions to first and second order of the warp bubble
velocity. By doing so we have been able to safely ignore the
causality problems associated with ``superluminal'' motion, and so
have focussed attention on a previously unremarked feature of the
``warp drive'' spacetime.  If it is possible to realize even a
weak-field warp drive in nature, such a spacetime appears to be an
example of a ``reaction-less drive''. That is, the warp bubble
moves by interacting with the geometry of spacetime instead of
expending reaction mass, and the spaceship (which in linearized
theory can be treated as a finite mass object placed within the
warp bubble), is simply carried along with it. We have verified
that in this case, the ``total amount'' of energy condition
violating matter (the ``net'' negative energy of the warp field)
must be an appreciable fraction of the positive mass of the
spaceship carried along by the warp bubble. This places an
extremely stringent condition on the warp drive spacetime, namely,
that for all conceivably interesting situations the bubble
velocity should be absurdly low, and it therefore appears unlikely
that, by using this analysis, the warp drive will ever prove to be
technologically useful. Finally, we point out that any attempt at
building up a ``strong-field'' warp drive starting from an
approximately Minkowski spacetime will inevitably have to pass
through a weak-field regime. Since the weak-field warp drives are
already so tightly constrained, the analysis of this work implies
additional difficulties for developing a ``strong field'' warp
drive.

Finally, in Section \ref{secIV}, we have analyzed some solutions
to the Einstein field equation that generate closed timelike
curves. Far from attempting at an exhaustive search of spacetimes
that possess closed timelike curves, we have focussed on two
particularly notorious properties of these geometries, namely,
solutions with a tipping over of the light cones due to a rotation
about a cylindrically symmetric axis and solutions that violate
the energy conditions of general relativity. Closed timelike
curves are troublesome as they apparently violate causality, and
it is not clear that even an eventual theory of quantum gravity
will provide us with an answer. However, as stated by Kip Thorne,
time travel in the form of closed timelike curves, is more than a
justification for theoretical speculation, it is a conceptual tool
and an epistemological instrument to probe the fundamental
foundations of general relativity and to extract some eventual
clarifying views.

\acknowledgments The author was funded by Funda\c{c}\~{a}o para a
Ci\^{e}ncia e a Tecnologia (FCT)--Portugal through the grant
SFRH/BPD/26269/2006.

\end{document}